\documentclass[12pt]{article}

\RequirePackage[OT1]{fontenc}
\RequirePackage{elsst-book}
\RequirePackage[authoryear]{natbib}
\RequirePackage{amsthm}
\RequirePackage{amssymb}

\usepackage{graphicx}
\usepackage{rotate}
\usepackage{psfrag}
\usepackage{a4}
\usepackage[]{natbib}

\theoremstyle{plain}

\theoremstyle{definition}

\theoremstyle{remark}

\begin{document}

\newcommand{\bq}{\ensuremath{{\bf q}}}
\renewcommand{\cal}{\ensuremath{\mathcal}}
\newcommand{\bqp}{\ensuremath{{\bf q'}}}
\newcommand{\bbq}{\ensuremath{{\bf Q}}} 
\newcommand{\bp}{\ensuremath{{\bf p}}}
\newcommand{\bpp}{\ensuremath{{\bf p'}}}
\newcommand{\bk}{\ensuremath{{\bf k}}}
\newcommand{\bx}{\ensuremath{{\bf x}}}
\newcommand{\bxp}{\ensuremath{{\bf x'}}}
\newcommand{\by}{\ensuremath{{\bf y}}}
\newcommand{\byp}{\ensuremath{{\bf y'}}}
\newcommand{\bxpp}{\ensuremath{{\bf x''}}}
\newcommand{\rmd}{\ensuremath{{\rm d}}}
\newcommand{\intk}{\ensuremath{{\int \frac{d^3\bk}{(2\pi)^3}}}}
\newcommand{\intq}{\ensuremath{{\int \frac{d^3\bq}{(2\pi)^3}}}}
\newcommand{\intqp}{\ensuremath{{\int \frac{d^3\bqp}{(2\pi)^3}}}}
\newcommand{\intp}{\ensuremath{{\int \frac{d^3\bp}{(2\pi)^3}}}}
\newcommand{\intpp}{\ensuremath{{\int \frac{d^3\bpp}{(2\pi)^3}}}}
\newcommand{\intx}{\ensuremath{{\int d^3\bx}}}
\newcommand{\intxp}{\ensuremath{{\int d^3\bx'}}}
\newcommand{\intxpp}{\ensuremath{{\int d^3\bx''}}}
\newcommand{\drho}{\ensuremath{{\delta\rho}}}
\newcommand{\rhoh}{\ensuremath{{\widehat{\rho}}}}
\newcommand{\fh}{\ensuremath{{\widehat{f}}}}
\newcommand{\phih}{\ensuremath{{\widehat{\phi}}}}
\newcommand{\thetah}{\ensuremath{{\widehat{\theta}}}}
\newcommand{\etah}{\ensuremath{{\widehat{\eta}}}}
\newcommand{\0}{\ensuremath{{(\bk,\omega)}}}
\newcommand{\x}{\ensuremath{{(\bx,t)}}}
\newcommand{\xp}{\ensuremath{{(\bx',t)}}}
\newcommand{\xtp}{\ensuremath{{(\bx',t')}}}
\newcommand{\xtpp}{\ensuremath{{(\bx'',t')}}}
\newcommand{\xttpp}{\ensuremath{{(\bx'',t'')}}}
\newcommand{\xtpn}{\ensuremath{{(\bx',-t')}}}
\newcommand{\xtppn}{\ensuremath{{(\bx'',-t')}}}
\newcommand{\xn}{\ensuremath{{(\bx,-t)}}}
\newcommand{\xpn}{\ensuremath{{(\bx',-t)}}}
\newcommand{\xppn}{\ensuremath{{(\bx',-t)}}}
\newcommand{\xpp}{\ensuremath{{(\bx'',t)}}}
\newcommand{\xxp}{\ensuremath{{(\bx,t;\bx',t')}}}
\newcommand{\Crr}{\ensuremath{{C_{\rho\rho}}}}

\newcommand{\Crf}{\ensuremath{{C_{\rho f}}}}
\newcommand{\Crt}{\ensuremath{{C_{\rho\theta}}}}
\newcommand{\Cff}{\ensuremath{{C_{ff}}}}
\newcommand{\Cffh}{\ensuremath{{C_{f\fh}}}}
\newcommand{\Ct}{\ensuremath{{\dot{C}}}}
\newcommand{\Ctt}{\ensuremath{{\ddot{C}}}}
\newcommand{\Crrp}{\ensuremath{{\dot{C}_{\rho\rho}}}}
\newcommand{\Crfp}{\ensuremath{{\dot{C}_{\rho f}}}}
\newcommand{\Crtp}{\ensuremath{{\dot{C}_{\rho\theta}}}}
\newcommand{\Cffp}{\ensuremath{{\dot{C}_{ff}}}}
\newcommand{\Crrpp}{\ensuremath{{\ddot{C}_{\rho\rho}}}}
\newcommand{\thetab}{\ensuremath{{\overline{\theta}}}}
\newcommand \be  {\begin{equation}}
\newcommand \bea {\begin{eqnarray} \nonumber }
\newcommand \ee  {\end{equation}}
\newcommand \eea {\end{eqnarray}}

\title{\bf How markets slowly digest changes in supply and demand}

\author{Jean-Philippe Bouchaud$^{*}$, J. Doyne Farmer$^{\dagger,\star}$, Fabrizio Lillo$^{\ddagger,\dagger}$}

\address{{$*$ Science \& Finance, Capital Fund Management, 6 Bvd Haussmann},
{75009 Paris, France}\\
$\dagger$Santa Fe Institute, 1399 Hyde Park Rd., Santa Fe NM 87505, USA\\
$\star$LUISS Guido Carli, Viale Pola 12, 00198 Roma, Italy\\
$\ddagger$Dipartimento di Fisica e Tecnologie Relative, Viale delle Scienze I-90128, Palermo, Italy\\$~$\\
}

\maketitle

\bigskip

\bigskip

\bigskip

$ $\\
$ $\\
$ $\\
$ $\\

\begin{abstract}
In this article we revisit the classic problem of tatonnement in price formation from a microstructure point of view, reviewing a recent body of theoretical and empirical work explaining how fluctuations in supply and demand are slowly incorporated into prices.  Because revealed market liquidity is extremely low, large orders to buy or sell can only be traded incrementally, over periods of time as long as months.  As a result order flow is a highly persistent long-memory process. Maintaining compatibility with market efficiency has profound consequences on price formation, on the dynamics of liquidity, and on the nature of impact.  We review a body of theory that makes detailed quantitative predictions about the volume and time dependence of market impact, the bid-ask spread, order book dynamics, and volatility.  Comparisons to data yield some encouraging successes. This framework suggests a novel interpretation of financial information, in which agents are at best only weakly informed and all have a similar and extremely noisy impact on prices. Most of the processed information appears to come from supply and demand itself, rather than from external news.  The ideas reviewed here are relevant to market microstructure regulation, agent-based models, cost-optimal execution strategies, and understanding market ecologies.
\end{abstract}

\newpage

\setcounter{tocdepth}{3}
\tableofcontents

\section{Introduction}

\label{intro}

\subsection{Overview}

In this article we discuss the slow process by which markets ``digest" fluctuations in supply and demand, reviewing a body of work that suggests a new approach to the classic problem of {\it tatonnement}, the dynamic process through which markets seek to reach equilibrium.  The foundation of this approach is based on several empirical observations about financial markets, the most important of which is long-memory in the fluctuations of supply and demand.  This is exhibited in the placement of trading orders, and corresponds to long term, slowly decaying positive correlations in the initiation of buying vs. selling.  It is observed in all the stock markets studied so far at very high levels of statistical significance.  It appears that the primary cause of this is the incremental execution of large hidden trading orders.     The fact that the long-memory of order flow must coexist with market efficiency (at least in a statistical sense) has a profound influence on price formation, causing dynamic adjustments of liquidity that are strongly asymmetric between buyers and sellers.   

This has important consequences for market impact.  (By market impact we mean the average response of prices to trades; liquidity refers to the scale of the market impact)\footnote{
Market impact is closely related to the demand elasticity of price, and is typically measured as the return associated with a transaction as a function of volume.  Liquidity (as we will use it here) measures the size of the price response to a trade of a fixed size, and is inversely proportional to the scale of the impact. If trading a given quantity produces only a small price change, the market is liquid, and if it produces a large price change it is illiquid.}.
We discuss theoretical work predicting the average market impact as a function of both volume and time.   The asymmetric liquidity adjustments needed to maintain compatibility between the long-memory of order flow and market efficiency can equivalently be interpreted in terms of the temporal response of market impact, leading to a slow decay of market impact with time.  

This work also has important consequences about the interpretation and effect of information in financial markets.   In particular, the explanation for market impact that we develop here from the standard view in the finance literature, which holds that the shape of the impact function is determined by differences in the information content of trades.  The body of work reviewed here instead assumes that the impact of trades depends only on their predictability, e.g. that highly predictable trades have little impact, as originally postulated by \cite{Hasbrouck88}.  We argue that this is a much simpler explanation, that produces stronger predictions, is more plausible from a theoretical point of view, and is more in line with what is observed in the data.

The implications of this work range upward from microstructure, i.e. at the level of individual price changes, to patterns of price formation on timescales that can be measured in months.  At the microstructure level this work makes several predictions, such as the relationship between market impact, the bid-ask spread, and volatility.  It also make predictions about the impact of large trades executed over long times, as well as the effect such trades may have in causing clustered volatility.

\subsection{Organization of the paper}

In the remainder of the introduction we discuss the motivation and scope of the work described here, and discuss our approach to creating a theory for market microstructure, which is somewhat unusual within economics.  In Section~\ref{marketStructure} we discuss the institutional aspects of the markets that form the basis of our empirical studies and define some of the terms that will be used throughout the paper.  In Section~\ref{information} we lay out some of the main conceptual issues, discussing the concept of information in finance and its relationship to market efficiency, and the important role that liquidity (or more accurately, the lack of liquidity) plays in forming markets.  We critique so-called ``noise trader" models and present an alternative point of view.  In Section~\ref{longMemory} we present the empirical evidence for long-memory in order flow, develop a theory for its explanation based on strategic order splitting, and present evidence that this theory is correct.  In Section~\ref{impactVolume} we describe the different types of impact and review the empirical evidence.  In Section~\ref{impactTemporal} we develop a theory for market impact for each of the different types of impact.  In Section~\ref{spread} we discuss the problem of explaining the behavior of the bid-ask spread, and compare theory and empirical observations.  Section~\ref{liquidityVolatility} discusses the close relationship between liquidity and volatility.  In Section~\ref{orderBook} we discuss models for the order book, which can be regarded as models for liquidity.  Section~\ref{execution} discusses the problem of trading in an optimal manner in order to minimize execution costs.  Section~\ref{ecologyemp} describes recent attempts to characterize trading ecologies of market behavior at short time scales, and Section~\ref{conclusions} presents our conclusions.

\subsection{Motivation and scope}

Markets are places where buyers meet sellers and the price of exchanged goods are fixed.  As originally observed by Adam Smith, during the course of this apparently simple process remarkable things happen.  The information of  diverse buyers and sellers, which may be too complex and textured for any of them to fully articulate, is somehow incorporated into a single number, the price.  
One of the powerful achievements of economics has been the formulation of simple and elegant equilibrium models that attempt to explain the end results of this process without going into the details of the mechanisms through which prices are actually set.  

There has always been a nagging worry, however, that there are many situations in which broad-brush equilibrium models that do not delve sufficiently deeply into the process of trading and the strategic nature of its dynamics, may not be good enough to tell us what we need to know; to do better we will ultimately have to roll up our sleeves and properly understand how prices change from a more microscopic point of view.  Walras himself worried about the process of {\it tatonnement}, the way in which prices settle into equilibrium.   While there are many proofs for the existence of equilibria, it is quite another matter to determine whether or not a particular equilibrium is stable under perturbations, i.e. whether prices that are initially out of equilibrium will be attracted to an equilibrium.   This necessarily requires a more detailed model of how prices are actually formed.  There is a long history of work in economics seeking to create models of this type (see e.g. \cite{Fisher83}), but many would argue that this line of work was ultimately not very productive, and in any case it has had little influence on modern mainstream economics.

A renewed interest in dynamical models that incorporate market microstructure is driven by many factors.  In finance, one important factor is growing evidence suggesting that there are many situations where equilibrium models, at least in their current state, do not explain the data very well.  Under the standard model prices should change only when there is news, but there is growing evidence that news is only one of several determinants of prices, and that prices can stray far from fundamental values (\cite{Campbell89,Roll84b,Cutler89,Bouchaud08})\footnote{
See \cite{Engle05} for a dissenting view.}.
Doubts are further fueled by a host of studies in behavioral economics demonstrating the strong boundaries of rationality.  Taken together this body of work calls into question the view that prices always remain in equilibrium and respond instantly and correctly to new information.

The work reviewed here argues that trading is inherently an incremental process, and that because of this, prices often respond slowly to new information.  The reviewed body of theory springs from the recent empirical discovery that changes in supply and demand constitute a long-memory process , i.e. that its autocorrelation function is a slowly decaying power law (\cite{Bouchaud04,Lillo03c}).  This means that supply and demand flow in and out of the market only very gradually, with a persistence that is observed on timescales of weeks or even months.  We argue that this is primarily caused by the practice of order splitting, in which large institutional funds split their trading orders into many small pieces.   Because of the heavy tails in trading size, there are long periods where buying pressure dominates, and long periods where selling pressure dominates.  The market only slowly and with some difficulty ``digests" these swings in supply and demand.  In order to keep prices efficient, in the sense that they are unpredictable and there are not easy profit making opportunities, it has to make significant adjustments in liquidity.  Understanding how this happens leads to a deeper understanding of many properties of market microstructure, such as volatility, the bid-ask spread, and the market impact of individual incremental trades.  It also leads to an understanding of important economic issues that go beyond market microstructure, such as how large institutional orders impact the price, and in particular how this depends on both the quantity traded and on time.  It implies that the liquidity of markets is a dynamic process with a strong historical dependence.
 
The work reviewed here by no means denies that information plays a role in forming prices, but it suggests that for many purposes this role is secondary.   In the last half of the twentieth century finance has increasingly emphasized information and de-emphasized supply and demand.  The work we review here brings forward the role of fluctuations in supply and demand, which may or may not be exogenous.  As we view it, it is useful to begin the story with a quantitative description of the properties of fluctuations in supply and demand.  Where such fluctuations come from doesn't really matter; they could be driven by rational responses to information or they could simply be driven by a demand for liquidity.  In either case, they imply that there are situations when order arrival can be very predictable.  Orders contain a variable amount information about the hidden background of supply and demand.  This affects how much prices move, and therefore modulates the way in which information is incorporated into prices.  This notion of information is internal to the market.  In contrast to the prevailing view in market microstructure theory, there is no need to distinguish between ``informed" and ``uninformed" trading to explain important properties of markets, such as the shape of market impact functions or the bid-ask spread\footnote{
Since modern continuous double auction markets are typically anonymous, it is hard to see how the identity of traders could play an important role in the size of the price response to trades.  See the discussion in Section~\ref{information}.}.

We believe the work here should have repercussions on a wide gamut of questions: 
\begin{itemize}
\item At a very fundamental level -- how do we understand why prices move, how information is reflected in prices and what fixes the value of the volatility?; 
\item At the level of price statistics -- what are the mechanisms leading to price jumps and volatility clustering?; 
\item At the level of market organisation -- what are the optimal trading rules to insure immediate liquidity and orderly flow to investors?; 
\item At the level of agent-based models -- what are the microstructural ingredients necessary to build a realistic agent-based model of price changes?
\item At the level of trading strategies and execution costs -- what 
are the consequence of empirical microstructure regularities on transaction costs and implementation shortfall?
\end{itemize}
We do not wish to imply that these questions will be answered here, only that the word described here bears on all of them.  We will return to discuss the implications in the conclusions.

\subsection{Approach to model building}

Because this work reflects an approach to model building that many economists will find unfamiliar, we first make a few remarks to help the reader understand the philosophy behind this approach.  Put succinctly, our view is that the enormous quantities of data that are now available fundamentally change the approach one should take to building economic theories about financial markets.

In recent years the computer has made it possible to automate markets, has enabled an explosion in the amount of recorded data, and makes it possible to analyze unprecedented quantities of information.  Financial instruments are now typically standardized, stable entities that are traded day after day by many thousands of market participants.  Modern electronic markets offer an open and transparent environment that allows traders across the world to get real time access to prices, and most importantly for science, makes it possible to save detailed records of human decision making.  The past decades have seen an explosion in the volume of stored data. For example, the total volume of data related to US large caps on, say Oct 2, 2007, was 57 million lines, approximately a gigabyte of stored data.  The complete record of world financial activity is more than a terabyte per day.  Each market has slightly different rules of operation, making it possible to compare market structures and how they affect price formation, and most important of all, to look for patterns of behavior that are common across all market structures.  The system of world financial markets can be viewed as a huge social science experiment in which profit-seekers spend large quantities of their own money to collect enormous quantities of data for the pleasure of scientists.

With so much data it becomes possible to change the style in which economics is done.   When one has only a small amount of noisy data statistical testing must be done with great care and it is difficult to test and reject competing models unless the differences in their predictions are very large.
%
Data snooping is a constant worry.  In contrast, with billions of data points if an effect is not strong enough to leap out of the noise it is unlikely to be of any economic importance.   
Even more important is the effect this has on the development and testing of theories.  With a small data set inference requires strong priors.  This fosters an approach in which one begins with pure theory and tests the resulting models only after they are fully formulated; there is less opportunity to let the data speak for itself.  Without great quantities of data it is difficult to test a theory in a fully quantitative manner, and so predictions of theories are typically qualitative.

The work reviewed in this article takes advantage of the size of financial data sets by strongly coupling the processes of model formation and data analysis.   This begins with a search for empirical regularities, i.e.  behaviors that under certain circumstances follow consistent quantitative laws.   Even though such effects do not have the consistency of the laws of physics, one can nonetheless be somewhat more ambitious than simply trying to establish a set of ``stylized facts".  An attempt is made to describe regularities in terms that are sufficiently quantitative so that theories have a clear target, and can thus sensibly make strongly falsifiable  predictions.  A key goal of such theories is of course to understand the necessary and sufficient conditioned for regularities.  

The approach for building theory described here is phenomenological.  That is, it does not attempt to derive everything from a set of first principles, but rather simply tries to connect diverse phenomena to each other in order to simplify our description of the world.   Many economists will be uncomfortable with this approach because it often lacks ``economic content", i.e. the theories that are developed do not invoke utility maximization.  In this sense this body of work lies somewhere between pure econometrics and what is usually called a theory in micro-economics.  Even though the models infer properties of agent behavior, and connect them to market properties such as prices, there is no attempt to derive the results from theories that maximize preferences, contenting ourselves with weaker assumptions, such as market efficiency.  Given all the empirical problems surrounding the concept of utility, we view this as a strength rather than a weakness.  

The work described here is still in an early stage, and is very much in flux; many of these results are quite new and indeed our own view is still changing as new results appear.

\section{Market structure}
 \label{marketStructure}
 
All of the work described here is based on results from studying stocks from the London, Paris, New York (NYSE and NASDAQ) and Spanish stock markets.  These markets differ in their details, but they all do at least half of their trading (and in some cases all their trading) through a continuous double auction.  ``Auction" indicates that participants may place quotes (also called {\it orders}) stating the quantities and prices at which they are willing to trade; ``continuous" indicates that they can update, cancel or place new quotes at any time, and ``double" indicates that the market is symmetric between buyers and sellers\footnote{
There are some small exceptions to symmetry between buying and selling, such as the uptick rule in the NYSE, but these are relatively small effects.}.  

There are some important differences in the way these markets are organized.  The NYSE is unusual in that each stock has a designated specialist who maintains and clears the limit order book.  The specialist can see the identity of all the quotes and can selectively show them to others.   The specialist can also trade for his own account but has regulatory obligations to ``maintain an orderly market".  The London Stock Exchange in contrast, has no specialists.  It is completely transparent in the sense that all orders are visible to everyone, but completely anonymous in the sense that there is no information about the identity of the participants, and such information is not disclosed even to the counterparties of transactions.  The Spanish Stock Market is unusual in that membership codes for quotes are publicly displayed.  Thus these exchanges are generically similar but have their own peculiar characteristics.

Markets also differ in the details of the types of orders that can be placed.  For example, the types of orders in the London Stock Exchange are called ``limit orders",``market orders with limiting price", ``fill-or-kill", and ``execute \& eliminate".  In order to treat these different types simply and in a unified manner we simply classify them based on whether an order results in an immediate transaction, in which case we call it an \emph{effective market order}, or whether it leaves a limit order sitting in the book, in which case we call it an \emph{effective limit order}. Marketable limit orders (also called crossing limit orders) are limit orders that cross the opposing best price, and so result in at least a partial transaction. The portion of the order that results in an immediate transaction is counted as an effective market order, while the non-transacted part (if any) is counted as an effective limit order; thus in this case a single action by the participant gets counted as two separate orders.  Note that we typically drop the term ``effective", so that e.g. ``market order" means ``effective market order".   Similarly a limit order can be removed from the book for many reasons, e.g. because the agent changes her mind, because a time specified when the order was placed has been reached, or because of the
institutionally-mandated $30$ day limit on order duration. We will
lump all of these together, and simply refer to them as
``cancellations".

In addition to continuous double auctions the London Stock Exchange has what is called the off-book market and the New York Stock Exchange has what is called the upstairs market.  These are both bilateral exchanges, in which members of the exchange can interact in person or via telephone to arrange transactions.  Such transactions are then reported publicly at a later time.  With exceptions noted in the text all the results obtained are from the continuous markets.

\section{Information, liquidity \& efficiency}\label{information}

The aim of this section is to motivate the empirical study of microstructure in a broader economic context, that of the information content of prices and the mechanisms that can lead to market efficiency.  We discuss several fundamental questions concerning how markets operate.  The discussion here sets the stage for the detailed quantitative investigations that we report in the following sections.  Since one of our main subjects here is market impact, we review and critique the standard model for market impact, which is based on informed vs. uninformed trading.

\subsection{Information and fundamental values}

It is often argued that there is a fundamental value for stocks, correctly  known to at least
some informed traders, who buy underpriced stocks and sell overpriced stocks.  By doing so they make
a profit and, through the very impact of their trades,
drive back the price toward its fundamental value. This mechanism is the cornerstone of the theory of efficient markets, and is often used to justify why prices are unpredictable.   In such a framework, the fundamental value of a stock can only change with unanticipated news. The scenario is then the following:  A piece of news becomes available, market participants work out how this changes the price of the stock, and trade accordingly. After a (supposedly fast) phase of `tatonnement', the price converges to its new equilibrium value, and the process repeats itself.  To explain deviations from this picture one can add a suitable fraction of uninformed trades to add some high frequency noise.  

Is this picture fundamentally correct to explain why prices move and to account for the observed value 
of the volatility? Judging from the literature, it looks as if a majority of academics still believe that this 
story is at least a good starting point (but see, for example \citet{Lyons01}). Recent empirical microstructure studies open the way to testing in detail the basic tenets and the overall plausibility of the standard equilibrium picture. We hope to convince the reader that the story is in fact significantly different.  That is, we argue that an alternative way of looking at events provides superior explanatory power based on a simpler set of hypotheses.  Before discussing at length the microstructural evidence for a change of paradigm, we would like at this stage to make several general comments that will be 
relevant below; first on the very notion of fundamental value and information, and second on
various orders of magnitude and time scales involved in the problem.  

Is the fundamental value of a stock or a currency a valid concept, in the sense that it can
be computed, at least as a matter of principle, with arbitrary accuracy with all information known at time $t$?
The number of factors influencing the fundamental value of a company or of a currency is so large that there should be, at the very least, an irreducible intrinsic error. All predictive tools used by traders, either based on economic ratios, earning forecasts, etc., are based on statistical models detecting trends or mean-reversion, are obviously noisy and sometimes even biased. For example, financial experts are known to be on the whole rather bad at
forecasting the next earning of a company (see e.g. \cite{Guedj05}). News are often ambiguous and not easy to interpret.  But if we accept the idea of an intrinsically noisy fundamental value with some band of width $\Delta$ within which the price can almost freely wander, the immediate question is: how large is the uncertainty $\Delta$? Is it very small, say $10^{-3}$ in relative terms, or quite a bit larger, say $100\%$, as suggested by \cite{Black86}?   If Black is right (which we tend to believe) and the uncertainty in the fundamental value is large, then the information contained in a trade is noisy, and the amount of information contained in any given trade is necessarily small.  Analysis of price impact makes it clear that the standard deviation of impacts is very large compared to their mean, suggesting that this is indeed the case.

\subsection{Market efficiency}

Market efficiency is one of the central ideas in finance and appears in many guises.  A standard definition of market efficiency (in the informational sense) is that the current price should be the best predictor of future prices, i.e. that prices should be a martingale.  Another closely related notion is arbitrage efficiency, which in its weakest form states that it should not be possible to make a profit without taking risks; in a stronger form it says that two strategies with the same risk should make the same profits, at least once their usefulness for inclusion in a portfolio is taken into account.  Steve Ross, among others, has advocated that market efficiency (rather than equilibrium) should be the core postulate for financial theory (\cite{Ross04}).  

We agree with this point of view, at least in so far as it does not imply believing in allocative efficiency, i.e. that 
prices correctly reflect the underying value of the assets. Strictly speaking a market is allocatively efficient if it is Pareto optimal, in the sense that there is no alternative allocation of prices and holdings that makes someone better off without making someone worse off.   This is related to whether or not prices are set at their ``proper" values.  It is entirely possible to imagine a market in which prices are unpredictable and yet in which there is no sense in which prices are set correctly.  That is, once we depart from neoclassical equilibrium a market might be informationally efficient yet allocatively inefficient.

A closely related point is that there are two very different possible explanations for market efficiency.  (1) The standard view in economics is that perfect efficiency reflects perfect information processing.  Traders process each new bit of information as it arrives and prices immediately go to their new equilibrium values.   This is, however, by no mean the only explanation for why prices can be a martingale.  An obvious alternative is the standard one that explains randomness in many other fields, such as fluid turbulence:  (2) Markets are too complicated to be predictable.  Under this explanation prices move randomly because investor behavior is complicated, based on many hidden factors, so to an external observer it is ``as if" individual investors are just flipping coins.  The correct explanation is likely to be a mixture of both effects.  On one hand markets are inherently complicated, but on the other hand, whatever predictability is left over is substantially removed by arbitrageurs.  Under this synthetic view, which we take here, one can simply associate an impact with trades, treat all investors as more or less the same, and adjust the expected impact as needed to preserve efficiency based on factors that derive from the predictability of trades.

Finally we want to emphasize that while we believe that market efficiency is a very useful concept and provides an excellent starting point for developing theories, it is inherently contradictory, and is at best an approximation.  Markets can only be informationally efficient at first order but must necessarily be inefficient at second order.   This was originally pointed out by Milton Friedman, who noted that without informed traders to push prices in the right direction, there is no reason that markets should ever be efficient.  If markets were truly efficient then informed traders should make the same profits as anyone else, and there would be no motivation for them to remain in the market.  Thus markets cannot be fully efficient.  

Even if for many purposes it can be a good approximation to assume that they markets are efficient, there are other situations where deviations from efficiency can be quite important.  Understanding how markets evolve from inefficient to efficient states, predicting the necessary level of deviations from efficiency that must persist in steady state, and understanding their role in how markets function remains an area of investigation that is still largely not understood. This is relevant for our discussions on incorporating information into prices because when we speak about information we must have traders to process that information and trade based on it.  It is precisely the market impact of these traders that moves prices.  Thus while on one hand market impact is a friction, it can also be viewed as the factor that maintains efficiency, and so it is essential to properly understand it.

\subsection{Trading and information}

Informational efficiency means that information must be properly incorporated into prices.  Under assumptions of rationality, when all traders have the same information, prices should move more or less automatically, with very little trading (\cite{Milgrom82,Sebenius83}).  But of course that's not true -- people don't have the same information, and even if they did, real people are likely to take different views about what the information means.  The empirical fact that there is so much trading supports this (\cite{Shiller81}).  \cite{Grossman80} developed an equilibrium model in which traders have different information, that shows that in this situation trading and price movements are informative (see also \cite{Grossman89}).  If I know that you are rational, and I know that you have different information than I have, when I see you trade and the price rises I can infer the importance of your information and thus I should change my own valuation.  

Intuitively the problem with this view is that even small deviations from rationality and perfect information can lead to incorrect prices and instabilities in the price process.  Suppose, for example, that you and I both overestimate how much information the other has.  Then when I see you trade I change my valuation too much.  When I see you buy, I also buy, but I buy more than I should.  To make this slightly more quantitative, let the initial price be $p_0$ and suppose that after agent A observes new fundamental information the price rises by $f$, which might or might not be the correct fundamental level.  After agent A trades the new price becomes $p_1 = p_0 + f$.  Agent $B$ sees the price rise by $f$, and assuming that agent $A$ has more information than he really does, he buys and causes the price to rise to $p_2 = p_0 + af$.  Then B sees the price rise more than $f$, so he buys, driving it to $p_3 = p_0 + a^2 f$, and so on.  This process is clearly unstable if $a > 1$.  The agents either need to know the value of $a$ exactly or they need to be able to adapt $a$ based on information that is not contained in the price.  It is difficult to understand how they can do this since by definition if they are not rational, not only do they not have full information, they do not know how much information they have, and they thus cannot know a priori the proper value of $a$.  Under deviations from rationality, deviations from fundamentals are inevitable. For a beautiful model where copy-cats lead to such instabilities, see \cite{Curty06}.

In its extreme version, this is just the kind of scenario that occurs during a bubble (see \cite{Bouchaud98} for an explicit model of this).  Any reasonable investor who lived through the millennium technology bubble experienced this problem.  Even though high prices seemed difficult to rationalize based on values, prices kept going up.  This led many sanguine investors to lose confidence in their own valuations, and to hang on to their shares much longer than they thought was reasonable.  If they didn't do this they experienced losses as measured relative to their peers.  Under this view, bubbles stem from the problem of not knowing how much information price movements really contain, and the feedback effects that occur when most people think they contain more information than they really do.  This point of view differs from that in the standard literature on rational bubbles.  As we argue below, while not entirely different, there are important contrasts between this view and the standard rational expectations/noise trader models.

\subsection{Different explanations for market impact \label{kindsOfImpact}}

Why is there market impact? We will distinguish three possibilities:

\begin{enumerate}
\item
{\it Trades convey a signal about private information.}  This idea, discussed in the previous section, was developed by \cite{Grossman80}.  The arrival of new private information causes trades, which cause other agents to update their valuations, which changes prices.  In this case it is fair to say that trades cause price changes, since even if there happens to be no information, unless this is common knowledge the observation of a trade is still interpreted as information, which causes the price to change.
\item
{\it Agents successfully forecast short term price movements and trade accordingly.}  This can result in measurable market impact even if these agents have absolutely no effect on prices at all.  If an agent correctly forecasts price movements and trades based on this forecast, when this agent buys there will be a tendency for the price to subsequently rise.  In this case causality runs backward, i.e. because the price is about to rise, agents are more likely to trade in anticipation of it, but a trade based on no information will have no effect.
\item
{\it Random fluctuations in supply and demand.}  Even in the standard market clearing framework, if a given agent increases her demand while other agents keep theirs constant, when the market clears that agent buys and the price rises.  Fluctuations in supply and demand can be completely random, unrelated to information, and the net effect regarding market impact is the same.  In this sense impact is a completely mechanical -- or better, statistical, phenomenon.  As we will see in Appendix 1, the meaning of this can be subtle and may depend on the market framework.
\end{enumerate}

All three of these result in identical short term market impact, i.e. a positive correlation of trading volume and price movement, but are
conceptually very different. If some traders really have an information on the ``true'' price at some time in the future (say the end of the day, after the market closes), then the observation of an excess of buy trades allows the market to guess that the price will move up and to change the quotes accordingly (see Section \ref{gm-section} on the Glosten-Milgrom model). In this sense, information is progressively included into prices, as a function of the observed order flow. In this picture, as emphasized in \cite{Hasbrouck07}, ``orders do not {\it impact} prices. It is more accurate to say that orders {\it forecast} prices.'' But if the mechanical interpretation is correct, correlation between price changes and order flow is a tautology. If prices move only because of trades, ``information revelation'' may merely be a self-fulfilling prophecy which would occur even if the fraction of informed traders is zero. The only possible differences between these picutres come about in the temporal behavior of impact, which we will discuss in Section~\ref{impactTemporal}.

\subsection{Noise trader models and informed vs. uninformed trading}

In behavioral finance the problem of irrational investors is typically coped with by introducing ``noise trader" models, in which some agents (the noise traders) are stupid while others are completely rational (\cite{Kyle85,Delong90,Shleifer00}).   Noise trading could be driven by the need for liquidity (here meaning the need to raise capital for other reasons), it could be driven by the desire to reduce risk, or it could be ``irrational behavior", such as trend following.   The assumption is made that such investors lack the skill or information processing ability to collect and/or make full use of information.  The rational investors, in contrast, are assumed to correspond to skilled professionals.   Their trading is perfect, in the sense that they know everything.  Examples of what they must know includes the strategies of all the noise traders,  and the fraction of capital traded by noise traders as opposed to rational investors.  In such models prices can deviate from fundamental values due to the action of the noise traders and the desire of the rational agents to exploit them as much as possible, but the rational agents always keep them from deviating too much.  In such models the rational traders make ``informed" trades while the noise traders make ``uninformed" trades.

There are several conceptual problems with noise trader models that are clear {\it a priori}.  No one can seriously dispute that traders must have different levels of skill, but is the noise trader approach the right way to model this?  While it might be fine to model a continuum of skill levels as ``low" and ``high", the idea of identifying the ``high" level with perfect rationality postulates a level of skill at the top end that is difficult to imagine.   The panoply of strategies used by real traders is large, and financial professionals (and even private investors) are sufficiently secretive about what they do, that it is difficult to imagine that even the most skilled traders could fully understand everyone else's behavior.  

Another problematic issue is the operational problem of measuring information.   For example, under the theory that urgency is a proxy for informativeness, empirical work on the subject has often defined an informed trade as one that is executed by a market order, and defined an uninformed trade as one that is represented by a limit order.  This goes against the fact that many of the most successful hedge funds make extensive use of limit orders\footnote{
We are basing this on personal conversations with market practitioners and so can only place a lower bound:  We know many people working in many sophisticated trading operations and all of them at least partially use limit orders.  We suspect the correct statement is that ``most", or even ``nearly all" successful hedge funds use limit orders for at least a substantial part of their trading.}.
The only alternative is to use data that contains information about the identity of the agents making the trades.  Such data does indeed confirm that professionals perform better than amateurs (\cite{Barber04}), but as mentioned above, there is no demonstration that this means they are rational, and other than stating that professionals make larger profits, it is impossible to determine whether or not  professionals are good enough to be considered rational.  (On this point, see also \cite{Odean99}). 

\subsection{A critique of the noise trader explanation of market impact}

One of the most important questions to ask about any theory is what it explains that is not explained by a simpler alternative.   Noise trader models have been proposed to explain why market impact is a concave function of trading volume.  The empirical evidence for this will be discussed in detail in Sections~\ref{impactVolume} and \ref{impactTemporal}; in any case, it is a well established empirical fact that the market impact as a function of trade size has a decreasing derivative.  This can be alternatively stated as saying that the price impact per share decreases with the total size of the trade. The standard explanation for this is that it is due to a mixture of informed and uninformed trading. If more informed traders use small trade sizes and less informed traders use large trade sizes, then small trades will cause larger price movement per share than large trades.

There are several problems with this theory:
\begin{itemize}
\item
A concave market impact function is observed in all markets that have been studied, including many such as the London and Paris markets where the identity of orders is kept completely anonymous.  This rules out any explanation that depends on trades made by some agents communicating more information about prices than others, and leaves only the possibility that some traders are able to anticipate short term price movements better than others -- see the discussion in Section~\ref{kindsOfImpact}.
\item
The model is unparsimonious in the sense that it requires the specification of a function that states the information that traders have as a function of the size of the trades that they use.
\item
The model is difficult to test because it requires finding a way to specify the information that different groups of traders have {\it a priori}.  One proposal is to do this based of the average profits of different groups of traders.  This suffers from the problem that the time horizon for market impact is typically very different than the time horizon on which traders attempt to make profits.  A fund manager who intends to a buy a stock and hold it for three years may make the trade to take up that position in a single day.  While this manager might have great skill in predicting stock price movements on a three year time horizon she may have no skill at all on a daily horizon.  Thus in a large fraction of cases, even under large variations in trader skill, impact may have little correlation with profits.
\item
If it is indeed advantageous to use small trades then since this is a trivial strategy, one would think that everyone would quickly adopt it and the effect would disappear.  In fact in the last five years or so there has been a huge increase in algorithmic trading, in which brokers automatically execute large trades for clients by cutting them up into small pieces.  One would therefore think that in modern times the concavity should have diminished or even been eliminated entirely.  There is little evidence for this - the impact continues to be highly concave.
\end{itemize}

Thus we have argued above that the theory is implausible, but even more important, that it makes weak and untestable predictions.  The prediction of concavity requires a set of assumptions that are complicated to specify and impossible to measure.  The predictions are purely qualitative, and it is not obvious how they might be extended to other properties of impact, such as its temporal behavior.


\subsection{The liquidity paradox -- price are not in equilibrium}

We will argue here that liquidity is an important intermediary that modulates the effect of information.  We are defining liquidity in terms of the size of the price response to a trade of a given size.   High liquidity implies a small price response.  Since trades carry information, then if the size of trades in response to a given level of information remains constant, as the liquidity varies the price response to information varies with it. 

Under the assumption that trading is an intermediate step in the response of prices to information, one can conceptually decompose it into two terms.
\begin{equation}
\label{liquidityAndVolume}
\Delta p = \mathcal{T}(I)/\lambda,
\end{equation}
where $\lambda$ is the liquidity and $\mathcal{T}(I)$ is the response of trades to information $I$.  Variations in the liquidity do not tell the full story about the response of prices to information -- to do that one would also need to understand $\mathcal{T}(I)$.  Nonetheless, as we argue here, the effects of varying liquidity are substantial, and they have the huge advantage of being easily measurable\footnote{
Of course liquidity may also depend on information, and indeed in Section~\ref{impactTemporal} we will develop this connection.}.
In contrast, since information is difficult to measure, $\mathcal{T}(I)$ is difficult to measure. Furthermore, the above equation should be interpreted rather loosely: as we shall see below, 
impact is in fact neither linear nor permanent. 


A very important empirical fact that is crucial to understand how markets operate is that even ``highly liquid" markets are in fact not that liquid. Take for example a US large cap stock. Trading 
is extremely frequent: a few thousand trades per day, adding up to a daily volume of roughly $0.1$ -- $ 1 \%$ of total market capitalization. Trading is even more frantic on futures and Forex markets. However, the volume of buy or sell limit orders typically available in the order book at a given instant of time is quite small: only the order of $1 \%$ of the traded daily volume, i.e. $10^{-4}$ -- $10^{-5}$ of the market cap for stocks.  Of course, this number has an intraday pattern and fluctuates in time, and it can reach much smaller values in liquidity crises. 

The fact that the outstanding liquidity is so small has an immediate consequence: trades must be fragmented. The theoretical motivations for this were originally discussed by \cite{Kyle85}.  It is not uncommon that investment funds want to buy large fractions of a company, often exceeding several percent.  One possibility is to arrange upstairs block trades but this lacks transparency and can be costly.  If trading occurs through the continuous double auction market, the numbers above suggest  that  to buy $1 \%$ of a company requires at least the order of $100$ -- $1000$ individual trades.  This is under the unrealistic assumption that each individual trade completely empties the order book -- more realistically each trade consumes only a fraction of the order book, and the number of trades is even larger.  But since a thousand trades corresponds to roughly the whole daily liquidity, it is clear that these trades have to be diluted over several days, since otherwise the market would be completely destabilized.  Thus an informed trader cannot use her information immediately, and has to trade into the market little by little.

But why is liquidity, as measured by the number of standing limit orders, so low? Both for similar and for opposite reasons. Too large a buy limit order from an ``informed" trader would give her away and raise the price of the sellers.  Too large a limit order from a liquidity provider would put her at risk of being `picked-off' by an informed trader.  There is a kind of hide and seek liquidity game taking place in organized markets, where buyers and sellers face a paradoxical situation:   Both want to have their trading done as quickly as possible, but both try not to show their hands and reveal their intentions.  As a result, markets operate in a regime of vanishing {\it revealed liquidity}, but large {\it latent liquidity}; this leads to a series of empirical regularities that we will present below. 

From a conceptual point of view, however, the most important conclusion of this qualitative discussion is that prices are typically not in equilibrium, in the traditional Marshall sense.  That is, the true price is very different than it would be if it were set so that supply and demand were equal as measured by the honest intent of the participants, as opposed to what they actually expose.  As emphasized above, the volume of individual trades is much smaller than the total demand or supply at the origin of the trades. This means that there is no reason to believe that instantaneous prices are equilibrium, efficient prices that reflect all known information. Much of the information is necessarily latent, withheld due to the small liquidity of the market, and only slowly revealing itself (see \cite{Lyons01} for similar ideas).  At best, the notion of equilibrium prices can only make sense over a long time scale; high frequency prices are necessarily soiled by a significant amount of noise.

\subsection{Time scales and market ecology}\label{ecology}

Consider again the case of a typical US large cap stock, say Apple, which (as of Nov. 2007) had a daily turnover of around 8B\$.  There are on average 6 transactions per second, and on the order of 100 events per second affecting the order book. These are extremely small time scales compared to the typical time for public news events, in which a hot stock like Apple might be mentioned by name every few hours during a period of fast information arrival.  Perhaps surprisingly, the number of large jumps in price is much higher.  For example, if we define a jump as a one minute return exceeding three standard deviations, there are the order of ten such jumps per day, reflecting the very heavy tailed distribution of high frequency returns (\cite{Bouchaud08}).  More often than not such jumps occur in the absence of any identified news.  It is obviously a particularly important question to understand the origin and the mechanisms leading to these jumps.  The difference
between the frequency of news and the frequency of jumps already suggests that something else must be at work, such as fluctuations in liquidity, that may have little or nothing to do with external news entering the market.

What is the typical time scale of the round trip trades of investors? This depends very much on the style of trading -- traditional long-only funds have investment horizons on the scale of years, while more aggressive long-short stat-arbs have time scales of weeks or days, sometimes even shorter. Some empirical results support the existence of a broad spectrum of investment horizons 
(see Section \ref{lmf} and \ref{hidordemp}). The optimal frequency of a trading strategy is a trade-off between the expected profit and the friction and transaction costs. Since the fraction of costs grows with the trading volume large investment funds cannot trade too quickly. This, again, is directly related to the small prevailing liquidity. So it is reasonable to think that information based trading decisions have intrinsic frequencies ranging from a few days to years.  As we have already emphasized, for large investors a single decision may generate many more trades:  A decision to buy or sell may persist for days to months, generating a series of small trades. Again, the important message is that low frequency, large volume investment decisions imply high frequency, small volume trades, and that high frequency prices cannot be equilibrium prices.
 
There is however a potentially viable high-frequency strategy called {\it market-making} that consists in providing instantaneous liquidity to buyers and sellers and trying to eke out a profit from the bid-ask spread. As originally shown by \cite{Glosten85}, the difficulty is to avoid losses due to adverse price moves.  Since market makers are offering either to buy or to sell, they are giving a free option to others who might have better information.  The profitability of market-making strategies depends both on the spread, which is beneficial, and on the long-term impact of trades, which is detrimental.  This intuition will be made more precise and discussed in detail
in section~\ref{spread}. On some exchanges market-making is institutionalized, with certain obligations and advantages bestowed to those who take the burden
of providing liquidity.  Markets have become, however, more and more electronic, with an open orderbook allowing each investor to behave either as
a liquidity provider by posting limit orders, or as a liquidity taker by issuing market orders. Depending on market conditions (for example,
the instantaneous value of the spread), investors can choose either type of order. There is both empirical and anecdotal evidence that some
players implement high frequency, market-making like strategies. This contribution to order flow is often described as ``uninformed''. 
Although this flow differs from longer horizon trades, which are supposed to be economically informed, these market-making strategies routinely 
use sophisticated short-term prediction tools and exploit any profitable high-frequency signals.
The above simplified separation of market participants into two broad classes, speculators/liquidity hunters that trade at medium to low frequencies 
and market-makers/liquidity providers at high frequencies is both realistic and useful to understand the {\it ecology} of financial markets 
(\cite{Handa96, Farmer02,Wyart06,Lillo07}). The competition between these two categories of traders allows one to make sense of a number of empirical facts, we believe much more usefully than noise trader models.
In Section \ref{ecologyemp} we present some recent empirical results on the characterization of a market ecology.

\subsection{The volatility puzzle \label{volPuzzle}}

Given that markets are ecological systems where participants have a broad distribution of time horizons from seconds to years, it is perhaps not surprising to see long-memory effects in financial markets, e.g. in trading volume, volatility and order flow. What is {\it a priori} surprising, however, is that despite the fact that high frequency prices cannot possibly be in equilibrium because of lack of liquidity, and despite the fact that it should take time for the market to interpret a piece of news and agree on a new price, the average volatility is remarkably constant on a wide range of different timescales.  As measured by autocorrelation,  prices are remarkably efficient down to the fastest timescales.  We have argued that news arrival happens on much longer timescales.  Given that this is true, how can prices remain so efficient, at least with respect to linear models, even on very fast time scales?

One possible explanation for this is might be that public information as evidenced on news feeds is only a small part of the available information.  Instead, suppose there are many sources of private information, which agents are continually processing.  As they make their decisions they trade.  Given that heavily traded stocks average many trades per second, this would suggest that a truly staggering amount of information is being processed.  We find this explanation implausible.

The alternative is to that there is an information processing cascade from fundamental information on slow time scales to technical information on fast time scales.  As we have argued above, fundamental information enters at a relatively slow rate, and then is processed and incorporated into prices.  Under this view high frequency strategies play an important role.   Such strategies do not process external information directly, but rather serve the role of digesting that information and keeping the price stream unpredictable.  Such strategies are not processing fundamental information, but rather are acting as technical trading strategies, processing information contained in the time history of prices, trading volume and other information that is completely internal to the market.  The ability to substitute information in a time history for state information is well supported in dynamical systems theory (\cite{Packard80,Takens81,Casdagli91}).  Thus we argue that in the ecology of financial markets, high frequency strategies are fed by lower frequency strategies through an information cascade from longer to shorter timescales, and from fundamental to technical information, finally resulting in white noise on all scales. 

This also suggests that microstructural effects may influence the value of the volatility, as suggested by \cite{Lyons01}, e.g.  ``microstructure implications may be long-lived'' and ``are relevant to macroeconomics''. We will comment on the relation between microstructure and volatility in Section~\ref{liquidityVolatility}. This relation is also relevant for the regulator, who might attempt to alter the microstructural organisation of markets in order to reduce the volatility.

\subsection{The Kyle model}

A classic model noise trader model for market impact, which is a natural a point of comparison  is due to \cite{Kyle85}.  This model assumes that there are three types of traders:  Noise traders who make random trades, market makers who set prices in order to guarantee efficiency, and an insider who has access to superior information.  Under the most general version of the model the noise traders and insider trade continuously from a starting time until a final liquidation time, at which point everyone is paid the liquidation price for their holdings.  The insider has superior information about the final liquidation price $p_\infty$, and an infinite bank, which she uses to maximize  profits at the expense of the noise traders.

The optimal amount that the investor should trade is easily found to be proportional to the difference $p_\infty-p_t$.  With the assumption of a linear and permanent impact, in Kyle's notation the price evolution is given by:
\be\label{Kyle}
p_{t+1}-p_t = \lambda \left[\Phi_t  + \xi_t\right] + \eta_t; \qquad \Phi_t = \beta [p_\infty - p_t] 
\ee
where $\Phi_t$ is the signed demand of the investor, $\lambda$, $\beta$ are coefficients, 
and $\xi_t$ is the noise trader demand coming from all other market participants, and $\eta_t$ a noise term accounting for possible changes of prices not induced by trading (news, etc.). The above equation can easily be solved, and leads to an exponential relaxation of the initial price towards $p_\infty$ plus a bounded noise term.

The impact in this model can be regarded as essentially mechanical.  There is an apparently permanent change in price which is linearly proportional to the total amount that the noise traders and insider trade.  We say ``apparently permanent" because, since there is a final liquidation time, what happens past this point is undefined.  Note that in this model the price will move toward $p_\infty$ regardless of whether it is the correct price; all that is necessary is that insider {\it believe} it is the correct price.  A random assignment of beliefs about $p_\infty$ will result in a corresponding random set of impacts.  Thus, referring to our discussion of the different explanations for market impact in Section~\ref{kindsOfImpact}, while the Kyle model is built in the spirit of explanation (1), that trades convey a signal about private information, it is equally consistent with (3), random fluctuations in supply and demand.

The assumption of a final liquidation price can naively lead to erroneous conclusions. For example, this model suggests that one can easily manipulate the price. However, in the absence of a liquidation price where a transaction with a counterparty can be realized without impact, things are not so trivial: as soon as the investor wants to close his position, he will again mechanically revert the price back to its initial value and take losses.  (To see this note that in a single round trip the investor will buy at a high price and sell at the original price).  The above impact model, Eq. (\ref{Kyle}), although very often used in agent based models of price fluctuations (two of us have also developed similar ideas, i.e.  \cite{Bouchaud98,Farmer02})), is far too naive to represent the way real markets operate, at least at the tick by tick level.

Thus we see that while the Kyle model provides a good starting point for understanding why there should be market impact, and why it is useful to trade into a position incrementally, it falls short of making realistic predictions about impact.  We feel that the key elements that need to be extended are: (1) Removing the final liquidation price, (2) eliminating the infinite bank of the insider and replacing it with the more realistic assumption of a finite, predetermined trading size, and (3) eliminating the distinction between the insider and the noise trader.  The aim of the following sections is to explain in detail how to construct a model generalizing Eq. (\ref{Kyle}), using an approach based on robust facts observed in empirical data and consistency arguments. We will find that impact is in general {\it non-linear} and {\it transient} -- or equivalently, as explained in section \ref{historyDependent},  {\it history dependent}.  It is only after a properly 
defined ``coarse graining'' procedure that such an impact model can possibly make sense.

\section{Large fluctuations and long-memory of order flow}
\label{longMemory}

From a mechanical point of view price formation process is the outcome of (i) the flow of orders arriving in the market and (ii) the response of prices to individual orders.   Since price dynamics are reasonably well described by a Brownian motion one might naively assume that this would be true for order flow as well.  In fact this is far from the truth.  As we will explain in detail in this section, order flow is a highly autocorrelated long-memory process.  As a consequence, to maintain market efficiency the price response to orders must strongly depend on the past history of order flow.  This has profound conseqences for the way in which markets incorporate information.

\subsection{Empirical evidence for long-memory of order flow}

\begin{figure}[ptb]
\begin{center}
\includegraphics[scale=0.33]{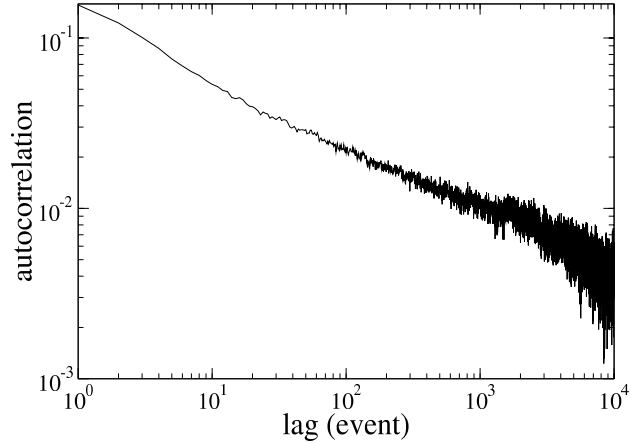}
\end{center}
\caption{Autocorrelation function of the time series of signs of orders that result in immediate trades (effective market orders) for the stock Vodafone traded on the London Stock Exchange in the period May 2000 - December 2002, a total of $5.8 \times 10^5$ events.}
\label{vod-acf}
\end{figure}

We discuss here the statistical properties of order flow by considering the time series of signs of orders. Specifically, consider the symbolic time series obtained in event time by replacing buy orders with $+1$ and sell orders with $-1$, irrespective of the volume of the order.  We reduce these series to $\pm 1$ rather than analyzing the signed series of order sizes directly in order to avoid problems created by the large fluctuations in order size\footnote{
Fluctuations in order size are heavy tailed and have long-memory themselves, so statistical averages based on them converge only slowly.  The essential behavior is captured by the series of signs.}. 
This reduction can be done for market orders, limit orders, or cancellations, all of which show very similar behavior\footnote{
Long memory is also observed if the signs of all orders, including both limit and market orders, are taken together.  In contrast, if one assigns a cancellation of a buy order a negative sign, corresponding to the fact that the only nonzero price movements it can produce are downward, then the combined sequence of signs for market orders, limit orders, and cancellations does not show long-memory.}.
We denote with $\epsilon_i$ the sign of the $i^{th}$ market order.  Figure \ref{vod-acf}  shows the sample autocorrelation function of the market order sign time series for Vodafone (VOD) in the period 1999-2002 in double logarithmic scale.  
The figure shows that the autocorrelation function for market order signs decays very slowly. The autocorrelation function is still above the statistical noise level even after $10^4$ transactions, which for this stock corresponds to roughly $10$ days. This result indicates that if one observes a buy market order now, based on this information alone there is some non-vanishing predictability of the market order signs two weeks from now. 

We also note that the autocorrelation function shown in Fig. \ref{vod-acf} is roughly linear in a double logarithmic scale over more than $4$ decades\footnote{
The noisy behavior for large $\tau$ comes from the fact that for large lags the statistical errors are remaining roughly constant while the signal decreases, so the relative size of the fluctuations becomes larger.}.
This suggests that a power-law relation $C_\tau \sim \tau^{-\gamma}$ might be a reasonable description for the sample autocorrelation function\footnote{
$f(y) \sim g(y)$ means that there exists a constant $K\ne0$ such that $\lim_{y\to\infty} f(y)/g(y) = K$.}.

Stochastic processes for which the autocorrelation function decays asymptotically as a power-law with an exponent smaller than one are called long-memory processes \cite{Beran94}. 
A precise definition of long-memory processes can be given in terms of the autocovariance function $\Gamma_\tau$. We define a process as long-memory if in the limit
$\tau\to \infty$
\begin{equation}
\Gamma(\tau) \sim \tau^{-\gamma} L(\tau),
\label{LMdef}
\end{equation} 
where $0<\gamma<1$ and $L(\tau)$ is a slowly varying
function\footnote{$L(x)$ is a slowly varying function (see Embrechts et al., 1997)
if $\lim_{x \to \infty} L(tx)/L(x) = 1$  $\forall t$.  In the definition above, and
for the purposes of this paper, we are considering only positively
correlated long-memory processes. Negatively correlated
long-memory processes also exist, but the long-memory processes we will
consider in the rest of the paper are all positively correlated.} at
infinity.  The degree of long-memory is given by the
exponent $\gamma$; the smaller $\gamma$, the longer the memory. 
The integral of the autocovariance (or autocorrelation) function of a long memory process diverges.
Long-memory can also be discussed in terms of the Hurst exponent $H$,
which is simply related to $\gamma$.  For a long-memory process
$H=1-\gamma/2$ or $\gamma=2-2H$. Short-memory processes have $H =
1/2$, and the autocorrelation function decays faster than $1/\tau$.  A
positively correlated long-memory process is characterized by a Hurst
exponent in the interval $(0.5,1)$.  The use of the Hurst exponent is
motivated by the relationship to diffusion properties of the
integrated process.  For normal diffusion, where by definition the
increments do not display long-memory, the standard deviation
asymptotically increases as $t^{1/2}$, whereas for diffusion processes
with long-memory increments, the standard deviation asymptotically
increases as $t^H L(t)$, with $1/2 < H < 1$, and $L(t)$ a slow-varying
function. In econometrics of financial time series many variables have the long-memory property. For example, it is widely accepted that the volatility of prices (\cite{Ding93}) and stock market trading volume (\cite{Lobato00}) are long memory processes. Models of long-memory processes include fractional Brownian noise
(\cite{Mandelbrot68}) and the ARFIMA process introduced by
\cite{Granger80} and \cite{Hosking81}.

As Figure \ref{vod-acf} suggests, and as discovered by \cite{Bouchaud04} and \cite{Lillo03c}, order flow is also described by a long-memory process.  The long-memory of order flow is very robust, and is consistently observed for every stock that has so far been examined.  Lillo and Farmer tested for long-memory in a panel of 20  highly capitalized stocks traded at the London Stock Exchange using Lo's modified R/S test (\cite{Lo91}), which is known to be a strict test for long memory.  They found that even on short samples, in most cases the hypothesis of long-memory could not be rejected.   The value of $H$ observed in the London Stock Exchange was generally about $H \approx 0.7$, which corresponds to $\gamma=0.6$. Bouchaud {\it et al.} (2004) measured a larger interval of $\gamma$ values in the Paris Stock Exchange, ranging from $0.2$ to $0.7$.  Long-memory has also measured long-memory for an assortment of stocks in the NYSE (these results are mentioned in \cite{Lillo03c}, but have not been published in detail).

\subsection{On the origin of long-memory of order flow}

What causes long-memory in order flow? The presence of persistent time correlations in the order flow suggests two possible classes of explanations:  The first type of explanation is that this is a property of the order flow of each investor, independent of the behavior of other investors, as proposed by \cite{Lillo05b}. The second type of explanation is that investors herd in their trading though an imitation process that involves an interaction between them, as proposed by \cite{Lebaron07}.  It is of course possible that both effects operate at once, but in any case one would like to know their relative magnitude.  

We believe that the evidence gathered so far strongly favors the first explanation.  More explicitly, we believe the dominant cause is the strategic behavior of large investors who split their orders into many small pieces and execute them incrementally.  The evidence from this comes from two sources.  One is the agreement of the properties of the order flow with theory, and the other is additional evidence based on data that gives information about the identity of participants.  We summarize both of these here.

\subsection{Theory for long-memory in order flow based on strategic order splitting}\label{lmf}

\cite{Lillo05b} have hypothesized  that the cause of the long-memory of order flow is a delay in market clearing. To make this clearer, imagine that a large investor decides to buy ten million shares of a company. It is unrealistic for her to simply state her demand to the world and let the market do its job.  There are unlikely to be sufficient sellers present, and even if there were, revealing the intention to buy a large quantity of shares will very likely push the price up substantially.  Instead the large investor keeps her intentions as secret as possible and trades incrementally over an extended period of time, possibly through intermediaries.  As already discussed, the strategic reasons for doing this were made clear by \cite{Kyle85}, who investigated a model in which an insider with information about a final liquidation price tries to maximize profits.  In simple terms, the motivation is that by splitting the hidden order into small pieces the investor is able to execute much of the hidden order at prices that do not reflect the full price movement that it will eventually cause. 

Our perspective differs from Kyle's in that we assume that the size of the order, which we call the {\it hidden order}, is given at the outset when the initial trading decision is made.  We believe that the size of such orders is largely determined by the fund manager a priori, and is influenced by a combination of the funds under management and the timescale of the strategy, which is typically much longer than the time scale for completing the trade.  The other key differences is that we do not assume a final liquidation price, and we do not make a distinction between informed traders and noise traders.  When taken together these differing assumptions create key differences in the predictions of the model in comparison with Kyle.

In several studies based on data giving the identity of hidden orders, about a third of the dollar value of such institutional trades took more than a week to complete (\cite{Chan93,Chan95,Vaglica07}).  This conflicts with the standard model of market clearing presented in textbooks, which assumes that agents fully state their supply and demand and that prices are set so that supply and demand are evenly matched.   The fact that large orders are kept secret and executed incrementally implies that at any given time there may be a substantial imbalance of buyers and sellers.  Effective market clearing is delayed, by variable amounts that depend on fluctuations in the size and signs of the unrevealed hidden orders. 

We now describe a recently proposed simple model of order flow that postulates the independence of trading activity of investors, and which is able to reproduce the long-memory properties of order flow (\cite{Lillo05b}).  In the simplest version of the model, assume that at any time there are $K$ hidden orders present in the market. Initially the size $V$ of these hidden orders are drawn from a distribution $P(V)$ and the sign $\epsilon_i$ is randomly chosen. For simplicity we assume that $V$ is an integer number. We indicate with $V_i(t)$ the volume of hidden order $i$ that has not yet been traded at time $t$. At each timestep $t$ an existing hidden order $i$ is chosen at random with uniform probability, and a unit volume of that order is traded, so that $V_i(t+1) = V_i(t) - 1$.  This generates a revealed order of unit volume and sign $\epsilon_i$.  A hidden order $i$ is removed if $V_i(t+1) = 0$, i.e. when the hidden order is completely traded. When this happens a new hidden order is created with a random sign and a new size. 

It is possible to find  a closed expression for the autocorrelation function of the trade sign $C_\tau$ as a function of the hidden order size distribution $P(V)$. The asymptotic behavior of $C_\tau$ can be obtained through a saddle point approximation. If the hidden order size is asymptotically Pareto distributed, i.e.
\begin{equation}
P(V)\sim\frac{\alpha}{V^{\alpha+1}},
\end{equation}
then the autocorrelation function of order sign behaves asymptotically as (\cite{Lillo05b})
\begin{equation}
C_\tau\sim\frac{K^{\alpha-2}}{\alpha}\frac{1}{\tau^{\alpha-1}}.
\end{equation}
Thus the model makes the falsifiable prediction that the exponent $\gamma$ of the power-law asymptotic behavior of the autocorrelation of order sign is determined by the exponent $\alpha$ of the power-law asymptotic behavior of the hidden order size distribution through 
\begin{equation}
\alpha = \gamma + 1.
\end{equation}
Since we observe that $\gamma\simeq 0.5$, this model predicts that $\alpha=1.5$.

It is worth noting that \cite{Lillo05b} also introduced a more general model where the number of hidden orders is not constant in time. Specifically, at each time $t$ a new hidden order is generated with probability $0 < \lambda < 1$ if $K(t) > 0$, or probability one if $K(t) = 0$. Although this model is not solved analytically, numerical simulation shows that the relation between the exponent of the autocorrelation of order sign and the exponent of order size distribution is the same as in the simpler model where the number of hidden orders is fixed.  
 
This model for the origin of correlation in order flow is in principle empirically testable. The main difficulty arises from the lack of large and comprehensive databases of the hidden orders of investors.  There are two ways to check the consistency of the theory.
The first one is to compare the distribution of trade sizes in block markets to the autocorrelation function of order signs in order book markets.  In block markets trades are made bilaterally and the identity of counterparties is known.  Brokers do not like order splitting and strongly discourage it.  Thus block markets can be considered a crude proxy for observing the distributional properties of hidden orders\footnote{
The exception is that it is possible to split an order and trade with multiple brokers.}. 
In the next section we discuss evidence that suggest that block trade volume is indeed asymptotically power law distributed with an exponent $\alpha\simeq 1.5$.
For comparison\footnote{
The error bars in computing both $\gamma$ and $\alpha$ are substantial, as can be seen by computing them for sub-samples of the data, and the close agreement observed by \cite{Lillo05b} between $\gamma$ and $\alpha - 1$ is probably fortuitious.  Unfortunately there are still not good statistical methods for assigning confidence intervals for exponents of power laws, particularly when the observations have long-memory, but the errors can be roughly assessed by examining sub-samples.}
the average measured values of $\gamma$ for LSE stocks is $\gamma = 0.57$, close to $\hat{\gamma} = 0.59$ as predicted by $\hat{\gamma} = \alpha -1$. 
The second supporting evidence comes from a study of the Spanish stock exchange by \cite{Vaglica07} who have inferred hidden orders using data with membership codes. This study will be discussed in Section \ref{hidordemp}.  

\subsection{Evidence based on exchange membership codes}

Empirical testing is difficult due to the fact that it is not easy to collect data on the behavior of individual investors.  Nonetheless, partial information about the identity of participants can be obtained by making use of data that identifies the broker or the member of the exchange who executes the trade, which we will simply call the {\it membership code}.  There are many stock markets, such as the LSE, the Spanish Stock Exchange, the Australian Stock Exchange, and the NYSE, where it is possible to obtain data containing this information. It is important to stress that knowing the membership code is not the same as knowing the individual participant, since the member may either trade on its own account or may act as a broker for other trades, or may do both at once.  Nonetheless, several recent papers have demonstrated that it is possible to extract useful information about the identity of individual traders using such information, e.g. showing that there are consistent behaviors that are persistent in time associated with particular membership codes, that such behaviors can be organized into a taxonomic tree, and that it is possible to detect the presence of large institutional trades (\cite{Lillo07,Zovko07,Vaglica07}).

\begin{figure}[tbp]
\begin{center}
\includegraphics[scale=0.5]{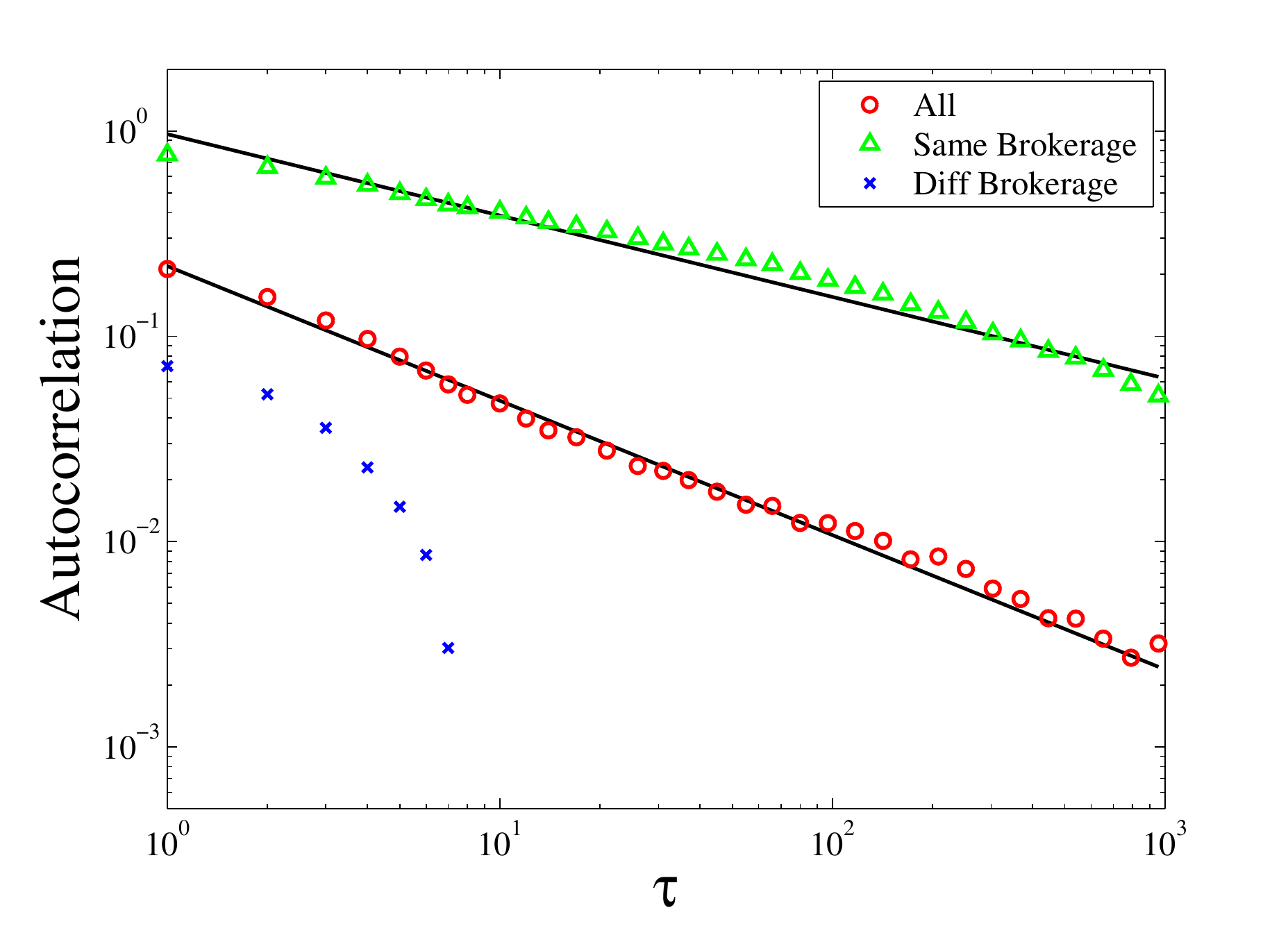}
\end{center}
\caption{Autocorrelation of signs vs. transaction lag for transactions with same membership code, different membership code, and all transactions irrespective of membership code, plotted on double logarithmic scale. The investigated stock is AstraZeneca (AZN) traded at LSE in the period 2000-2002.}
\label{longMemoryFig}
\end{figure}

Gerig et al. have used membership codes of the London Stock Exchange to test the hypothesis of the theory presented in \cite{Lillo05b}.  The autocorrelation function of market order signs is computed by considering realized orders placed by the same membership code or by different membership codes separately. Figure \ref{longMemoryFig} shows the autocorrelation function of market order signs with the same membership code, different membership codes, and all transactions irrespective of membership code. 
The red circles are the autocorrelation function irrespective of the membership code and, as anticipated above, it well fitted by a power law. When only transactions with the same membership code are considered (green triangles) the autocorrelation is still power law with a slightly smaller exponent. Moreover for a fixed lag $\tau$ the autocorrelation function with the same membership code is one order of magnitude larger than the autocorrelation function irrespective of the membership code. Finally, when only transactions with different membership code are considered the autocorrelation function decays very rapidly to zero and it is clearly not consistent with power law behavior.   Under the assumption that most investors use only a few brokers to execute a given hidden order, this plot strongly supports the hypothesis that the long memory of signs is due to the presence of investors that place many revealed orders of the same sign and that there is no clear sign of herding behavior among different investors. It is in principle possible that herding happens between investors using the same broker, but not between investors with different brokers; however the reasons why this would occur are unclear and it seems implausible that it could explain such a dramatic difference.

\subsection{Evidence for heavy tails in volume}\label{heavyvol}

The theory developed above makes it clear that the distribution of trading volume play a key role in shaping many properties of the market, including the long-memory of order flow, which we will show in turn has important consequences for market impact.  In recent years there has been a debate about the statistical properties of trading volume. This is partly due to the fact that markets have different structures and one should be careful in specifying which volume is considered in the analysis.
\cite{Gopikrishnan00} originally observed that volume of trades at the NYSE are asymptotically power law distributed. Specifically, they claimed that for large volumes the probability distribution scales as
\begin{equation}
P(V>x)\sim x^{-3/2}.
\label{halfcubic}
\end{equation}
This law has been termed the ``half cubic" law. The NYSE, as many other financial markets, employs two parallel markets which provide alternative methods of trading, called the on-book or "downstairs" market, and the off-book or "upstairs" market. 
Orders in the on-book market are placed publicly but anonymously and execution is completely automated. The off-book market, in contrast, operates through a bilateral exchange mechanism, via telephone calls or direct contact of the trading parties. The anonymous nature of the on-book market facilitates order splitting, i.e. large orders are split in smaller pieces and traded incrementally. On the other hand the off-book market is a block market, where large orders can be traded in a single transaction. The NYSE data used by \cite{Gopikrishnan00} includes a mixture of order book trades and block trades. Since the typical size of block trades is much larger than the size of orders traded in the order book, the size of block trades dominates the tail of volume distribution. 

\begin{figure}[ptb]
\begin{center}
\includegraphics[scale=0.33,angle=-90]{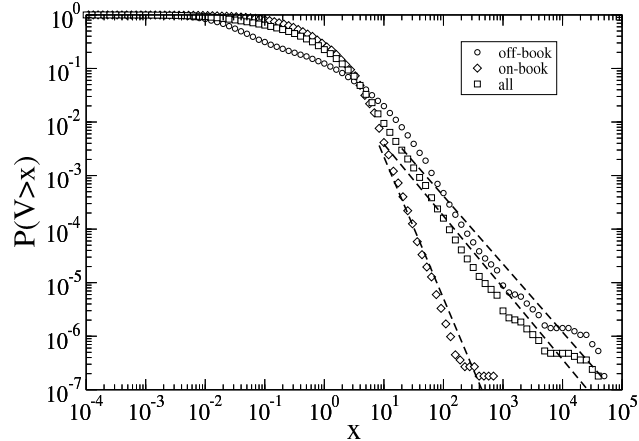}
\end{center}
\caption{Volume distributions of off-book trades (circles), on-book trades (diamonds), and the aggregate of both (squares). We show this for a collection of $20$ different stocks, normalizing the volume of each by the mean volume before combining. The dashed black lines have the slope found by the Hill estimator and are shown for the largest one percent of the data. Adapted from \cite{Lillo05b}.}
\label{voldistfig}
\end{figure}

This can be seen more clearly in a market (or database) where it is possible to separate block trades from order book trades.  In figure~\ref{voldistfig} (from \cite{Lillo05b}) we show the cumulative distribution function of trading volume of off-book trades, on-book trades, and the aggregate of both for a collection of $20$ LSE stocks.
The distribution of block trades is consistent with the power-law hypothesis of Eq.~\ref{halfcubic} with an exponent  close to $1.5$, whereas the distribution of order book trades is not consistent with the half-cubic law, and instead has a much thinner tail (see also \cite{Farmer04} and \cite{Plerou04}).

\section{Summary of empirical results for diverse types of market impact}
\label{impactVolume}

The relation between the transacted volume and the consequent expected price shift is called the {\it price impact} or alternatively the {\it market impact} function.  Letting $R$ be a price return associated with a trade of size $V$, the market impact a time $l$ after the trade occurred is
\[
\mathcal{I}(V, l) = E[ R | V, l].
\]
For many purposes it is useful to separate the dependence on volume from the dependence on time.    One can make the hypothesis that the impact function can be written as a product of two functions, i.e.
\[
\mathcal{I}(V, t) = \mathcal{S}(V) \mathcal{R}(l)
\]
In this section we will primarily discuss the dependence on volume, saving the discussion of time dependence for Section~\ref{impactTemporal}.  

We are intentionally being vague at this stage about the definition of the return $R$ and the volume $V$; defining these more precisely is one of the main points of this chapter.  The way in which the market impact behaves depends on the market structure as well as on what one means by ``return" and ``volume".   Many studies have empirically investigated market impact with a range of different results; we will argue that in many cases these differences stem from differences in what is being studied. The important distinctions that should be made are:

\begin{itemize}
\item First of all, one can consider market impact of an individual transaction vs. an aggregate of many transactions. Aggregation here means that the market impact is conditioned on a given number of trades or to a given interval of time.  We will discuss the volume dependence of individual impact in Section~\ref{impact1} and the time dependence in Sections~\ref{temporal1} - \ref{historyDependent}, and we will study the properties of aggregate impact in Sections~\ref{empiricalAggregate} and \ref{aggregatemod}.  
\item
A second important aspect is the type of market exchange where the transactions take place. As we have said above, most financial markets have upstairs or block markets as well as downstairs or orderbook markets.  In the downstairs market trades are made by placing orders in a limit order book, and it is quite common to aggressively split large trading orders into many small pieces.  The upstairs market trades are arranged bilaterally between individuals.   As a result of the different market structures the impacts can be quite different.   
\item
A third factor that must be kept in mind is that large trading orders, which we will call {\it hidden orders}, are typically split into small pieces and executed incrementally.   This is in contrast to {\it realized orders}, which are the actual orders that are traded, e.g. the pieces into which hidden orders are split.  For realized orders the impacts may be part of a larger process of order splitting that is invisible with the data that we have here.   The impacts of hidden orders may be quite different than those of realized orders.  The impacts of individual orders behave much like those of individual transactions, see the next bullet.  We will discuss the impact of hidden orders in Section~\ref{hiddenOrderImpact}.
\item
Finally, even if we have discussed market impact in terms of transacted volume, other events in the market have an impact on price. Specifically, in double auction market limit orders and cancellations can have a market impact that is different from the impact of a market order. 
\end{itemize}

In the following we will discuss the empirical regularities in these different types of market impact.

\subsection{Impact of individual transactions}\label{impact1}

We now discuss the impact of individual transactions in limit order book markets, whose volume we will denote by $v$
Many studies have examined the market impact for a single transaction, and all have observed a concave function of the transaction volume $v$, i.e. one that increases rapidly for small $v$ and more slowly for larger $v$.
The detailed functional form, however, varies from market to market and even period to period.   Early studies by Hasbrouck (Hasbrouck91) and Hausman, Lo and MacKinlay (Hausman92) found strongly concave functions, but did not attempt to fit functional forms.  \cite{Keim96} also observed a concave impact function for block trades.  Based on Trades and Quotes (TAQ) data for a set of $1000$ NYSE stocks the concavity of the market impact was interpreted by \cite{Lillo03d} using the functional form
\begin{equation}
E[r|v]=\frac{\epsilon v^{\psi}}{\lambda}
\label{priceImpact}
\end{equation}
\begin{figure}[ptb]
\begin{center}
\includegraphics[scale=0.3]{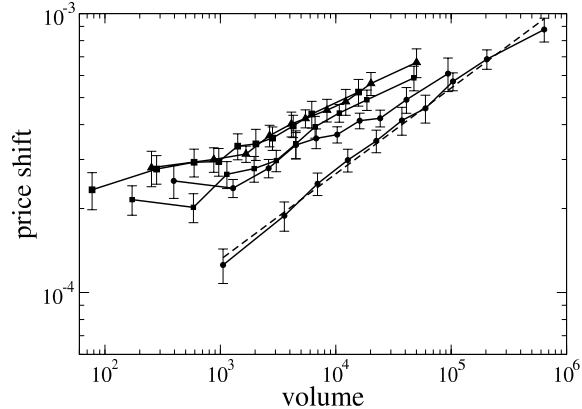}
\caption{
Market impact function of buy market orders for a set of 5 highly capitalized stocks traded
in the LSE, specifically AZN (filled squares), DGE (empty squares), LLOY
(triangles), SHEL (filled circles), and VOD (empty circles). Trades of
different sizes are binned together, and the average size of the
logarithmic price change for each bin is shown on the vertical
axis. The dashed line is the best fit of the market impact of VOD with
a functional form described in Eq.\ref{priceImpact}.  The value of the fitted
exponent for VOD is $\psi=0.3$.}
\label{impact}
\end{center}
\end{figure}
The exponent $\psi(v)$ is approximately $0.5$ for small volumes and $0.2$ for large volumes.  Even normalizing the volume $v$ by daily volume, the liquidity parameter $\lambda$ varies for different stocks; there is a clear dependence on market capitalization $M$ that is well-approximated by the functional form $\lambda \sim M^\delta$, with $\delta \approx 0.4$.  Potters and Bouchaud (2003) analyzed stocks traded at the Paris Bourse and NASDAQ and found that a logarithmic form gave the best fit to the data.  For the London Stock Exchange, \cite{Farmer04} and \cite{Farmer05} found that for most stocks Eq. (~\ref{priceImpact}) was a good approximation with $\psi = 0.3$, independent of $V$.  \cite{Hopman02} studied market impact on a thirty minute timescale in the Paris bourse for individual orders and found $\psi \approx 0.4$, depending on the urgency of the order.  Thus all the studies find strongly concave functions, but report variations in functional form that depend on the market and possibly other factors as well.
Figure~\ref{impact} shows the price impact of buy market orders for 5
highly capitalized LSE stocks, i.e. AZN, DGE, LLOY, SHEL, and VOD.  
The price impact is well fit by the relation $E[r| v] \propto v^{0.3}$.

\subsection{Impact of aggregate transactions \label{empiricalAggregate}}
 
Studies of aggregated market impact have produced variable results, reaching different conclusions that we will argue depend substantially on the time scale for aggregation.   The BARRA market impact model, an industry standard, uses the TAQ data aggregated on a half hour time scale (\cite{Torre97}).  They compare fits using Equation~\ref{priceImpact} and find $\psi \approx 0.5$; they obtain similar results using individual block data.  \cite{Kempf99} studied data for futures on the DAX (the German stock index) on an five minute time scale and found a very concave functional form.  \cite{Plerou02} studied data from the NYSE during 1994-95 ranging from 5 to 195 minute time scales and fit the market impact function with a hyperbolic tangent.   They noted that at shorter time scales this functional form did not work well for small $v$; $\tanh (v)$ is linear for small $v$, but at short time scales (e.g. 5 or 15 minutes) they observed a nonlinear impact function, becoming more linear as they went toward longer time scales.   \cite{Evans02} studied foreign exchange rate transactions data for DM and Yen against the dollar at the daily scale over a four month period.   They used the number of buyer initiated transactions minus the number of seller initiated transactions as a proxy for the signed order flow volume $v$, and found a strong positive relationship to concurrent returns.     
\cite{Chordia04} study impact for stocks in the S\&P 500 at a daily time scale and perform linear regressions, but do not compare to other functional forms.  For the Paris bourse \cite{Hopman02} measures aggregate order flow as $ \sum_i \epsilon_i v_i^\psi$, where the sum is taken over fixed time intervals.  At a daily scale he finds he gets the best linear regression against contemporary daily returns with $\psi \approx 0.5$.  He also documents that the slope of the regression decreases with increasing time scale.  Finally, as discussed in more detail below, \cite{Gabaix03,Gabaix06} have made extensive studies of data from the New York, London and Paris stock markets on a fifteen minute time scale, and find exponents $\psi \approx 0.5$.

\begin{figure}[ptb]
\begin{center}
\includegraphics[scale=0.33]{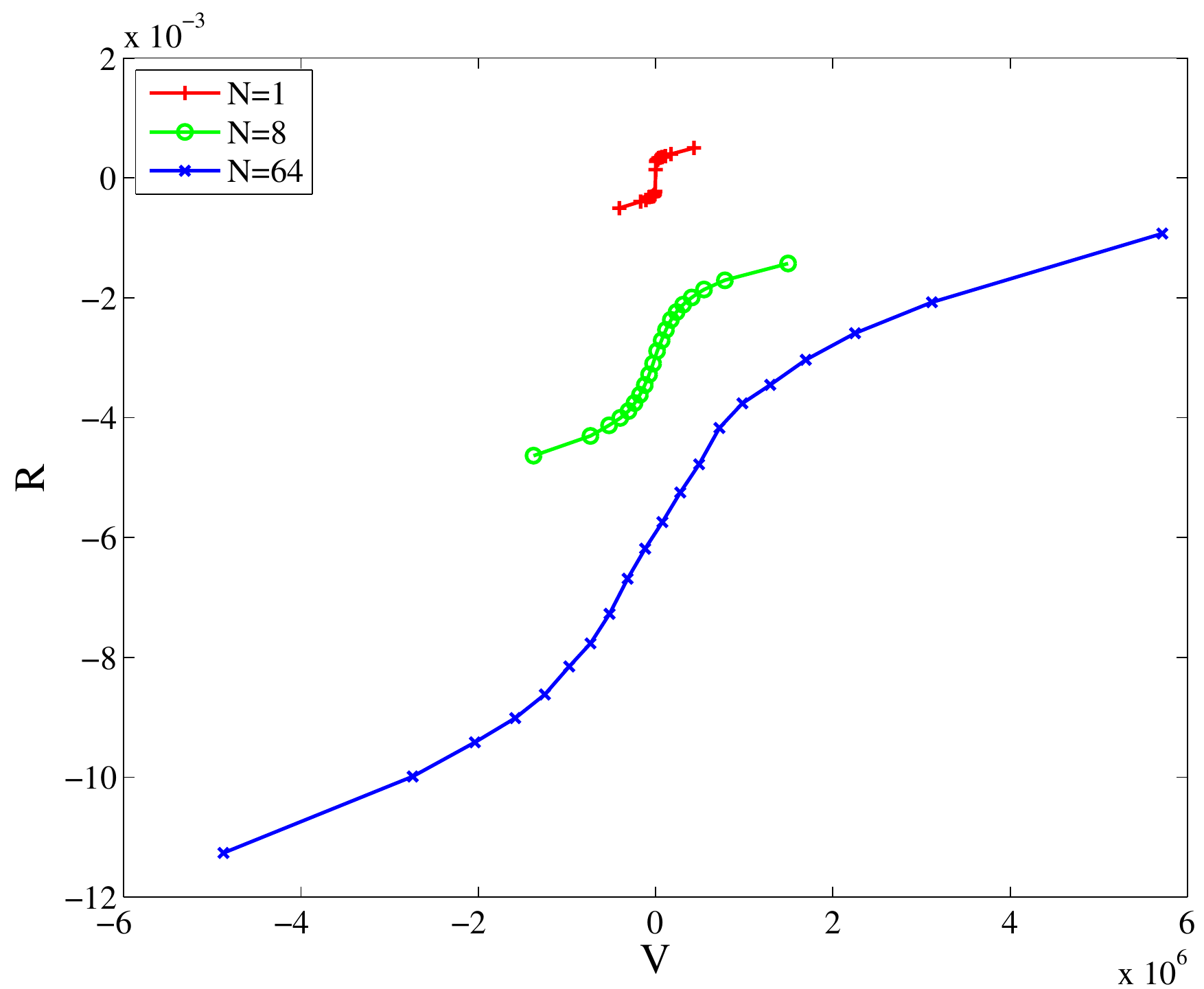}
\includegraphics[scale=0.33]{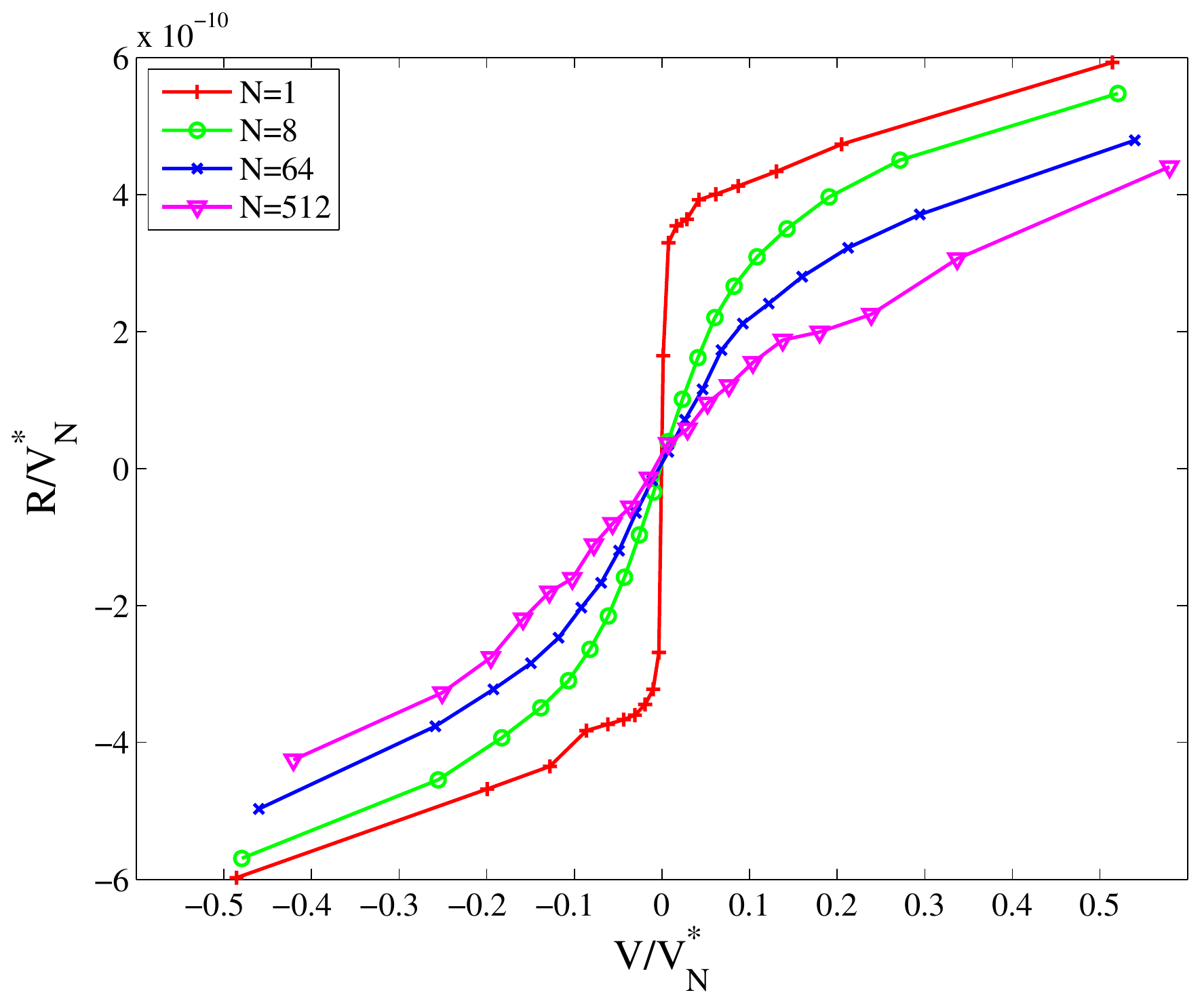}
\caption{Aggregate market impact $R(Q,N)$ for the LSE stock Astrazeneca for 2000-2002.  In (a) we plot the shifted aggregate return $R(Q,N) + R_0$ vs. the aggregate signed volume $Q$ for three values of $N$. The arbitrary constant $R_0$ is added to aid visualization; its values are $R_0 = \{0, -3 \times 10^{-3}, -6 \times 10^{-3}\}$ for $N = 1, 8$ and $64$ respectively.  In (b) for each $N$ we rescale both the horizontal and vertical axes by $Q^*_N = Q_N^{(95)} - Q_N^{(5)}$, where $Q_N^{(5)}$ is the $5\%$ quantile and $Q_N^{(95)}$ is the $95\%$ quantile of $Q$.
}
\label{priceImpactFig}
\end{center}
\end{figure}

What is the origin of these differences in the observed functional form of the aggregate market impact? 
Part of the difference comes certainly from the fact that these studies consider different markets, different assets, and different time periods. However another important difference across studies is the time scale of aggregation. There is no reason why the aggregate market impact over a 10 minute time interval should have the same functional form of that over a one hour time interval or over an interval that is defined by 30 trades. 

To have an idea of how the market impact changes its shape with aggregation scale consider a specific example.  
 Let $v_t$ be the volume of transaction happening at time $t$ (in event time).  Let $r_t = \log ( p_{t+1}/p_t)$ be the corresponding log-return, where $p_t$ is the price of transaction $t$.   For a sequence of $N$ successive transactions beginning at time $t$, let $Q_{N} = \sum_{i=1}^N \epsilon_{t+i}v_{t + i}$ be the aggregate volume and $R_{N} = \sum_{i=1}^N r_{t + i}$ be the aggregate return.  The average market impact conditioned on volume is
\begin{equation}
R(Q,N) = E[R_{N} | Q_{N} = Q],
\end{equation}
i.e. it is the expected return associated with a signed volume fluctuation $Q$.  We write $R(Q,N)$ to emphasize that this can depend both on the signed trading volume imbalance $V$ and the number of transactions $N$.
In Figure~\ref{priceImpactFig} we show empirical estimates for the market impact for the stock AZN, which is traded on the London Stock Exchange, from \cite{Lillo08}.  In Figure~\ref{priceImpactFig} we show the market impact for different values of $N$ with offsets added to the vertical axis to aid visualization.  As one would expect, the scale increases with $N$.  The shape of $R(Q,N)$ also changes, becoming more linear with increasing $N$.  This is illustrated more clearly in Figure~\ref{priceImpactFig}(b), where we rescale the horizontal and vertical axes using a rescaling factor based only on $Q_{N}$.  The renormalization makes the increasing linearity clearer.  As $N$ increases the market impact near $Q = 0$ becomes linear, and the size of the region that can be approximated as linear grows with increasing $N$.  It also illustrates a surprising feature:   The slope of the linear region decreases with $N$.  These same basic features (increasing linearity and decreasing slopes) hold for all the stocks in our sample, in both the New York and London Stock Exchanges.
This result shows that the shape and the scale of the aggregate market impact change with the aggregation scale.  At short time scales the function is significantly nonlinear, but at large aggregation scales the market impact becomes close to linear, and the slope of the impact decays with the aggregation scale. For this reason it is in general misleading to compare aggregate impact curves with different scales, unless one has a theory for how the market impact depends on aggregation scale. This also shows why the studies mentioned above found different forms of the market impact. In Section \ref{aggregatemod} we present some models that help to explain the behavior of aggregate impact observed in real data.

\subsection{Hidden order impact}

Because data for hidden orders, which are sometimes also called trading packages, are difficult to obtain, there are only a few studies (\cite{Chan93,Chan95,Almgren05,Vaglica07}).  These studies show that hidden orders can be extremely long, involving thousands of realized trades spread over periods of many weeks or even months.  As reviewed in Section~\ref{hidordemp}, the most recent study by Vaglica et al. confirms that hidden orders obey a power law distribution of size, which as we argue in Section~\ref{impactTemporal} plays an important role in determining their impact.

The theoretical considerations for treating hidden orders are quite different than those for individual orders, and they also very different from those of aggregated anonymous orders.  The reason is because such orders come from the same agent, creating bursts of orders in the order flow which are all of the same sign.  As we argued in Section \ref{lmf}, this generates strong correlations in order flow that have to be compensated for, as discussed in Section \ref{impactTemporal}.  The volume dependence of hidden order impact is intimately connected to the temporal aspects, and so we save the development of the theory for hidden order impact for the next section.

\subsection{Upstairs market impact}

Market impact in the upstairs market has been studied by \cite{Keim96}.  As in other cases they find empirically that market impact is concave.  They explain this based on a model for the difficulty of finding counterparties for trading.  Ultimately, as pointed out by \cite{Gabaix06}, upstairs market impact should match hidden order impact, for the simple reason that the upstairs market is competing with the downstairs market, and if costs in the upstairs market are too high they have the option of splitting their trades up in the downstairs market.  This is convenient because it implies that a theory for either market automatically gives a theory for the other.


\section{Theory of market impact}
\label{impactTemporal}

In this section we develop theoretical explanations for both the volume dependence and the temporal dependence of market impact.  As stressed in the previous section, there are several distinct types of impact that require a different approach to their analysis.  We begin in Section~\ref{whyConcave} by explaining why the impacts associated with individual trades are so concave, arguing that the dominant cause is selective liquidity taking.  Then in Sections~\ref{temporal1} - \ref{historyDependent} we develop a theoretical approach to understanding the temporal behavior of impacts associated with individual trades.  We show that the long-memory of order flow and market efficiency play a crucial role, which one can take into account one of two ways.  One can either assume a fixed impact, in which case the future contribution to the impact of each trade must decay to zero with time, or one can assume a varying but permanent impact, which implies asymmetry liquidity.  We show that these two approaches are equivalent.   In Section~\ref{empimp} we present empirical results supporting these ideas.   In Section~\ref{hiddenOrderImpact} we develop a theory for the impact of hidden orders, i.e. linked sets of trades made by large investors.  Finally in Section~\ref{aggregatemod} we develop a theory for the aggregate impact of successive trades and show that it does a good job of explaining the empirical results of Section~\ref{empiricalAggregate}.

\subsection{Why is individual transaction impact concave?\label{whyConcave}}

Let us consider first the impact of individual transactions. Several different theories have been put forth to explain why market impact for single transactions is concave. 
These can be grouped into three classes:  (1) Size dependent informativeness of trades (e.g. due to stealth trading, as postulated by (\cite{Barclay93})), (2) average depth vs. price in the limit order book (\cite{Daniels03}) and (3) selective liquidity taking (\cite{Farmer04b}). 

The standard reason given for the concavity of market impact is that it reflects the informativeness of trades.  If small trades carry almost as much information as large trades, then the price changes caused by small trades should be nearly as big as those for large trades.  For example, this could be due to ``stealth trading", i.e. because informed traders keep their orders small to avoid revealing their superior knowledge  (\cite{Barclay93}).   Hypothesis (2), due to \cite{Daniels03}), is that it reflects the accumulation of liquidity in the limit order book.  I.e., the depth in the order book as a function of the price will determine the market impact for a market order as a function of its size.  
Hypothesis (3) is that this is due to selective liquidity taking, i.e. that liquidity takers submit large orders when liquidity is high and small orders when it is low (see \cite{Farmer04b}), \cite{Weber04}, and \cite{Hopman02}).  

\begin{figure}[tbp]
\begin{center}
\includegraphics[scale=0.5]{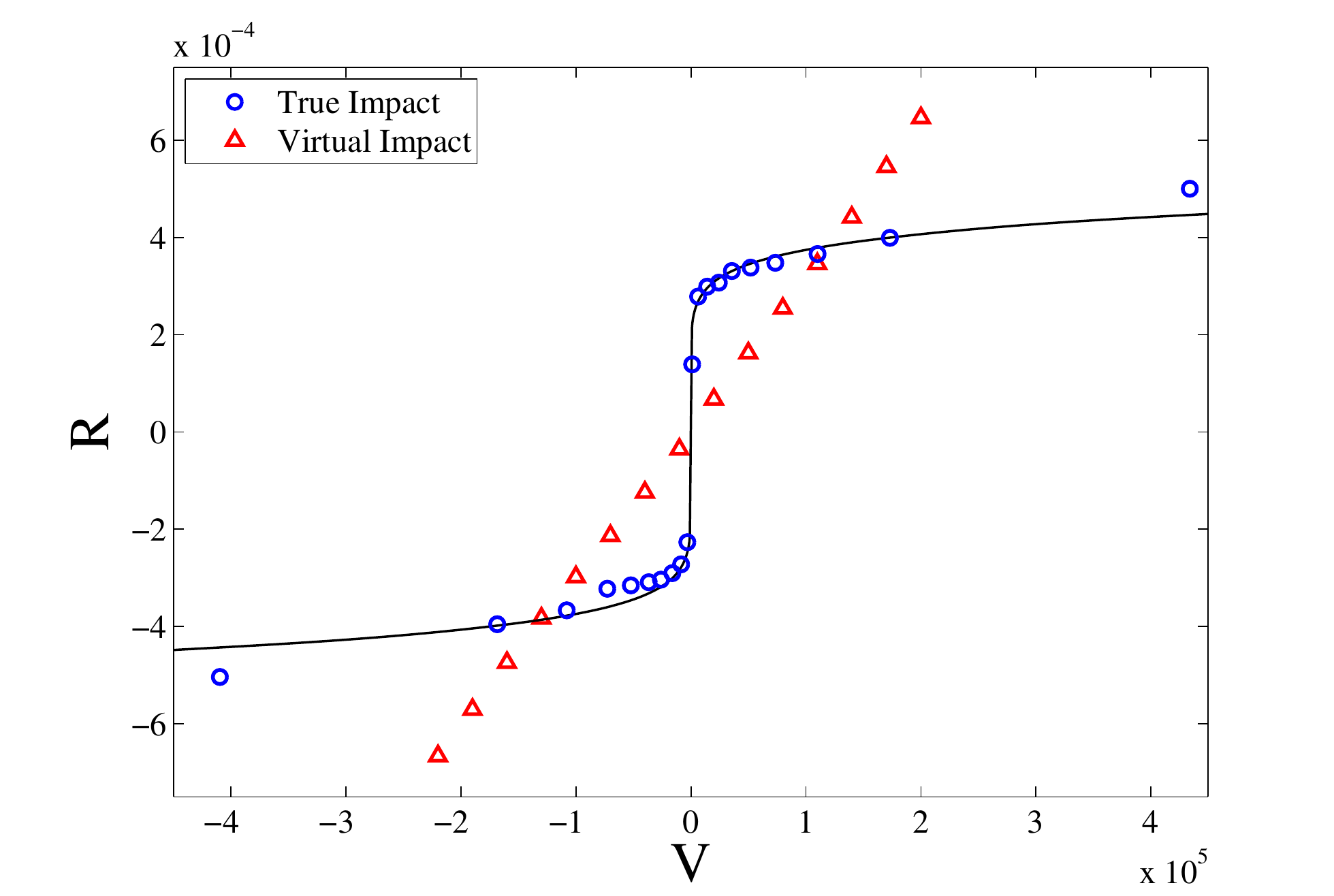}
\end{center}
\caption{Comparison of virtual to true market impact.  True impact is shown in blue circles, virtual impact in red triangles.  The fitted curve for true impact (solid black) is of the form $f(v) = Av^\psi$, with $\psi = 0.3$.}
\label{virtualImpact}
\end{figure}

Theory (2) is easily ruled out by computing the average virtual market impact as a function of volume.  This is defined as the average price change that would instantaneously occur for an effective market order of size $v$ (\cite{Weber04}), \cite{Farmer07}).  In Figure~\ref{virtualImpact} we show the virtual impact for AZN, computed by hypothetically submitting orders for a range of different values of $v$ and measuring the immediate price response.  This is done for each time when real effective market orders were submitted.  The resulting price response is a direct probe of the depth of the limit order book.  The fact that the mechanical impact is linear to very good degree of approximation makes it clear that this is not the cause of the concavity of the real market impact function.

The selective liquidity taking (hypothesis (3)) means that agents condition the size of their transactions on liquidity, making large transactions when liquidity is high and small transactions when it is low.   As shown by \cite{Farmer04b}, for LSE stocks it is rare that a trade penetrates more than one price level\footnote{See Table 2 of Farmer et al.}.  For example, for Astrazeneca approximately $87\%$ of the market orders creating an immediate price change have a volume equal to the volume at the opposite best. Moreover, approximately $97\%$ of the market orders creating an immediate price change have a volume that is either equal to the opposite best or larger than this value but smaller than the sum of volume at the second best opposite price.  This means that to a good approximation the market impact can be written in the very simple form
\begin{equation}
E[r|v] = P(+ | v) E[r],
\label{singleImpactModel}
\end{equation}
where $P(+ | v)$ is the probability that a trade of size $v$ generates a nonzero return, i.e. the probability that $v \ge \Phi_b$, where $\Phi_b$ is the volume offered or bid at the opposite best price.  $E[r]$ is the expected return given that there is a nonzero return, which is of the order of the bid-ask spread (see Section \ref{spread} for more precise statements).  This demonstrates that trading orders that penetrate the opposite best are rare.  This is because agents do not like to suffer price degradation more than the opposite best, and so condition the size of their orders on what is being offered there. 

We have now to explain why $P(+|v)$ is a concave function. An explanation in terms of selective liquidity taking is the following. Suppose that the volume at the best is drawn from a distribution $P_b(\Phi_b)$ and suppose that the liquidity taker draws the volume $v$ she would like to trade from another distribution and independently from $\Phi_b$. If $v<\Phi_b$ she places a market order of size $v$, whereas if $v>\Phi_b$ she places a market order of size $\Phi_b$. What is the probability $P(+|v)$ under this simple model? A straightforward calculation shows that $P(+|v)= \int_0^v P_b(\Phi_b) d\Phi_b$, i.e. it is equal to the cumulative distribution of the volume at the best. This is  an increasing and concave function of $v$ that could be used to fit the empirical $P(+|v)$.  Under this model the shape of the market impact is explained by $P(+ | v)$, i.e. by the conditioning of trading orders on the liquidity that is offered.  In other words, theory (3) does a good job of explaining, at least qualitatively, the data.

It is a matter of interpretation, however, whether this is also consistent with theory (1), i.e. that smaller trades are proportionately more informative than larger trades.  From one point of view, one can simply say that the market impact {\it defines} the informativeness of trades.  If so, then it is obviously consistent.  However, if it means that price changes are a response to the new information contained in trades, then the evidence presented above is inconsistent with theory (1).  In the LSE the quoted volume is visible to all, and so except for occasional latency problems, in which the quote changes just before a trade is placed, the trader is aware of the quote when she places the trade.  The fact that the size of the trade is strongly correlated with the size of the best quote implies that the size of the trade carries little new information.  This does not mean that the trade is based on inferior information -- it just means that other market participants do not learn much from its size when it occurs.  It is the conditioning of trade size on best quotes that drives concavity, and not because the smaller trades are nearly as ``informed" as the larger trades.

\subsection{A fixed permanent impact model \label{temporal1}}


In the previous section we described how midquote prices react on average to market orders of a given volume $v$. The above discussion was restricted to the immediate impact, i.e. the impact that is felt immediately after a trade is completed.  In general this can have both temporary and permanent components.  In this section we will discuss the impact of individual transactions, i.e. the average midquote price change between just before the $n$th trade and just before
the $n+1$th trade. It is an empirical fact that this immediate impact, defined as $E[r_n|\epsilon_n v_n]$, is non zero and can be written as
$E[r|\epsilon v]=\epsilon f(v)$, where $f$ is a function that grows with $v$. Clearly, it is important to understand if and how this immediate impact  evolves with time (which we will measure in terms of the sequence number of the trades).  Is the impact of a trade permanent or transient? Is it fixed or is it variable? How does it depend on the past order flow history?

The simplest situation is that of a usual random walker, where position at any time is the sum over all past steps -- however far in the past they 
might be. In financial language, this corresponds to the case where the impact of a transaction is permanent, which translates into the 
following equation for the midquote price $m_n$ at time $n$:
\be\label{basicimpact}
r_n=m_{n+1}-m_n = \epsilon_n f(v_n;\Omega_n) + \eta_n,
\ee
where $\eta_n$ is an additional random term describing price changes not directly attributed to trading itself, for example the impact of news where quotes could instantaneously jump without any trade. We will assume here that $\eta_n$ is independent on the order flow and we set  $E[\eta]=0$ and $E[\eta^2]=\Sigma^2$. 
We have included a possible dependence of the impact on the instantaneous state $\Omega_n$ of the order book. We expect such a 
dependence on general grounds: a market order of volume $v_n$, hitting a large queue of limit orders, will in general impact the
price very little. On the other hand, one expects a very strong correlation between the state of the book $\Omega_n$ and the the 
size of the incoming market order: large limit order volumes attract larger market orders. 

The above equation can be written as:
\be \label{RW}
m_n = \sum_{k < n} \epsilon_k f(v_k;\Omega_k) + \sum_{k < n} \eta_k,
\ee
which makes explicit the non-decaying nature of the impact in this model: $\epsilon_k \partial m_n/\partial v_k$ (for $k < n$) 
does not decay as $n-k$ grows. 
This simple model  makes the following predictions for
the lagged impact function ${\cal R}_\ell$ and the lagged return variance ${\cal V}_\ell$:
\be \label{Rdef}
{\cal R}_\ell \equiv E[\epsilon_n \cdot (m_{n+\ell}-m_n)] = E[f]; \, {\cal V}_\ell \equiv E[(m_{n+\ell}-m_n)^2]=\left(E[f^2] + \Sigma^2\right) \ell,
\ee
i.e. constant price impact and pure price diffusion, close to what is indeed observed empirically on small tick, liquid contracts. 
However if we consider the autocovariance of price returns we find that 
\be
E[ r_n r_{n+\tau}] \propto E[ \epsilon_n \epsilon_{n+\tau}] \sim \tau^{-\gamma}
\ee
which means that price returns are strongly autocorrelated in time. This fact would violate market efficiency because price returns would be easily predictable even with linear methods. We therefore come to the conclusion that the empirically observed long memory of order flow is incompatible with the random walk model above if prices are efficient  (\cite{Bouchaud04,Lillo03c,Challet07}). In other words one of the assumptions of the random walk model above must be relaxed. Among the various possibilities we will relax either the assumption that  price impact is permanent or the assumption that price impact is independent of the order flow. As we will see these two possibilities are related one to each other, but for the sake of clarity we will present them in two different subsections. 

\subsection{The MRR model}\label{mrrsec}

In order to illustrate the above concepts, let us discuss a slight variant of a model due to Madhavan, Richardson and Roomans (\cite{Madhavan97b}), 
which helps define various quantities and hone in on relevant questions. The assumptions of the model 
are (i) that all trades have the same volume $v_n=v$ and (ii) the $\epsilon_n$'s are generated by a Markov process with 
correlation $\rho$, which means that the expected value of $\epsilon_n$ conditioned on the past only depends on $\epsilon_{n-1}$ and is
given by:
\be
E \left[ \epsilon_n | \epsilon_{n-1} \right] = \rho \epsilon_{n-1},
\ee
The case $\rho=0$ corresponds to independent trade signs, whereas $\rho > 0$ describes positive autocorrelations of trade signs. Note
that in this model, correlations decay exponentially fast, i.e.
\be
C_\ell = E[ \epsilon_i \epsilon_{i+\ell} ] = \rho^\ell.
\ee
which, as we discussed in Section~\ref{longMemory}, does not conform to reality. 

The {\sc mrr} model postulates that the mid-point $m_n$ evolves 
only because of unpredictable external shocks (or news) and because of the surprise component in the order flow. 
This postulate of course automatically removes any predictability in the price returns and ensures efficiency. 
Under the assumption that the surprise component of the order flow at the $n$th trade is given by $\epsilon_n -  \rho \epsilon_{n-1}$, one writes the following evolution equation for the price\footnote{
The assumption that prices respond linearly to the order flow is a very strong assumption.}.
\be
m_{n+1} - m_n = \eta_n + \theta [\epsilon_n -  \rho \epsilon_{n-1}],
\ee
where $\eta$ is the shock component, and the constant $\theta$ measures the size of trade impact.

These equations make it possible to compute several important quantities such as the lagged impact function defined above (Eq. (\ref{Rdef})).
One may write:
\be
m_{n+\ell}-m_n = \sum_{j=n}^{n+\ell-1} \eta_j + \theta  \sum_{j=n}^{n+\ell-1} [\epsilon_j-  \rho \epsilon_{j-1}],
\ee
The full impact function is found to be constant, equal to: 
\be\label{RellMRRsimp}
{\cal R}_\ell = \theta (1-\rho^2), \quad \forall \ell
\ee
We can also define the `bare' impact of a single trade $G_0(\ell)$, which measures the influence of a single trade at time $n-\ell$
on the the mid-point at time $n$. In terms of $G_0(\ell)$, the mid-point is therefore written as:
\be\label{G0def}
m_n = \sum_{j=-\infty}^{n-1} \eta_j +  \sum_{j=-\infty}^{n-1} G_0(n-j-1) \, \epsilon_j,
\ee
is here found to given by $G_0(\ell=0)=\theta$ and $G_0(\ell \geq 1)=\theta (1-\rho)$: a part $\theta \rho$ of the impact instantaneously decays
to zero after the first trade, whereas the rest of the impact is permanent. The instantaneous drop of part of the impact compensates the sign correlation
of the trades. Finally, the volatility, within this simplified version of the {\sc mrr} model, reads:
\be
{\cal V}_\ell = \theta^2 (1-\rho^2) \ell.
\ee

\subsection{A transient impact framework} \label{transient}

Compared to the above simplifying assumptions of the MRR model, the data shows that (i) the volumes $v$ of the incoming market orders are very broadly distributed, with a power-law tail (see section \ref{heavyvol}); (ii) the sign time series $\epsilon_n$ has long range correlations $C_\ell$ that decays again as
a power-law $\sim c_0\ell^{-\gamma}$ with $\gamma < 1$, defining a long memory process. The smallness of $\gamma$ makes the correlation function $C_\ell$ non-summable: the average 
relaxation time is infinite, whereas the correlation time of the Markovian sign process in the above {\sc mrr} model is finite, equal to $(1-\rho)^{-1}$. 

In this section we relax the assumption that impact of a single trade is permanent in time. Rather, we find that long range correlations in trades 
imply that the impact itself has to decay slowly with time. In the next section, we will discuss an alternative but equivalent model, where the
impact is permanent but asymmetric and history dependent.

\subsubsection{Transient impact and mean reversion}

What would happen if the impact of each trade was purely transient, for example an exponential decay in time? 
Eq. (\ref{basicimpact}) would now read: 
\be
m_n = \sum_{k < n} \alpha^{n-k-1} \epsilon_k f(v_k;\Omega_k) + \sum_{k < n} \eta_k,\qquad (0 \leq \alpha < 1).
\ee
The lagged impact and the return variance would then be given by:
\be
{\cal R}_\ell = \alpha^{\ell-1} E[f]; \qquad {\cal V}_\ell= 2E[f^2]\frac{1-\alpha^\ell}{1-\alpha^2} + \Sigma^2 \ell,
\ee
i.e. a short-time volatility $\approx E[f^2] + \Sigma^2 $ larger than its long-time value $\Sigma^2$, in which only the `news' component survives. The 
price would exhibit significant high frequency mean reversion: impact kicks it temporarily up and down, but the long term wandering
of the price is unrelated to trading. Of course, one could be in a mixed situation where the impact decays exponentially but towards a positive value, 
in which case the long term volatility still involves an impact component.  This conforms with conventional wisdom about efficient markets:  an increased 
value of high frequency volatility driven by the ``tatonnement" process, and a long term volatility made up both of unexpected news and long-term impact
of market orders, which translates private information into prices.  However, recall that this does not conform to observations, which show volatility very nearly constant across all time scales (see Section~\ref{volPuzzle}).

What is the relation between the average ${\cal R}_\ell$ and the impact of a single trade, that we call $G_0(\ell)$ henceforth? If trades were uncorrelated, the two 
quantities would be identical, but trade correlations, as we shall see below, change the picture in a rather interesting way.

\subsubsection{Mathematical theory of long term resilience}

The long term memory of trades is {\it a priori} paradoxical and hints towards a non trivial property of financial markets, which can be called {\it 
long-term resilience}. Take again Eq. (\ref{G0def}) with the assumption that single trade impact is lag independent: $G_0(\ell)=G_0$ and that 
volume fluctuations can still be neglected. The mid-price variance is easily computed to be:
\be
{\cal V}_\ell \equiv \langle (m_{n+\ell} - m_n)^2 \rangle = [\Sigma^2+G_0^2] \ell + 2 G_0 \sum_{j=1}^{\ell} (\ell - j) C_j.
\ee
When $\gamma < 1$, the second term of the rhs can be approximated, when $\ell \gg 1$, by $2c_0G_0 \ell^{2-\gamma}/(1-\gamma)(2-\gamma)$, which 
grows faster than the first term. In other words, the price would {\it super-diffuse}, or trend, at long times, with a volatility diverging with 
the lag $\ell$. This of course does not occur:  The market reacts to trade correlations so as to prevent the occurence of such trends. In fact, within
the present linear model, the impact to single trades must be transient, rather than permanent. 
Before explaining why and how this occurs in practice, let us first express 
mathematically how the efficiency of prices imposes strong constraints on the shape of the single trade impact function. For an
arbitrary function $G_0(\ell)$, the lagged price variance can be computed explicitly and reads: 
\be\label{diff}
{\cal V}_\ell = \sum_{0 \leq j < \ell} G_0^2(\ell-j) + \sum_{j > 0} \left[G_0(\ell+j)-G_0(j)\right]^2 + 2\Delta(\ell)+ \Sigma^2 \ell,
\ee
where $\Delta(\ell)$ is the correlation induced contribution:
\bea
\Delta(\ell) &=& \sum_{0 \leq j < k < \ell} G_0(\ell-j) G_0(\ell-k) C_{k-j}\\
\nonumber
&+& \sum_{0 < j < k} \left[G_0(\ell+j)-G_0(j)\right]\left[G_0(\ell+k)-G_0(k)\right]
C_{k-j} \\
&+& \sum_{0 \leq j < \ell} \sum_{k> 0} G_0(\ell-j) \left[G_0(\ell+k)-G_0(k)\right] C_{k+j}.
\eea
Assume that $G_0(\ell)$ itself decays at large $\ell$ as a power-law, $\Gamma_0 \ell^{-\beta}$. When $\beta, \gamma < 1$, the
asymptotic analysis of $\Delta(\ell)$ yields:
\be
\Delta(\ell) \approx \Gamma_0^2 c_0 I(\gamma,\beta) \ell^{2-2\beta-\gamma},
\ee
where $I > 0$ is a certain numerical integral. If the single trade impact does not decay ($\beta=0$), we recover the above superdiffusive result. But as
the impact decays faster, superdiffusion is reduced, until $\beta=\beta_c=(1-\gamma)/2$, for which $\Delta(\ell)$ grows exactly linearly with $\ell$ 
and contributes to the long term value of the volatility. However, as soon as $\beta$ exceeds $\beta_c$, $\Delta(\ell)$ grows sublinearly with $\ell$, 
and impact only enhances the high frequency value of the volatility compared to its long term value $\Sigma^2$, dominated by `news'. We therefore
reach the conclusion that the long range correlation in order flow does not induce long term correlations nor anticorrelations in the price returns 
if and only if the impact of single trades is transient ($\beta > 0$) but itself non-summable  ($\beta < 1$). This is a rather odd situation where the
impact is not permanent (since the long time limit of $G_0$ is zero) but is not transient either, because the decay is extremely slow. 
The convolution of this semi-permanent
impact with the slow decay of trade correlations gives only a finite contribution to the long term volatility. The mathematical constraint $\beta=\beta_c$
will be given more financial flesh below.

\begin{figure}[ptb]
\begin{center}
\includegraphics[angle=0,scale=0.38]{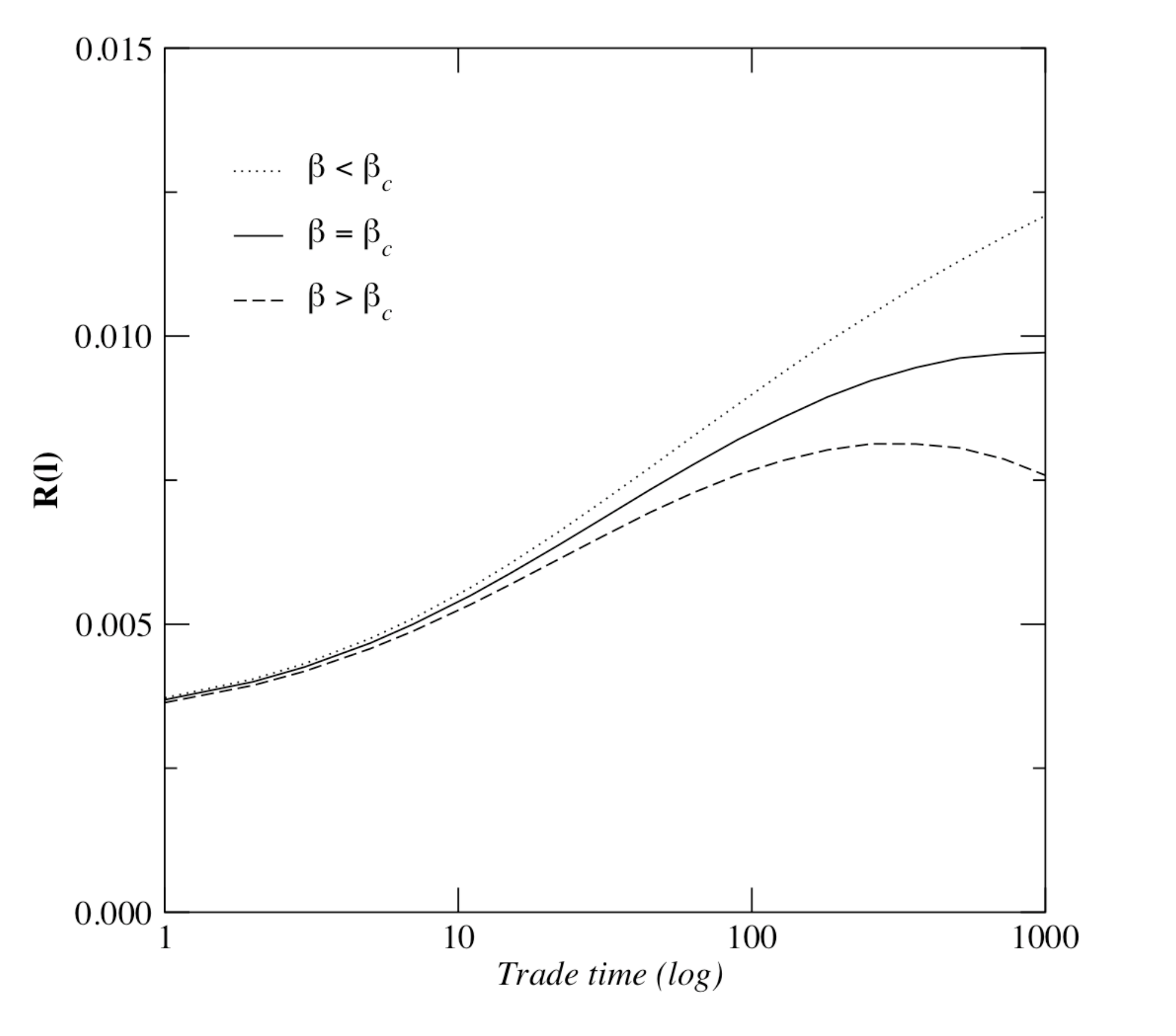}
\end{center}
\caption{Theoretical impact function ${\cal R}_\ell$, from Eq. (\ref{response}), and for values of $\beta$ close to $\beta_c$.
When $\beta=\beta_c$, ${\cal R}_\ell$ tends to a constant value as $\ell$ becomes large. When $\beta < \beta_c$ (slow decay of $G_0$), 
${\cal R}_{\ell \to \infty}$ diverges to $+\infty$, whereas for $\beta > \beta_c$, ${\cal R}_{\ell \to \infty}$ diverges to $-\infty$.}
\label{Fig9}
\end{figure}

Within the above framework, one can also compute the average impact function ${\cal R}_\ell$. From Eq. (\ref{Rdef}) one readily obtains:
\be \label{response}
{\cal R}_\ell = G_0(\ell)+ \sum_{0 < j < \ell} G_0(\ell-j) C_j + \sum_{j > 0} \left[G_0(\ell+j)-G_0(j)\right] C_j.
\ee
This equation can be understood as a way to extract the impact of single trades $G_0$ from directly measurable quantities, such
as ${\cal R}_\ell$ and $C_n$ -- see section \ref{empimp} and Appendix 2. From a mathematical point of view, 
the asymptotic analysis can again be done when $G_0(\ell)$ decays as $\Gamma_0 \ell^{-\beta}$. When $\beta+\gamma < 1$, one finds:
\be
{\cal R}_\ell \approx_{\ell \gg 1} \Gamma_0 c_0 \, \frac{\Gamma(1-\gamma)}{\Gamma(\beta)
\Gamma(2-\beta-\gamma)} \left[\frac{\pi}{\sin \pi \beta} - 
\frac{\pi}{\sin \pi (1-\beta-\gamma)}\right] \ell^{1-\beta-\gamma},
\ee
where we have explicitly given the numerical prefactor to show that it exactly vanishes when $\beta=\beta_c$, which means that in this particular
case one cannot satisfy oneself with the leading term. When $\beta < \beta_c$, one finds that ${\cal R}_\ell$ diverges to $+\infty$ for large $\ell$,
whereas for $\beta > \beta_c$, ${\cal R}_\ell$ diverges to $-\infty$, which is perhaps counter-intuitive, but means that when the decay of single 
trade impact is too fast, the accumulation of mean reverting effects leads to a negative long term average impact -- see Fig.~(\ref{Fig9}). 
When $\beta$ is precisely equal to $\beta_c$, ${\cal R}_\ell$ tends to a finite positive value ${\cal R}_\infty$: the decay of single trade 
impact precisely offsets the positive correlation of the trades.

In the above framework, volume fluctuations have been neglected. An extended version of the model, which is directly related to the discussion of the
next section, is presented in Appendix 2 (see also \cite{Bouchaud04}).

\subsection{History dependent, permanent impact \label{historyDependent}}

\subsubsection{Predictable order flow and statistical efficiency}\label{predictable}

An alternative interpretation of the above formalism is to assume that price impact is permanent, but history dependent as to ensure 
statistical efficiency of prices (\cite{Lillo03c,Farmer06,Gerig07}). Let us consider a generalized MRR model:
\be\label{MRRgen}
r_n=m_{n+1} - m_n = \eta_n + \theta (\epsilon_n -  \hat\epsilon_n),\qquad \hat\epsilon_n=E_{n}[\epsilon_{n+1}|I]
\ee
where $I$ is the information set available at time $n$. In line with our discussion in Section \ref{ecology}, we assume that in the market there are three types of traders. First of all there are directional traders (liquidity takers) which have large hidden orders to unload and, by placing many consecutive orders with the same sign, create a correlated order flow. The second group of agents are the liquidity providers, who post bid and offer and attempt to 
earn the bid-ask spread. The third group is made by noise traders, i.e. traders placing uncorrelated order flow. Anticipating the discussion in Section\ref{phasediagram}, it is indeed reasonable to assume that the strategies of the first two type of agents will adjust in such a way to remove any predictability of the 
midpoint change, or in other words that $E_{n-1}[r_n|I]=0$ as implied by Eq. (\ref{MRRgen}) above. 
This is a plausible first approximation, although one can expect (and indeed observes) deviations from strict unpredictability at high frequencies.

Within the above simplified model, in which we have neglected volume fluctuations (see Appendix 2 for an attempt to include them), there are only two possible outcomes. Either the sign of the $n$th transaction matches the sign of the predictor $E_{n}[\epsilon_{n+1}|I]$, or they are opposite. Let us call $r^+_n$ and $r^-_n$ the 
expected ex-post absolute value of the return of the $n^{th}$ transaction given that $\epsilon_n$ either matches or does not match the predictor.  If we indicate with $\varphi^+_n$ and ($\varphi^-_n$) the ex ante probability that the sign of the $n$-th transaction matches (or disagrees) with the predictor $\epsilon_n$, we can rewrite $E_{n-1}[r_n|I]=0$ as: 
\be
\varphi^+_n r^+_ n-\varphi^-_n r^-_ n=0.
\ee
Within the MRR model as above, this means
\begin{eqnarray}
r^+_ n=\theta(1-\hat\epsilon_n)\\
r^-_ n=\theta(1+\hat\epsilon_n).
\label{symmetryeq}
\end{eqnarray}  
This result shows that the most likely outcome has the smallest impact. We call this mechanism {\it asymmetric liquidity}: each transaction has a permanent impact, but the impact depends on the past order flow and on its predictability. The price dynamics and the impact of orders therefore depend 
on (i) the order flow process (ii) the information set $I$ available to the liquidity provider, and 
(iii) the predictor used by the liquidity provider to forecast the order flow.

\subsubsection{Equivalence with the transient impact model}

In the following we will consider the case where the information set available to liquidity providers is restricted to the past order flow. We call this information set {\it anonymous} because liquidity providers do not know the identity of the liquidity takers and are unable to establish whether or not two different orders come from the same trader. We assume also that the predictor used by liquidity takers to forecast future order flow comes from a linear model. In some cases, such as for an order flow generated according to the model presented in Section \ref{lmf}, this may not be an optimal predictor. However linear time series models are probably the most widely used forecasting tools.  Here we analyze a linear time series model based on the signs of executed transactions, and will assume a $K^{th}$ order autoregressive AR model of the form
\begin{equation}
\label{arModel}
\hat\epsilon_{n} =\sum_{i=1}^K a_i \epsilon_{n - i} ,
\end{equation}
where $a_i$ are real numbers that can be estimated on historical data using standard methods (see \cite{Lillo03c,Bouchaud04} and Appendix 2).  The MRR model corresponds to an AR(1) order flow, 
with $a_1=\rho$ and $a_k=0$ for $k>1$, with an exponential decay of the correlation. 

The resulting impact model, Eq. (\ref{MRRgen}) with a general linear forcast of the order flow is in fact
{\it equivalent}, when $K \to \infty$, to the temporary impact model of the previous section (see Appendix of \cite{Bouchaud04}). 
It is easy to show that one can rewrite the generalized MRR model in terms of a propagator as
\begin{equation}
m_n=m_{n-1}+\theta\epsilon_n+\sum_{i=1}^\infty[G(i+1)-G(i)]\epsilon_{n-i}+\eta_n, \qquad \theta=G(1).
\label{propagator}
\end{equation}
The equivalence is obtained with the relation:
\begin{equation}
\theta a_i={G(i+1)-G(i)}\quad {\mbox or} \quad G(i)=\theta[1-\sum_{j=1}^{i-1}a_j].
\end{equation}

\subsubsection{More general information models}

In the previous section we have seen that the fixed/temporary impact model is equivalent to the variable/permanent impact model under the additional assumptions that (i) the information set available to the liquidity provider is the set of the past order flow and (ii) that liquidity providers use a linear forecast model to predict the future order flow from the past and to adjust price response. These two assumptions of the variable/permanent impact model are far from general. In the following we discuss the more general situations where a different information set and forecast model can arise.

In most financial markets order flow is available in real time to all market participants and thus it is clear that any liquidity provider could use the past order flow time series to trade efficiently. However in some cases participants can make use of information other than the time series of order flow signs.  There are often indirect clues about the identity of orders such as the consistent use of particular round lots for orders that arrive at regular intervals.  Activity in block markets can also provide clues about the activity of large orders.  Another case is when a  trader is trying to execute her large order by a so-called ``slicing and dicing" algorithm. The liquidity provider could be able to detect the presence of this trader and therefore the liquidity provider has additional information to add to her information set.

The algorithm used by the liquidity provider to forecast the future order flow depends on the information set and on the degree of sophistication of the liquidity provider. Even if linear forecasting methods are widespread, they can lead to suboptimal predictions if the time series one is trying to forecast is strongly nonlinear. For example, in Section \ref{lmf} we have discussed a microscopically based order flow model which reproduces the correlation properties observed in the real order flow. This model (Lillo, Mike, and Farmer 2005) is clearly non-linear. Despite the fact that an optimal forecast method for this order flow model is not easily available, one can find suboptimal non-linear forecast models that outperform the linear forecast method. When one incorporates non-linear forecast models in the variable/permanent impact model the price dynamics will not be equivalent to the fixed/temporary model.

In conclusion, the variable/permanent model sets a general framework for describing the interaction between order flow and price dynamics. In a paper in progress Gerig {\it et al.} (2008) show how different assumptions on the information set and on the forecast method lead to different functional forms of the impact of hidden orders and on the dynamical properties of prices. 

\subsubsection{Mechanisms for Asymmetric Liquidity}

Let us rephrase in more intuitive terms the results established above.
Due to the small outstanding liquidity, order flow must develop temporal correlations. This is such an obvious empirical fact that 
high frequency traders/market makers quickly come to learn about it, and adapt to it. In the simple {\sc mrr} model where signs are exponentially 
correlated, the probability that a buy follows a buy is $p_+=(1 + \rho)/2$. The unconditional impact of a buy is $\theta$ (see Eq. \ref{surprise});
however, a second buy immediately following the first has a reduced impact equal to ${\cal R}^+_1=\theta(1-\rho)$. The second buy is 
not as surprising as the first, and therefore should impact the price less. A sell immediately following a buy, on the other hand, has
an {\it enhanced} impact equal to ${\cal R}^-_1=\theta(1+\rho)$, in such a way that the conditional average impact of the next trade
is zero: $p_+ {\cal R}^+_1 + (1-p_+) {\cal R}^-_1 \equiv 0$ \cite{Gerig07}. This is the  ``asymmetric liquidity" effect explained above  (\cite{Lillo03c,Farmer06,Gerig07}, see also \cite{Bouchaud04b} where it is called ``liquidity molasses''). This mechanism is expected to be present in general: because of the positive correlation in order flow, the impact of a buy following a buy should be less than the impact of a sell following a sell -- otherwise trends would appear. 

But what are the mechanisms responsible for asymmetric liquidity, and how can they fail (in which case markets cease to be efficient)? This is still an open empirical question which started to be investigated only recently. For example, \cite{Lillo03c} showed that when the order flow becomes more predictable the probability that a market order triggers a price change is larger for market orders with the unexpected sign than for those with the expected one. Moreover the same authors showed that the ratio between the volume of the market order and the volume at the opposite best is lower (higher) for market orders with expected (unexpected) sign.

Another  related  basic mechanism is ``stimulated refill'': buy market orders trigger an opposing flow of sell limit orders, 
and vice-versa (\cite{Bouchaud04b}). This rising wall of limit orders decreases the probability of further upward moves of the price, which is equivalent to saying that 
${\cal R}^+_1 < {\cal R}^-_1$, or else that the initial impact of the first trade reverts at the second trade.  This dynamical feedback between market orders and limit orders is  therefore fundamental for the stability of markets and for enforcing efficiency. It can be tested directly on empirical data. 
For example, \cite{Weber05} have found strong evidence for an increased limit order flow compensating market orders. 

Since such a dynamical feedback is so important to reconcile correlation in order flow with the diffusive nature of price changes, it is
worth detailing its intimate mechanism a little further, and insisting on cases where this feedback may break down. Recall our discussion
of the market ecology in section \ref{ecology} -- market participants can be, in a first approximation, classified as a function of their trading 
frequencies. Large latent demand arises from low frequency participants; the decision to buy or to sell can be considered as fixed over a time 
scale of a few hours or a few days, much longer than the average time between trades. These participants create long term correlations in the 
sign of the trades. Higher frequency traders try to make profit from microstructural effects and short time predictability. Even if institutionally designated market 
makers are no longer present in most electronic markets, these high frequency strategies are in fact akin to market making --
they make money from providing liquidity to lower frequency traders. This is why we often (incorrectly) call this category of participants 
``market makers''.\footnote{Of course, the above distinction between participants must be taken with a grain of salt: low frequency decisions
may be executed using smart high frequency algorithmic trading.  In this case, the same participant is at the same time a low frequency trader and
a market maker.} So one should think of two rather large latent supply and offer quantities that await for favorable conditions, both in
terms of price and quantity, to be executed on the market. Then begins a kind of hide and seek game, where each side attempts to guess the 
available liquidity on the other side. A ``tit-for-tat'' process then starts, whereby market orders trigger limit orders and limit orders 
attracts market orders.  A buy trade at the ask (say) is a signal that an investor is indeed
willing to trade at that particular price. But the seller who placed a limit order at the ask is also, by definition, willing to trade at that price. 
The natural consequence is that a flow of refill orders is expected to occur at the ask immediately after a buy trade (and at the bid after a sell).

In other words, optimized execution strategies that look for micro-opportunities impose strong correlations between market 
order flow of one sign and limit order flow of the opposite sign. Imagine a case where buy market orders eat up sell limit orders at the ask, with no 
refill. The ask then moves up one tick. By making the price more expensive the flow of buy market orders slows down and the probability that a sell limit
order reappears at the previous ask increases. Imagine now that the refill process is too intense; sell limit orders at the ask now pile up. This 
has two effects: (i) the probability of a large market order that executes a large volume in one shot increases; (ii) the large volume at the ask decreases 
the probability of further sell limit orders joining the queue because the priority of these new orders is low. Both cases (no refill or intense refill)
therefore induce a clear feedback mechanism ensuring local stability of the order book. 

The above mechanism can be thought of as a dynamical version 
of the supply-demand equilibrium, in the following sense: incipient up trends quickly dwindle because as the ask moves up 
the buy pressure goes down while the sell pressure increases. Conversely, liquidity induced mean-reversion -- that keeps the price low -- attracts more 
buyers and soon gives way. Such a balance 
between liquidity taking and liquidity providing is at the origin of the subtle compensation between correlation and impact explained above. 
It is interesting to notice that several other dynamical systems operate similarly, with a competition between two antagonist systems -- heartbeats is an
interesting example: the sympathetic and parasympathetic system act in opposition to speed up/slow down the cardiac rythm. 

One easily envisions that such a subtle dynamical equilibrium can quickly break down: for example, an upward fluctuation in buy order 
flow might trigger a momentary panic, with the opposing side failing to respond immediately. These liquidity micro-crisis are probably responsible for
the large number of price jumps; if the feedback mechanism changes sign, this can even lead to crashes. The tug-of-war is a vivid illustration of this
phenomenon. A major challenge of microstructure theory 
is to turn the above qualitative story into a quantitative model for heavy tailed return distributions and volatility clustering, with interesting
potential ideas on how to limit the occurence of these liquidity micro-crises. We are convinced 
that a consistent theory of hidden liquidity and stimulated refill is well within reach at this stage.

\subsection{Empirical results}\label{empimp}

The section reviews how the above ideas can be directly tested and measured on high frequency data.  

\begin{figure}
\begin{center}
\includegraphics[width=12cm,height=9.2cm,angle=0]{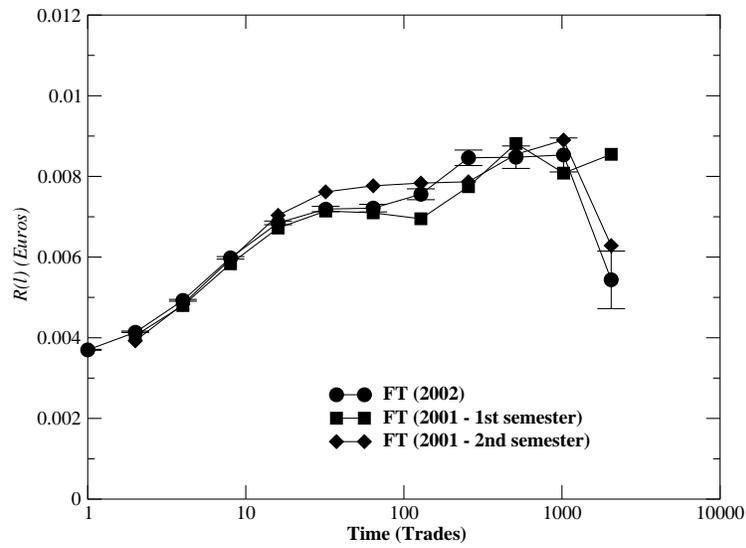}
\end{center}
\caption{Average empirical response function ${\cal R}_\ell$ for FT, during three different periods (first and second semester of 2001 and 2002). We have given error bars for the 2002 data. For the 2001 data, the $y-$axis has been rescaled such that ${\cal R}_1$ coincide with the 2002 result.
${\cal R}_\ell$ is seen to increase by a factor $\sim 2$ between $\ell=1$ and $\ell=100$.}
\label{FigRellFT}
\end{figure}

We start with the full impact function, defined by Eq. (\ref{Rdef}), which is easily measured -- at least when the lag $\ell$ is not too large. 
When $\ell$ becomes of the order of the number of daily trades or more, the error bar on ${\cal R}_{\ell}$ quickly becomes large. The main features
of ${\cal R}_{\ell}$ are however quite robust from stock to stock and also across different markets. For example, ${\cal R}_\ell$ for France Telecom 
in 2002 is shown in Fig.~(\ref{FigRellFT}). One sees a mild increase by a factor $\lambda \sim 2$ between $\ell=1$ and $\ell=1000$, before a saturation or maybe a decline
for larger lags. This behaviour is quite typical, in particular the roughly two-fold increase between small lags and large lags. So ${\cal R}_\ell$
reveals some non trivial temporal structure -- recall that ${\cal R}_\ell$ is constant within models where the  midpoint reacts to surprise in 
order flow. In an {\sc mrr} setting, the amplification factor $\lambda$ should be $1/(1-C_1)$, which in found to be in the range $1.2 -- 1.4$, still 
too small to explain $\lambda \sim 2$.

\begin{figure}
\begin{center}
\includegraphics[scale=0.5,angle=270]{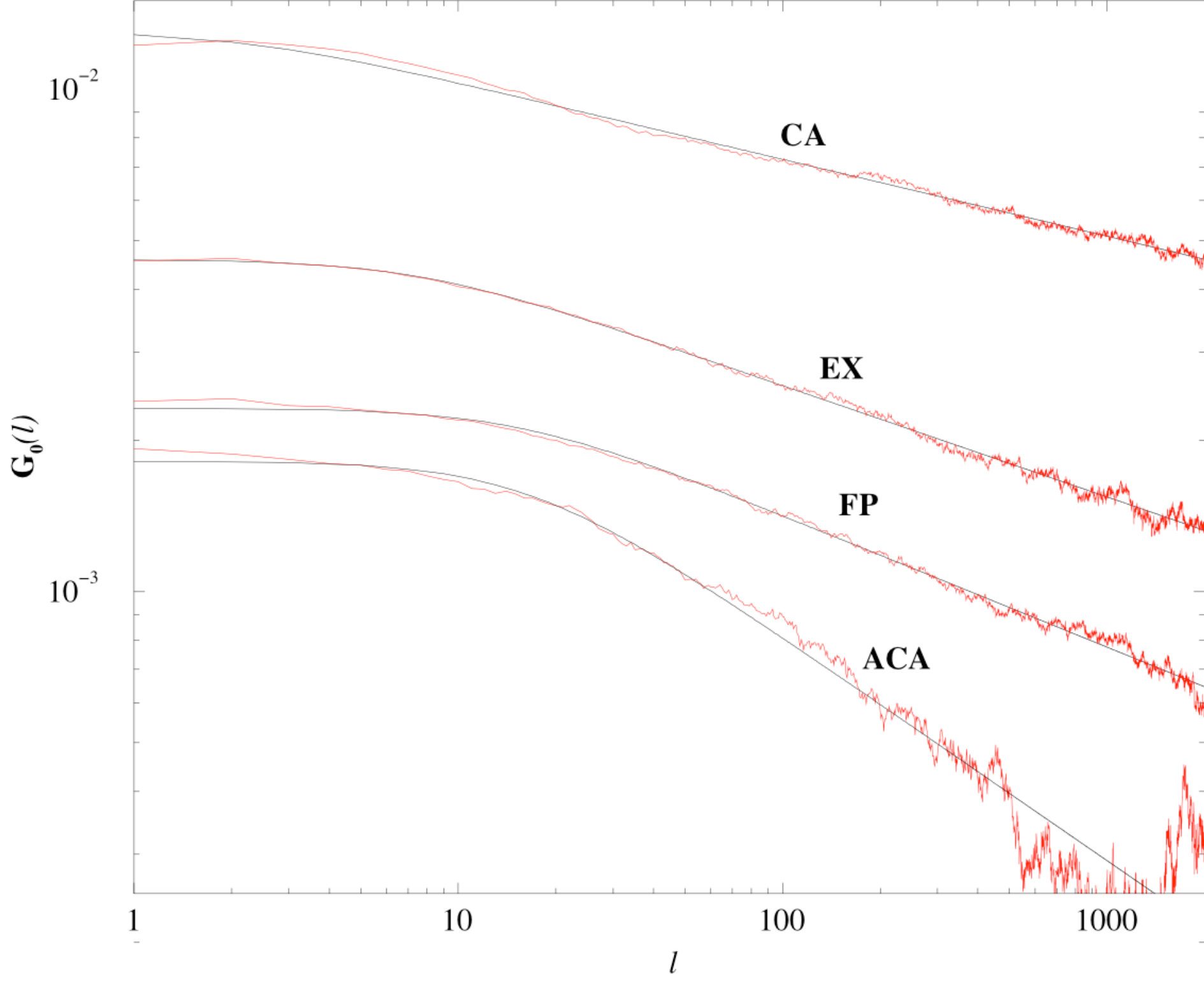}
\end{center}
\caption{Comparison betwen the empirically determined $G_0(\ell)$, extracted from ${\cal R}$ and
${\cal C}$ using Eq.(\ref{response}), and the power-law fit $G_0^f(\ell)=\Gamma_0/(\ell_0^2+\ell^2)^{\beta/2}$, for a selection of four stocks: ACA, CA, EX, FP.}
\label{FigG0ell}
\end{figure}

As noted above, one can in fact extract the theoretical impact of single trades $G_0(\ell)$ from the empirically measured impact ${\cal R}_\ell$ and 
the correlation between the sign of the trades $C_\ell$, using Eq. (\ref{response}). This was done in \cite{Bouchaud04b}, and indeed produces nice, power-law
decaying $G_0(\ell)$'s -- see Fig.~(\ref{FigG0ell}) for a few examples. Within the above restrictive theoretical framework, this provides a direct proof of the 
transient nature of the impact of single market orders and the long term resilience of markets. This is quite important as far as execution strategies are concerned -- see section \ref{execution}. 

We should however list a number of caveats. One is the assumption that the impact is time translation invariant, i.e. only the lag $\ell$ is 
relevant. This is clearly questionable, since strong intraday seasonality effects are expected. For example, there are indications that the trade 
sign correlation function $C_\ell$ for a given lag $\ell$ is quite different intraday and from one day to the next (\cite{Eisler08}). Similarly, we 
expect that the single trade impact should decay differently intraday and overnight. Second, we have to a large extent discarded the interesting 
correlations between the state of the order book $\Omega_n$, the incoming volume $v_n$ and the resulting impact (see Eq. (\ref{basicimpact}). 
All this complexity was replaced by an average description: $\epsilon_n f(v_n;\Omega_n) \longrightarrow \epsilon_n \ln v_n$. Certainly, a refined 
version is needed, in particular because the fluctuations of $f(v_n;\Omega_n)$ will contribute to the diffusion properties (see Eq. (\ref{diff})).
Finally, we have chosen from the start to give a special role to market orders, as if only those impact the price. But this is not true: obviously 
limit orders also impact the price. In fact, it is precisely the impact of limit orders
that offsets that of market orders and leads to a decay of the single trade impact $G_0(\ell)$. In other words, we have studied an effective model
in terms of market orders only, dumping into $G_0(\ell)$ the counter-acting effect of limit orders. A more symmetric version of the model, that 
treats market and limit orders on an equal footing, would be quite enticing (\cite{Eisler08}).

\begin{figure}
\begin{center}
\includegraphics[width=9.5cm]{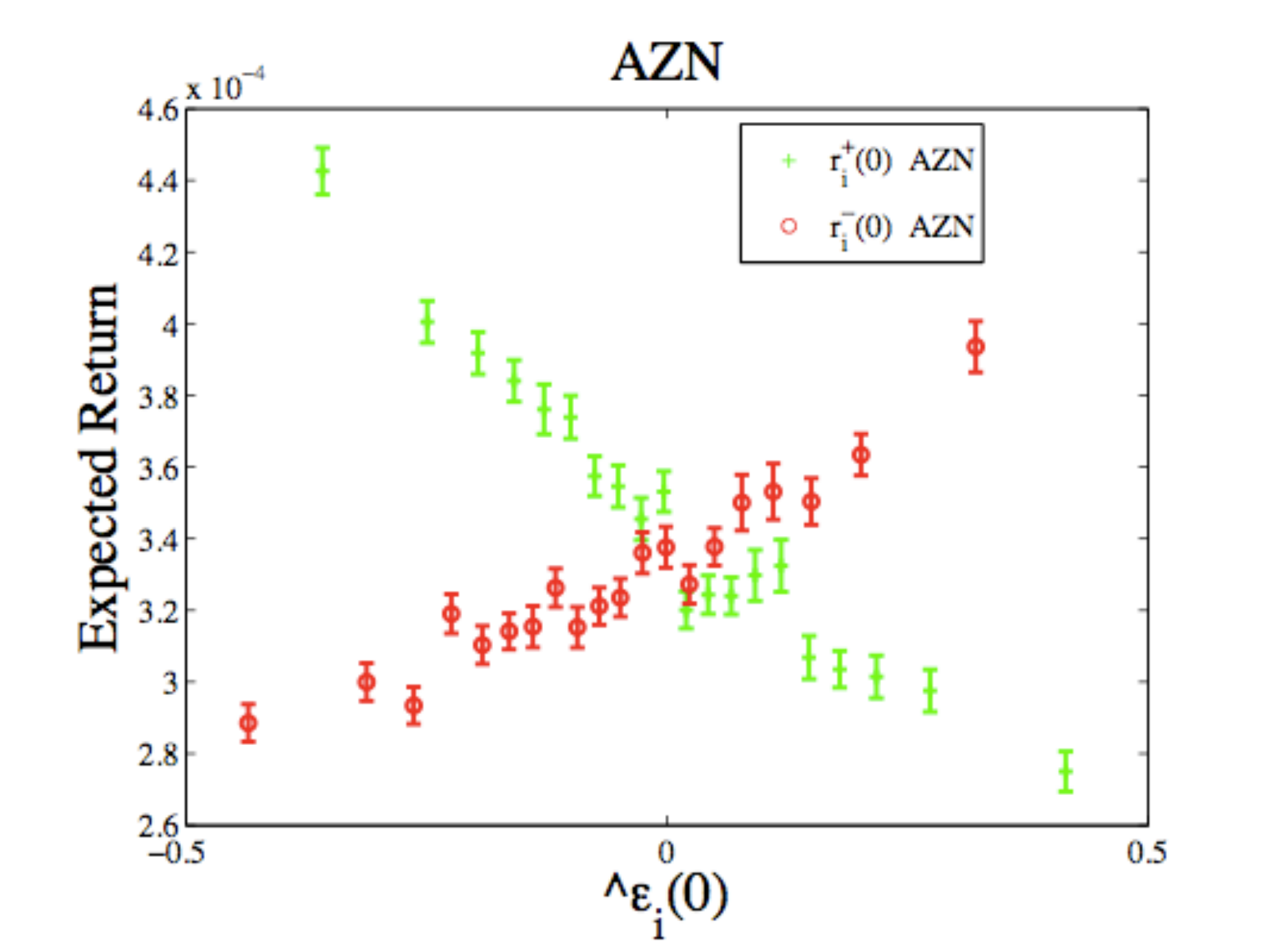}
\end{center}
\caption{The expected return  as a function of the sign predictor $\hat \epsilon$. The quantity $r^+$ ($r^-)$ refer to trades with a sign that is equal (opposite) to the one of the predictor. The data are binned in such a way that each point contains an equal number of observations. Error bars are standard errors. Adapted from \cite{Gerig07}.}
\label{symmetry}
\end{figure}

We now consider some empirical evidence for asymmetric liquidity. Figure \ref{symmetry} shows the behavior of the conditional returns $r^+$ and $r^-$ defined in Eq. \ref{symmetryeq} as a function of the sign predictor $\hat\epsilon$. The data we show in Fig. (\ref{symmetry}) is for Astrazeneca, 
a stock traded at the LSE. The sign predictor is the linear predictor defined in Eq. \ref{symmetryeq}. The larger the absolute value of $\hat\epsilon$, the stronger the predictability of the next market order sign. We have plotted the average value of the return conditioned to be in the direction of the predictor, $r^+$, and the average return when the sign of the predictor is wrong, $r^-$. We see that $r^-$ is indeed larger than $r^+$ and this difference increases with the predictability of the order flow. This is a clear evidence for asymmetric liquidity.  Note also that both $r^+$ and $r^-$ are approximately described by a linear function of the predictor $\hat\epsilon$. This is expected under the model described in Section \ref{predictable} (see Eq. \ref{symmetryeq}). However the slopes of $r^+$ and $r^-$ vs. $\hat\epsilon$ are different, challenging the implicit symmetric assumption (Eq. \ref{symmetryeq}) in the MRR model. Other evidence for the build up of the ``liquidity molasses'' accompanying the flow of market order can be found in \cite{Bouchaud04b} and \cite{Weber05}.

\subsection{Impact of a large hidden order \label{hiddenOrderImpact}}

We now want to calculate, within the above theoretical framework, the impact of an hidden order of size $N$. For simplicity, let us first assume that the hidden order is made of $N$ consecutive trades made by the same institution, though this remains ``hidden'' if trades are anonymous. 
Let us call $m_0$ the price at the beginning of the hidden order and  compute the average price $m_{N+t}$ observed $t$ transactions after the completion of the hidden order. Within the generalized MRR model with a linear predictor of the order flow, a straightforward calculation shows that
\begin{equation}
E[m_{N+t}]-m_0=\epsilon\theta\sum_{i=t+1}^{t+N}[1-\sum_{j=1}^{i-1}a_j]
\label{finalrev}
\end{equation}
For $t=0$ this expression gives the (temporary) total impact of the hidden order, while for $t>0$ we can calculate the  price reversion after the completion of the hidden order, and the permanent impact (if any) for $t\to \infty$.

The above result can be generalized to the case where there is only one hidden order active at a given time, which mixes with a flow of uncorrelated
orders with a constant participation rate $\pi$. The total time needed to execute the hidden order is then $T=N/\pi$. 
It is possible to show in this case that (Farmer, Gerig, Lillo, and Waelbroeck 2008):
\begin{equation}
E[m_{N}]-m_0= \epsilon\theta \sum_{i=1}^{N} \left( 1 - \sum_{k=1}^{i/\pi} a_k \right).
\label{impactx}
\end{equation}

Let us estimate the above formula in the case where the autocorrelation $C_\tau$ of order flow asymptotically decays
as a power law $C_\tau \sim \tau^{-\gamma}$ for large $\tau$.  There are several different ways of generating and forecasting long-memory processes. Here we assume that the participants observing public information model the time series with a FARIMA process.  It is known (\cite{Beran94}) that for large $k$ the best linear predictor coefficients of a FARIMA process satisfy $a_k \approx  k^{-\beta - 1}$ where $\beta = (1 - \gamma)/2$.  For large $k$ we can pass into the continuum limit and from Eq.~\ref{impactx} the impact is
\begin{equation}
E[m_{N}]-m_0 = \epsilon\theta\left[ 1+ \sum_{i=1}^{N-1} \left( 1 -  \left(1 - (n/\pi)^{-\beta} \right) \right) \right].
\end{equation}
Converting the sum to an integral gives
\begin{equation}
E[m_{N}]-m_0  \approx \epsilon\theta\left(1+\frac{2^{\beta-1}\pi^{\beta}}{1-\beta}[(2N-1)^{1-\beta}-1]\right)\sim \pi^{\beta} N^{1-\beta}.
\label{linearModelImpact}
\end{equation}
Thus, for a fixed participation rate, the market impact asymptotically increases with the length of the hidden order as $N^{1 - \beta}$.  A typical decay exponent for the autocorrelation of order signs is $\gamma \approx 0.5$ (\cite{Lillo03c,Bouchaud04}), which means that $\beta \approx 0.25$.  This means that according to the linear time series model the impact should increase as roughly the $3/4$ power of the order size.  An interesting property of this solution is that it depends on the speed of execution.  The size of the impact varies as $\pi^{\beta}$.  This means that the slower an order is executed, the less impact it has, and in the limit as the order is executed infinitely slowly the impact goes to zero. Note however that if the execution time $T=N/\pi$ is {\it fixed}, the impact become linear with $N$ but
decays as $T^{-\beta}$.

To investigate the reversion dynamics we make use again of the Eq. (\ref{finalrev}). We assume that the liquidity provider uses a FARIMA model to forecast order signs and for the sake of simplicity in the following we will assume that $\pi=1$, i.e. that there are no noise traders. Realistically the regression made by the liquidity provider on past signs will use a finite lag $K$, leading to:
\begin{equation}
\hat\epsilon_n=\sum_{i=1}^K a_i^{(K)}\epsilon_{n-i}
\end{equation}
where (\cite{Beran94}):
\begin{equation}
a_i^{(K)}=-{K\choose i}\frac{\Gamma(i-H+1/2)\Gamma(K-H-i+3/2)}{\Gamma(1/2-H)\Gamma(K-H+3/2)}
\end{equation} 
and $H=1/2-\beta$ is the Hurst exponent of the FARIMA process. 
It is possible to derive an analytical exact result for the permanent impact. In fact, from Eq.(\ref{finalrev}), one can obtain 
\begin{eqnarray}
E[m_\infty]-m_0=\epsilon\theta N(1-\sum_{j=1}^{K}a_j^{(K)})=\\
\epsilon\theta N\frac{4^{H-1}\sqrt{\pi}\Gamma[H]\sec[(K-H)\pi]}{\Gamma(3/2+K-H]\Gamma[2H-1-K]} \nonumber
\end{eqnarray}
By using the Stirling's formula and the reflection formula for the Gamma function one can show that for large $K$ the permanent impact scales as
\begin{equation}
E[m_\infty]- m_0 \sim \epsilon\theta\frac{N}{K^{\beta}}.
\end{equation}
If $K$ is infinite, then  $E[m_\infty]- m_0=0$, i.e. the impact is completely temporary. This can be shown in the mathematically equivalent propagator model \cite{Bouchaud04,Bouchaud04b}. For a FARIMA forecast model with finite $K$ (or equivalently if the sign autocorrelation function decays 
fast beyond time scale $K$), the permanent impact is non zero and is linear in $N$. Even if for large $K$ the permanent impact is small, the convergence to zero with the memory $K$ is very slow. 
%



Another interesting issue that can be discussed within the model is the decay of the impact immediately after the end of the hidden order (defined by Eq. (\ref{finalrev})). One finds that the initial drop for $t \ll N$ is in fact very sharp for $\beta < 1$:  $m_{N+t}-m_N \propto -t^{1-\beta}$, such that the slope of the decay is infinite when $t \to 0$ (in the continuous limit).

\subsection{Aggregated impact}\label{aggregatemod}

Impact is often measured not on a trade by trade level but rather on a coarse grained time scale, say five minutes or a day.  One then speaks of positive correlations between signed order flow and price returns. At the level of single trades, impact is strongly concave in volume and decays in time. How does this translate at a coarse-grained level? In Section \ref{empiricalAggregate} we have discussed this from an empirical point of view. Here we show how the impact theories we have developed so far make predictions about the impact function, following the approach of \cite{Lillo08}.

Suppose one aggregates the returns and volumes of $N$ consecutive trades (not necessarily from the same hidden order).  Using the same notation as in Section \ref{empiricalAggregate}, the total volume imbalance is $Q_N = \sum_{n=0}^{N-1} \epsilon_n v_n$. 
Conditioned to a particular value $Q_N=Q$, what is the average price return $R(Q)$ ? 
The answer to this question depends on the order flow and on the properties of the impact function. In the following we will consider two extreme cases. In the first case we consider an unrealistic model where the order flow is described by an independent identically distributed random process, 
and the impact is fixed and permanent. In the second case we will consider a correlated order flow and a fixed/temporary impact model.

\subsubsection{Independent identically distributed order flow} 

If the unconditional distribution of market order volume  and the functional form of the impact function are known, it is possible to find a closed expression for the impact $R(Q)$.
Consider a series of $N$ transactions with signed\footnote{
Only in this subsection we indicate with $v_i$ the signed and not the absolute value of volume.}
volumes $v_i$ corresponding to total return $R = \sum_{i=1}^{N} r_i$  and total signed volume $Q = \sum_{i=1}^{N} v_i$. 
The expected return given $Q$ can be written
\begin{equation}
R(Q, N) \equiv E[R |Q ]  = \int R P(R|Q,N)~dR=\frac{1}{P_{N}(Q)}\int R P(R,Q,N)~dR,
\label{ddd}
\end{equation}
where $P_{N}(Q)$ is the probability density for $Q$.   We assume that the $N$ individual price impacts $r_i$ due to the IID signed volumes $v_i$  are given by a deterministic function\footnote{The results remain the same if a noise term is added to the impact function.} $r_i=f(v_i)$.  Let the distribution of individual $v_i$ be $p(v_i)$.  Then the joint distribution of $v_i$ is $P(v_1, \ldots, v_{N}) = p(v_1) \ldots p(v_{N})$.
The integral above becomes
\begin{equation}
\int R P(R,Q,N)~dR= \int dv_1 \ldots dv_{N} p(v_1) \ldots p(v_{N}) \sum_{i=1}^{N} f(v_i) \delta(Q - \sum_{i = 1}^{N} v_i),
\label{ccc}
\end{equation}
where we introduced the Dirac delta function. 
By making use of  the integral representation of the Dirac delta function, after some manipulations it is possible to rewrite $R(Q,N)$ as
\begin{equation}
R(Q,N)=\frac{N}{2\pi}\frac{1}{P_{N}(Q)}\int d\lambda e^{(N-1) h(\lambda)} g(\lambda) e^{-i\lambda Q}
\label{superfinal}
\end{equation}   
where $h(\lambda)$ is the logarithm of the Fourier transform of the volume distribution and $g(\lambda)$ is the Fourier transform of the product of the volume distribution and the impact function. Moreover $P_{N}(Q)$ is the probability density that the total signed volume in the $N$ trades is $Q$. 

The functional form of the aggregate impact $R(Q,N)$ can be calculated by integrating this expression. 
It is possible to show that many of the properties of the solution are robust, independent of the details of the model.  For small values of $Q$ the aggregate impact $R(Q)$ is always linear with a slope which depends on $N$ and on the details of the volume distribution and of the impact function. For example if the impact function is a power law function $\epsilon |v|^\psi$ and the volume distribution decays asymptotically as $P(V)\sim V^{-\alpha-1}$, then for large $N$ the aggregate impact behaves for small $Q$ as
\begin{equation}
R(Q,N)\sim\frac{Q}{N^{\kappa}}
\end{equation} 
where $\kappa$ depends in a non trivial way on $\alpha$ and $\psi$ (see \cite{Lillo08}). For example,  if volumes have a finite second moment and the impact function is concave then $\kappa=0$; in constrast, if the second moment doesn't exist, and the impact function is sufficiently concave then $\kappa > 0$. The latter case agrees with what is seen in Fig. \ref{priceImpactFig}, where the slope of the aggregate impact decreases with $N$. Thus theories for the aggregate impact make falsifiable predictions connecting volumes, order flow and impact.

\subsubsection{Transient impact model} 

Within the model of section \ref{transient} above, the aggregate impact reads:
\be
R(Q,N) = \sum_{n=0}^{N-1} G_0(N-n) E[q_n|Q] + \sum_{m < 0} \left[G_0(N-m)-G_0(-m)\right] E[q_m|Q].
\ee
where $q_n=\epsilon_n \ln v_n$ and we assume that volumes are lognormally distributed (see Appendix 2).
Because trades are long ranged correlated, the second term is non-zero. But one can show it is subdominant when $N \gg 1$, so we discard it
in a first approximation. In the first term, one can compute $E[q_n|Q]=x$ using the fact that the $q_n$s are, within the model, Gaussian with rms $=s$. 
Noting also that typical values of $Q$ are of order $N^{1-\gamma/2} \ll N$, one finally finds:
\be
x \approx \frac{sQ}{{\cal I}N}, \qquad {\cal I}=2 \int_0^\infty {\rm d}u \,u \, e^{us-u^2/2}.
\ee
With $R(Q,N) \approx \Gamma_0 N^{1-\beta} x/(1-\beta)$ and the above relation between $\beta=(1-\gamma)/2$, we finally find the following result, written
in a suggestive scaling form:
\be
R(Q,N) = \sqrt{N} \frac{s\Gamma_0}{{\cal I}(1-\beta)} \left(\frac{Q}{N^{1-\gamma/2}}\right).
\ee
This means that by rescaling the return and the signed volume by their respective root mean square value, one obtains at large $N$ a limiting curve
which is a straight line. Whereas for small $N$ impact is strongly concave, impact becomes linear when $N \gg 1$. One can go one step further and
compute the leading non-linear correction in $Q$ when $N$ is large. One finds that it is negative, as a remnant of the small $N$ concavity, and
becomes noticeable at increasingly larger values of $Q \sim N$, as seen on empirical data -- see Fig (\ref{priceImpactFig}) below. 

The important conclusion of this model is that although the impact of individual trades is concave and decays in time, the compensating effect of correlated trades leads to a well defined {\it linear relation} between order imbalance and returns at an aggregated level. This is important because such a relation is often interpreted as a manifestation of the permanent component of the impact. 

Is this linear relation telling us that part of the trades have indeed {\it predicted} correctly the aggregated return (in Hasbrouck's words) -- see \cite{Hasbrouck07}? In light of the all the above results, it looks much more plausible to us that anonymous trades in fact statistically {\it induce} price changes, although in a quite non trivial and perhaps unexpected fashion.

\section{The determinants of the Bid-Ask spread}
 \label{spread}

In modern electronic markets, liquidity is {\it self-organized}, in the sense that any agent can choose, at any instant of time, to either 
provide liquidity or consume liquidity. The liquidity of the market is partially characterized by the bid-ask spread $S$, which sets the cost of an 
instantaneous round-trip of one share (a buy instantaneously followed by a sell, or vice versa).\footnote{Other determinants 
of liquidity discussed in the literature are the depth of the order book and market resiliency, see \cite{Black71,Kyle85}.} 
A liquid market is such that this cost is small. A question of both theoretical and practical crucial importance 
is to know what fixes the magnitude of the spread in the self-organized set-up of electronic markets, and the relative
merit of limit vs. market orders. In the economics literature (\cite{Ohara95,Biais97,Madhavan00,Glosten85}), the existence of the bid-ask spread 
is often attributed to three types of liquidity providing costs, \cite{Stoll78}: 
\begin{itemize}
\item (i) order processing costs (this includes the profit of the market maker); 
\item (ii) adverse selection costs: liquidity takers may have superior information on the future price of the stock, in which case the 
market maker loses money;
\item (iii) inventory risk: market makers may temporarily accumulate large long or short positions which
are risky. If agents are risk-sensitive and have to limit their exposure, this may add extra-costs.
\end{itemize}
A somewhat surprising conclusion of early econometric studies is that order processing costs account for a large fraction of the 
spread. This may make sense in illiquid markets where market makers exploit a monopolistic situation to open up large spreads, but
cannot be the correct picture in highly liquid, electronic markets in which market making is highly competitive. What we argue 
below is that the main determinant of the spread is in fact impact.

\subsection{The basic economics of spread and impact}

\subsubsection{The average gain of market makers}

What is the basic economics behind a trade, i.e. the encounter between a liquidity taker and one (or several) 
liquidity provider(s)?
Consider the sequence of all trades (not necessarily coming from the same hidden order). Let the $n$th trade have volume $v_n$ and sign $\epsilon_n$. 
The profit collectively made by liquidity providers on that given trade, marked to market at time $n+\ell$ is given by
\be
{\cal G}_L(n,n+\ell) =  v_n \epsilon_n \left[ (m_n + \epsilon_n \frac{S_n}{2}) - m_{n+\ell} \right],
\ee
where $S_n$ is the value of the spread at that moment in time. Think of a buy trade $\epsilon_n=+1$. The above equation compares the money received by the liquidity provider when the trade occurs ($v_n (m_n + \frac{S_n}{2})$) to its marked to market (midpoint) price at time $n+\ell$. 
Symmetrically, the profit made by the liquidity taker using market orders 
is ${\cal G}_L(n,n+\ell)=-{\cal G}_M(n,n+\ell)$. 
The above equation clearly shows that the profitability of market making comes from the spread ($+S_n/2$), while the losses are induced by market 
impact ($-\epsilon_n(m_{n+\ell}-m_n)$), which may or may not come from more informed traders (see below).  

Neglecting for simplicity volume fluctuations at this stage ($v_n \equiv v$), 
and using Eq. (\ref{Rdef}), we see that the average gain of the market maker in the absence of extra costs is given by:
\be\label{MMprofit}
E[{\cal G}_L](\ell) = v \left(E[\frac{S}{2}] - {\cal R}_\ell\right),
\ee
which shows explicitly that for a given total market impact ${\cal R}_\ell$, the spread $S$ should be larger than a minimum value for market making
strategies to be at all profitable on a time scale $\ell$ -- or else, for a given value of $S$, the impact function ${\cal R}_\ell$ should be
as small as possible. We recover here the idea that it is in the interest of liquidity providers to control the growth of ${\cal R}_\ell$ by 
tuning the liquidity asymmetry. 

In fact, the above reasoning neglects the cost of unwinding the market maker position, and a better estimate will be provided below. But the main message of the simple computation above is that the spread compensates for the impact of market orders. In the microstructure literature, this is refered to as `adverse selection'; as alluded to above, this implies that
market orders originate from better informed traders, with an information on the future price on average worth ${\cal R}_\ell$. But the same 
result would hold if impact was purely statistical, with no information content whatsoever. In fact, one could even revert the logic and
claim that it is the spread that determines the impact: if {\it some} trader accepted to pay $m_n + S_n/2$ for the stock, it is natural that the
market as a whole revises its fair price estimate from $m_n$ to $m_n + \alpha S_n/2$, where $\alpha \geq 0$ is a number measuring how trades 
influence the participants beliefs, leading to ${\cal R}_\infty= \alpha S/2$. The {\sc mrr} model with spread (see Section \ref{mrrspread}), in this context, assumes that market participants 
believe that the last traded price is indeed the correct price ($\alpha=1$). Clearly, in that model, the cost of a market order or the gain of a 
limit order are exactly zero. This leaves us, by the way, in the familiar but uncomfortable situation of the ``no trade theorem'': if the spread is such that the information content of a market order is compensated, why would the informed trader trade at all?

\subsubsection{How informed are the trades?}

So are some market orders informed? Can one finds convincing ex-post signatures of informed trades? A minimal definition of an informed trade is a 
trade that earns a profit significantly larger than the transaction costs (including both brokerage fees and market slippage). 
Introducing the 
signed return $r(n,n+\ell) \equiv \epsilon_n (m_{n+\ell}- m_{n})$, the profit of the $n$th market order on time scale $\ell$ is:
\be
{\cal G}_M(n,n+\ell) = v_n \left[r(n,n+\ell) - \frac{S_n}{2} \right].
\ee
Note that by definition the average of $r(n,n+\ell)$ is equal to the total impact ${\cal R}_\ell$, which is positive. If one averages the above 
equation over {\it all} trades, one in fact finds that $E[{\cal G}_M]$ is close to zero, which means that the spread compensates for the average impact, 
at least measured on short time scales $\ell$ (between a few seconds to a few days). 
More precisely, on liquid NYSE stocks in 2005 (when market makers were still present), one finds that $E[{\cal G}_M]$ is zero within error bars, which 
means that, after transaction costs, market orders lose money on average. The situation is slightly better on liquid PSE stocks in 2002, where one
finds $E[{\cal G}_M] = g E(S)/2$ with $g \approx 0.3$ (see Fig.~\ref{cac} below). This amounts to $3-5$ bp per trade, close to the transaction costs. So, on average, and although
market orders do impact prices, there does not seem to be much {\it short term} information in these orders -- at least judging from their ex-post profitability. The question of longer term information is of course left open here, simply because the statistics is not sufficient to judge the
average profitability of trades on long time scales, and also because long term drift effects cannot be neglected (on average, buy trades are 
profitable in the long run!).

\begin{figure}
\begin{center}
\includegraphics[scale=0.3]{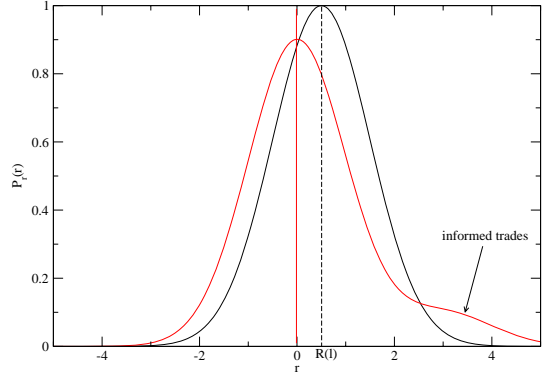}
\end{center}
\caption{Two extreme cases for the distribution $P(r_\ell)$ of signed returns $r_\ell$. (a) Black curve: nearly all trades are uninformed but impact prices, leading to a symmetric $P(r_\ell)$ around a non zero average impact. (b) Red curve: most trades are uninformed and do not impact prices, while some trades are informed and predict correctly the future return, leading to a thick tail in the $r_\ell > 0$ region.}
\label{cartoon}
\end{figure}

\begin{figure}
\begin{center}
\includegraphics[width=7cm,height=9.2cm,angle=270]{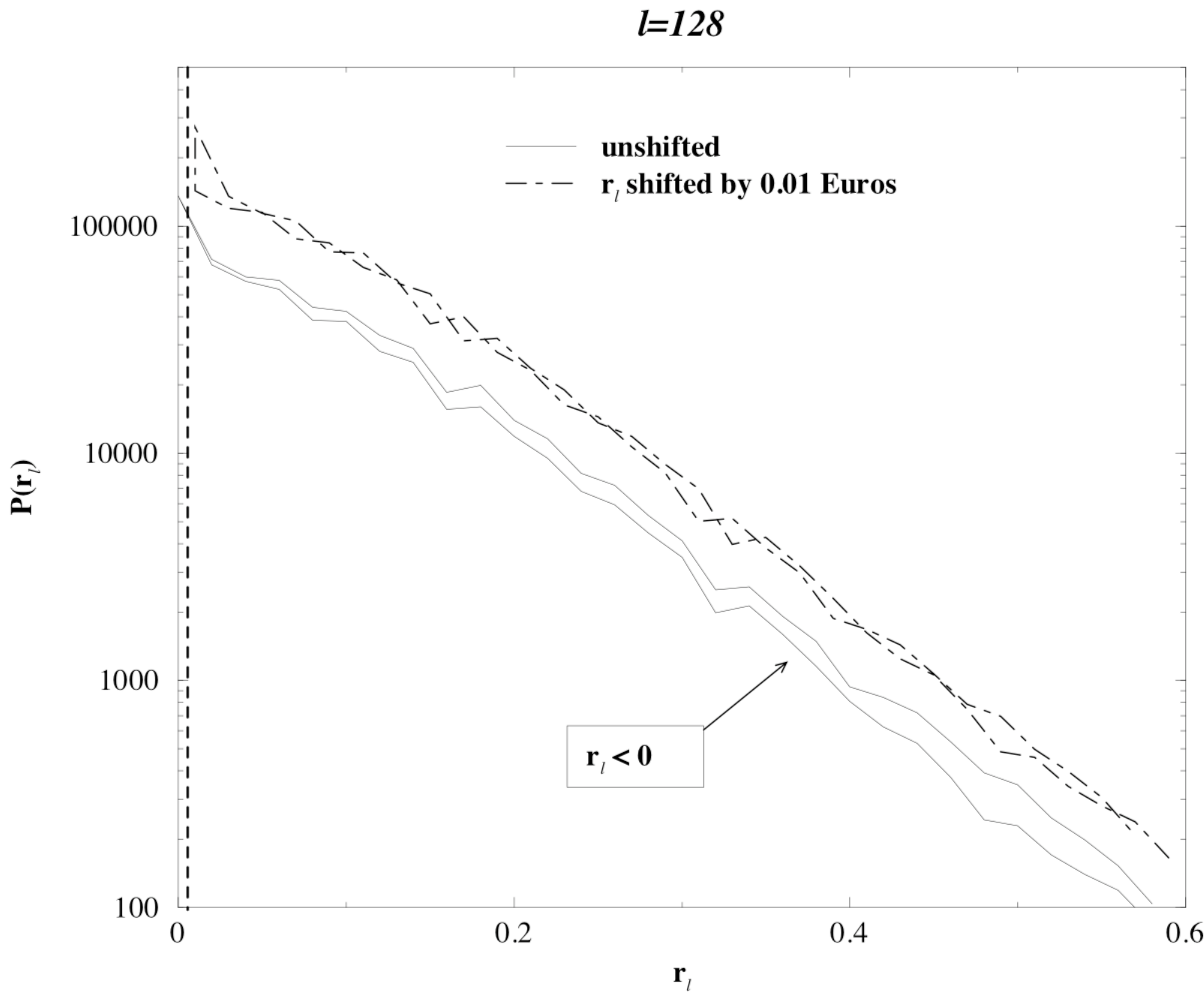}
\end{center}
\caption{Probability distribution $P(r_\ell)$ of the quantity
$r=(m_{n+\ell}-m_n).\varepsilon_n$ (in Euros), for $\ell=128$.
The data is again France Telecom during 2002.
The negative part of the distribution has been folded back to positive $r$ in order
to highlight the small positive assymetry of the distribution. The average value ${\cal R}_\ell =
E[r] \approx 0.01$ is shown as the vertical dashed line. The
dashed-dotted line corresponds to the distribution of $r-0.01$, for which no asymmetry of the type shown in Fig. (\ref{cartoon}) can be detected.
This curve has been shifted upwards for clarity.}
\label{FigPr}
\end{figure}

We can look in more detail at the full distribution of $r_\ell$, $P(r_\ell)$, which contains much more information. 
Note that its second moment $E[r^2|\ell]$ is very close to 
the volatility on scale $\ell$, which soon becomes much larger than ${\cal R}^2$ when $\ell$ increases. Concerning the shape of $P(r_\ell)$, 
two extreme scenarios could occur (see Fig.~\ref{cartoon} for a cartoon): 
\begin{itemize}
\item A small proportion of well informed trades {\it predict} the future price while a majority of trades are uninformed and 
do not impact the price at all. The distribution of $r_\ell$ should then be composed of a broad blob, 
symmetric around $r_\ell=0$, corresponding to uninformed trades, 
plus a hump (or more plausibly a broad shoulder) on the positive side, corresponding to well informed trades. The non zero value of $E[r_\ell]$ comes
from these informed trades. This is the scenario behind, for example, the Kyle model, or the Glosten-Milgrom model.
\item All trades are equally weakly informed or even not informed, but {\it all} statistically impact prices. 
In this case one expects a symmetric broad blob, but around the average impact $E[r_\ell]$.
\end{itemize}
Empirically, the distribution of $r_\ell$ is found to be very close to the second picture for $\ell$ corresponding to
intra-day time scales. In particular, no noticeable 
asymmetry (beyond the existence of a non zero value of $E[r_\ell]$) is observed on liquid stocks -- see Fig.~(\ref{FigPr}) for an example. 
This suggests that trades, on average, impact prices, but do not seem to `predict' future prices -- at least not on short time scales. The strong relation between order imbalance 
and price returns would then be tautological consequence of this impact (see section \ref{impactTemporal}), and not a signature of `true' information revelation. 

\subsection{Models for the bid-ask spread}

\subsubsection{The Glosten-Milgrom model}\label{gm-section}

One of the earliest theories of the spread that makes the above discussion is the sequential trade model of Glosten and Milgrom (\cite{Glosten85}). One assumes that market orders are either due (with some probability $q$) to informed traders, who know the end of day price $p_f$, or (with probability $1-q$) to noise traders. The value of $q$ is assumed to be known by the market maker, which is not necessarily very realistic (a similar assumption is made within the Kyle model). The end of day price $p_f$ can either be above ($p_>$) or below ($p_<$) the open price. The probabilities for either outcome at the start of the day are $\delta_+=\delta_-=1/2$ for simplicity. But as trading occurs, either at the bid or at the ask, the market maker updates in a Bayesian way the value of $\delta_+=1-\delta_-$: trades at the ask increase the value of $\delta_+$, while trades at the bid increase $\delta_-$. This leads to a certain update rule for $\delta_+$ as a function of the sign of the next trade, which we do not write here explicitly. Anticipating the value of $\delta_\pm$ after the next trade allows the market maker to position his quotes in such a way as not to have ex-post regrets. More precisely:                     
\be
a = \delta_+(+) p_> + \delta_-(+) p_<; \qquad b = \delta_+(-) p_> + \delta_-(-) p_<,
\ee
where $(\pm)$ refers to the sign of the next trade. This leads to the following prediction for the bid-ask $S_n$ after the $n$th trade:
\be
S_n = 4q \delta_+^{(n)}\delta_-^{(n)} (p_> -p_<)
\ee
where $\delta_\pm^{(n)}$ is the updated value of $\delta_\pm$ after $n$ trades (with $\delta_\pm^{(0)}=1/2$), and we have neglected terms of order $q^2$ which must be small if
this model is to be realistic. This model is by construction compatible with a random walk for the midpoint, with a volatility per trade $\sigma_1$ 
proportional to the bid-ask spread, as will be reported below. It also predicts that the bid-ask spread declines on average thoughout the day, since
the update rule drives $\delta_+$ either to zero or to one: as trading occurs, the market maker discovers more accurately which outcome is more likely.
A detailed comparison of this model with empirical data is given in \cite{Wiesinger08}. Here we simply note that as far as order of magnitude goes, 
the spread at the beginning of the day (when $\delta_\pm=1/2$) is typically $0.1 \%$ whereas the daily volatility fixes the order of magnitude of $p_> -p_<$ to typically $2 \%$, leading to $q \sim 0.05$. Within this framework, one finds again that the fraction of short time `informed trades' must be small.
One also finds that in this model the spread decays exponentially fast with time, at variance with the slow, power law relaxation that has been observed (see Section \ref{spreaddyn}).

\subsubsection{The MRR model with a bid-ask spread}\label{mrrspread}

The original {\sc mrr} model is in fact slightly different from the model described in Section \ref{mrrsec}. {\sc mrr} model rather assumes that it is the `true' fundamental price $p_n$, 
rather than the midpoint $m_n$, which is impacted by the surprise in order flow, and hence
\be
p_{n+1} - p_n = \eta_n + \theta [\epsilon_n -  \rho \epsilon_{n-1}].
\ee
{\sc mrr} then specify a rule for the bid and ask price, which in turn allows one to {\it compute} the midpoint $m_n$. Since market makers 
cannot guess the surprise of the next trade, they post a bid price $b_n$ and an ask price $a_n$ given by:
\be
a_n = p_n + \theta [1 -  \rho \epsilon_{n-1}] + \phi; \qquad b_n = p_n + \theta [-1 -  \rho \epsilon_{n-1}] - \phi,
\ee
where $\phi$ is the extra compensation claimed the market maker, covering processing costs and the shock component risk. 
The above rule ensures no {\it ex-post} regrets for the market maker: whatever the sign of the trade, the traded price is
always the `right' one. The midpoint $m \equiv (a+b)/2$ immediately before the $n$th trade 
is now given by:
\be
m_n = p_n - \theta \rho \epsilon_{n-1},
\ee
whereas the spread is given by $S=a-b=2(\theta+\phi)$. 

More generally, assuming that only the sign surprise matters, one can write, for arbitrary correlations between signs:
\be\label{surprise}
m_{n+\ell}-m_n = \sum_{j=n}^{n+\ell-1} \eta_n + \theta  \sum_{j=n}^{n+\ell-1} \left\{\epsilon_j- E_j[\epsilon_{j+1}]\right\},
\ee
where the last term is the conditional expectation of the next sign. In the Markovian case, $E_j[\epsilon_{j+1}]= \rho \epsilon_j$, and
we recover the above result. The impact function, in the general case, reads
\be\label{RellMRR}
{\cal R}_\ell = \theta \left[ 1 - C_\ell \right].
\ee
Using Eq. (\ref{MMprofit}), one sees that the long term profit of market makers is zero. However, due to correlations between trades, 
the long time impact is enhanced compared to the short term impact by a factor
\be
\label{enhance}
\lambda =\frac{1}{1-C_1} > 1. 
\ee

As discussed very generally above, spread and impact are two sides of the same coin. This is particularly clear within the {\sc mrr} model,
where the half-spread $S/2$ is set to be equal to the long term impact ${\cal R}_\infty=\theta$. This means that the profit of market makers is 
exactly zero (provided $\phi=0$), but also, as noted above, that the profit of putatively informed market orders is zero. 
The spread in the MRR model is
\be
S=2(\theta+\phi)=2({\cal R}_\infty+\phi)=2\lambda {\cal R}_1+2\phi
\ee
where $\lambda=(1-\rho)^{-1}$. An alternative, enlightening derivation is provided in Appendix 3.

One computes the mid-point volatility on scale $\ell$, defined as
\be
\sigma_\ell^2 = \frac{1}{\ell} \langle (m_{\ell+i}-m_i)^2 \rangle.
\ee				 
One finds a sum of a trade induced volatility $\theta^2 (1-\rho)^2$ and a `news' induced volatility $\Sigma^2$:
\be
\sigma_1^2 = \langle (m_{n+1} - m_n)^2 \rangle = \Sigma^2 + \theta^2 (1-\rho)^2 
\ee
and
\be
\sigma_\infty^2 =  \Sigma^2 + \theta^2 (1-\rho)^2 (1 + 2 \frac{\rho}{1 - \rho}) = \Sigma^2 + \theta^2 (1-\rho^2) \geq \sigma_1^2.
\ee
The {\sc mrr} model therefore leads to two simple relations between spread, impact and volatility per trade
\be\label{spreadmrr}
S = 2 \lambda {\cal R}_1 + 2 \phi; \qquad \sigma_1^2 = {\cal R}_1^2 + \Sigma^2,
\ee
where $\lambda=(1-\rho)^{-1}$ and $\phi$ is any extra compensation claimed by market makers. 
These relations will be generalized to more realistic assumptions and tested empirically in the next two sections. 

\subsection{Limit vs. market orders: the microstructure phase diagram}\label{phasediagram}

\subsubsection{Market order strategies}

\begin{figure}[tbp]
\begin{center}
\rotatebox{270}{\resizebox{7.0cm}{!}{\includegraphics{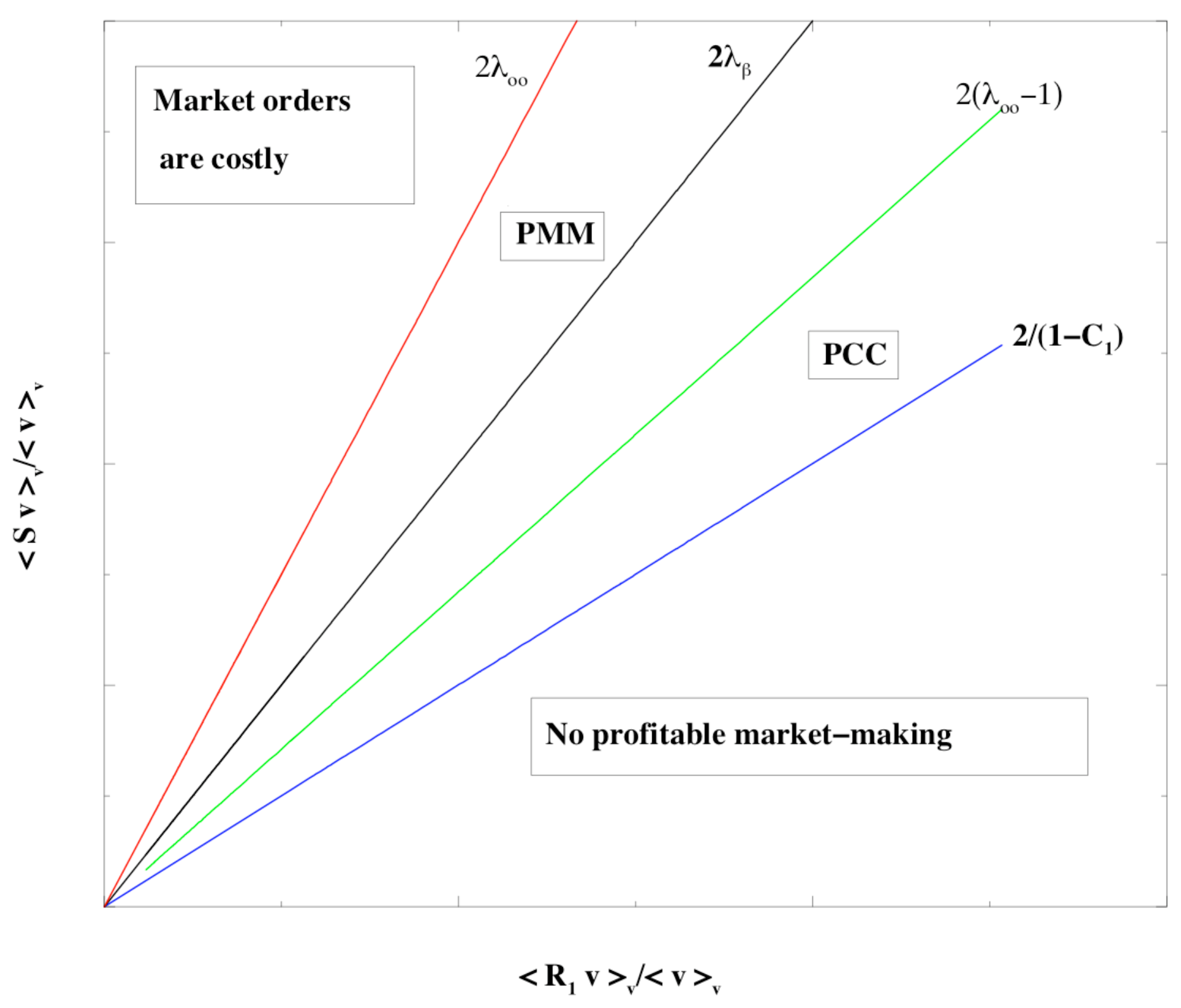}}}
\caption{\small{General ``phase diagram" in the plane $x={E[v {\cal R}_1(v)]}/{E[v]}$, $y={E[v S]}/{{E[v]}}$, showing several regions: 
(i) above the red line of slope $2 \lambda$, 
market orders are costly (on average) and market making is profitable; (ii) below the blue line of slope $\approx 2/(1-C_1)$, limit orders
are costly and no market-making strategy is profitable; (iii) above the black line of slope $2 {\overline \lambda}_\beta$, market making on time scale 
$\beta^{-1}$ (or faster) is profitable (PMM); (iv) below the green line of slope $2(\lambda-1)$, copy-cat strategies can be 
profitable (PCC). Since neither market orders nor liquidity providing should be systematically penalized for markets to ensure steady 
trading, we expect that markets should operate in the `neutral wedge' in between the blue and the red line. Competition between 
liquidity providers should push the market towards the blue line. Since copy-cat strategies should not be profitable either, the PCC green line cannot
lie above this blue line. Note that the blue, red and black lines all coincide within the {\sc mrr} model.
}}
\label{Diag}
\end{center}
\end{figure}

As we have mentioned above, the gain (or cost) of a given market order can be defined as $v_n [r(n,n+\ell) - \frac{S_n}{2}]$.
This definition in fact marks the trade to market after $\ell$ trades and is often referred to as the {\it realized spread} (\cite{Bessembinder03,Stoll00}). 
The volume weighted averaged gain (over a large number of trades) of market orders over a long horizon $\ell \gg 1$ is therefore:\footnote{Note that this definition neglects the fact that one single large 
market order may trigger transactions at several different prices, up the order book ladder, and pay more than the nominal spread. 
Nevertheless this situation is empirically quite rare on the markets we are concerned with, and 
corresponds to only a few percents of all cases \cite{Farmer04b}.}
\be\label{cc}
E[{\cal G}_M] \approx \lambda  \frac{E[v {\cal R}_1(v)]}{E[v]} - \frac{E[v S]}{2{E[v]}}.
\ee
In this expression we have introduced the volume dependent lagged impact function
\be
{\cal R}_\ell(v)=E[\epsilon_n(m_{n+\ell}-m_n)|v_n=v]
\ee
and we have used the above definition of the amplification factor $\lambda$: ${\cal R}_{\ell \gg 1} = \lambda {\cal R}_1$.
In the plane $x={E[v {\cal R}_1(v)]}/{E[v]}$, $y={E[v S]}/{{E[v]}}$ (which will repeatedly be used below), the condition 
$E[{\cal G}_M]=0$ defines a straight line of slope $2 \lambda$ separating an upper region where market orders 
are on average costly from a region where single market orders are favored: see the red line in Fig.~(\ref{Diag}). 
For large spreads, the positive average cost of market order would deter 
their use; limit orders would then pile up and reduce the spread.

Below the red line of slope $2 \lambda$ market orders have a negative cost, and 
one might be able to devise profitable strategies based solely on market orders. The idea would be to try 
to benefit from the impact term ${\cal R}_\infty$ in the above balance equation. The growth of ${\cal R}_\ell$ 
ultimately comes from the correlation between trades, i.e. the succession of buy (sell) trades that typically follow a given buy (sell) 
market order.  The simplest `copy-cat' strategy which one can rigorously test on empirical data is to imagine placing a market order with 
vanishing volume fraction (so as not to affect the subsequent history of quotes and trades), immediately following another market order. 
This strategy suffers on average from the impact of the initial trade, used as a guide to guess the direction of the market. 
Therefore, the profit ${\cal G}_{CC}$ of such a copy-cat strategy, marked to market after a long time and
neglecting further unwinding costs, is reduced to:
\be\label{cc3}
{\cal G}_{CC} = [\lambda-1]  \frac{E[v {\cal R}_1(v)]}{E[v]} - \frac{E[v S]}{2{E[v]}}.
\ee
By requiring that this gain is non-positive, one obtains a lower line in the plane $x,y$, of slope $2 (\lambda-1)$. 
Only below this green line can the above infinitesimal copy-cat strategy be profitable. 
We therefore expect markets to operate above this line and below the red line of slope $2\lambda$.

Note also that the long-time impact of an isolated market order, uncorrelated with the order flow, is given by $G_0(\ell \gg 1)$ 
which is small (see section \ref{temporal1}). These isolated market orders thus also have a positive 
cost equal to half the spread. The only way to benefit from the average impact ${\cal R}_\ell$ is to 
free-ride on a wave of orders launched by others, as in the above copy-cat strategy. Let us now take the 
complementary point of view of limit orders and determine the region of profitable market making strategies.

\subsubsection{An infinitesimal market making strategy}

We now compute the gain of a simple market making strategy which amounts to participating in a vanishing fraction of 
all trades through limit orders. The simplest strategy is to consider a market maker with a certain time horizon 
who provides an infinitesimal fraction $\varphi$ of the total available liquidity. As illustrated by Eq. (\ref{cc}), 
the cost incurred by the market maker comes from market impact: the price move is anti-correlated with 
the accumulated position. When the crowd buys, the price goes up while the market making strategy accumulates a short position which would be costly 
to buy back later, and vice-versa. 

We consider a {\it steady-state} market making strategy that avoids explicit unwinding costs. The strategy is such that tendered volume dynamically depends on the accumulated position, 
which insures that the inventory is always bounded. We choose the tendered fraction $\varphi$ to be given by $\varphi_i= \varphi_0(1 + \alpha V_i \epsilon)$, where $V_i$ is the (signed) position accumulated up to 
time $i^-$, and $\epsilon=+1$ for orders placed at the ask and $\epsilon=-1$ for orders placed at the bid. 
This mean-reverting strategy ensures that the typical position is always bounded. One can now use this strategy for
an arbitrary long time $T$; its profit \& loss is simply given by 
\be
{\cal G}_L = \sum_{i=0}^{T-1} \varphi_i \epsilon_i v_i (m_i +\epsilon_i \frac{S_i}{2}).
\ee
For large $T$ one can replace this expression by its average,
\be
{\cal G}_L = T E[ \varphi_i \epsilon_i v_i (m_i +\epsilon_i \frac{S_i}{2}) ],
\ee
with $O(T^0)$ corrections due to the residual position at $T$. This quantity has been computed in \cite{Wyart06}, and  depends
on the value of 
$\beta =1- \alpha \varphi_0 E[v]$ that fixes the typical time scale $\ell^*=(1-\beta)^{-1}$ of the market making strategy. 
When $\beta \to 0$
(fast market making), the gain per unit time and unit volume reduces to
\be
\frac{{\cal G}_L(\beta \to 0)}{T \varphi_0 E[v]} \approx \frac{E[v S]}{2E[v]} \left[1 - C_1 \right] - 
\frac{E[v {\cal R}_1(v)]}{E[v]},
\ee
whereas $\beta \to 1$, corresponding to slow market making, yields:
\be
\frac{{\cal G}_L(\beta \to 1)}{T \varphi_0 E[v]} = \frac{E[v S]}{2E[v]} -\frac{E[v {\cal R}_1(v)]}{E[v]}.
\ee
The competition between impact and spread is more favorable to limit orders when the strategy is fast ($\beta=0$)
than when it is slow ($\beta=1$). Imposing that there is a certain frequency $\beta$ such that the gain of 
market making strategies is zero leads to a linear relation between spread and impact, generalizing the above {\sc mrr} relation
Eq. (\ref{spreadmrr})
\be
\frac{E[v S]}{E[v]} = 2 \lambda_\beta \frac{E[v {\cal R}_1(v)]}{E[v]}.
\ee
Using the empirical shape of ${\cal R}_\ell$ and $C_\ell$, the slope $2\lambda_\beta$ 
is found to increase between $\approx 2/(1-C_1)$ and $2 \lambda$ when $\beta$ increases from zero to one. 
When $\beta \to  1$, $\lambda_\beta \to \lambda$ and 
the lower limit of profitability of very slow market making is precisely the red line of Fig.~(\ref{Diag}) where 
market orders become profitable. Faster strategies correspond to smaller values of $\lambda_\beta$, closer to $1/(1-C_1)$, 
leading to an extended region of profitability for market making. From the assumption that the above market making strategy for any 
value of $\beta$ should be at best marginally profitable (since one might find more sophisticated strategies, which take 
full advantage of the correlations between signs and volumes), we finally obtain the following bound 
between spread and impact: 
\be
\label{mmmg}
\frac{E[v S]}{E[v]}  \leq  \frac{2}{1-C_1} \frac{E[v {\cal R}_1(v)]}{E[v]},
\ee
defining the blue line of slope $2/(1-C_1)$ in the $x,y$ plane of Fig.~(\ref{Diag}). Consistently with the {\sc mrr} model, when 
$\lambda = 1/(1-C_1)$, the blue and red line of Fig.~(\ref{Diag}) exactly coincide. 
Using that fact that  ${\cal R}_1^{n+} \leq {\cal R}_1^{(n-1)+}$, a simple generalisation of the argument presented in Appendix 3 allows one 
to show directly that the cost of limit orders is indeed negative above the blue line.

Eqs.~(\ref{cc},\ref{mmmg}) and the resulting microstructural ``phase diagram'' of Fig.~(\ref{Diag}) are 
the central results of this section. The above analysis delineates, in the impact-spread plane, a central
wedge bounded from above by a slope $2 \lambda$ and from below by a slope $\approx 2/(1-C_1)$, within which both market 
orders and limit orders are viable. In the upper wedge, market orders would always be costly and would be substituted 
by limit orders. In the lower wedge, market making strategies, even at high frequencies, would never eke out any profit. 
Such a market would not be sustainable in the absence of any incentive to provide liquidity. But if the spread happened to fall in this
region, the enhanced flow of market orders would soon reopen the gap between bid and ask. In the {\sc mrr} model, this 
wedge reduces to a single line.

\subsubsection{Comparison with empirical data}

\begin{figure}[tbp]
\begin{center}
\rotatebox{270}{\resizebox{7.0cm}{!}{\includegraphics{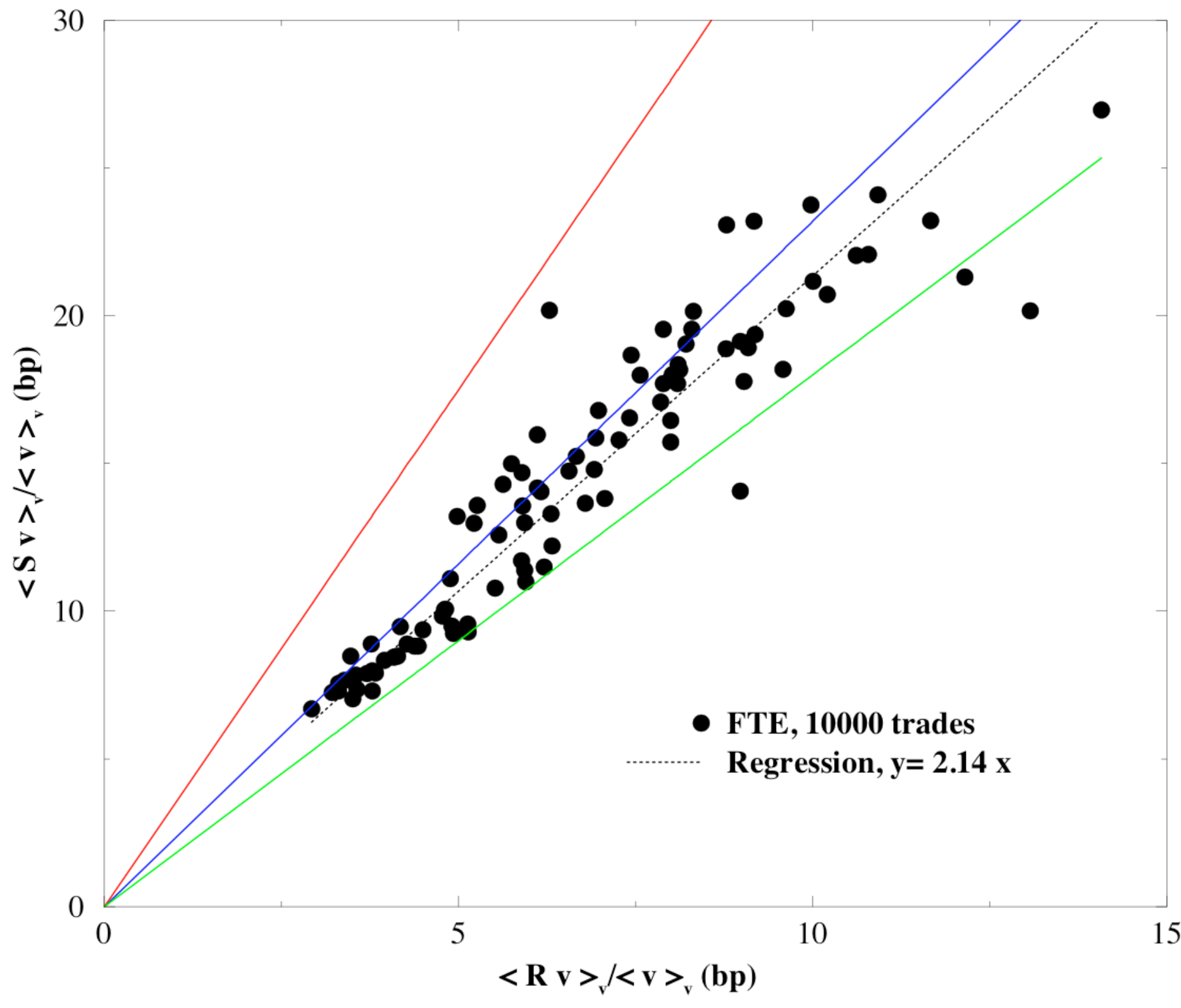}}}
\caption{\small{68 stocks of the Paris Stock Exchange in 2002. Each point corresponds to a pair 
($y=\langle v S \rangle/\langle v\rangle$, $x=\langle v {\cal R}_1 \rangle/\langle v\rangle$), computed by averaging 
over the year. Both quantities are expressed in basis points. We also show the different bounds, 
Eqs. (\ref{cc},\ref{cc3},\ref{mmmg}), 
and a linear fit that gives a slope of $2.86$, while $\langle 2/(1-C_1) \rangle \approx 2.64$. The correlation is $R^2=0.90$.}}
\label{cac}
\end{center}
\end{figure}

In conclusion of the above theoretical section, one expects electronic markets to operate in the vicinity of the blue line of Fig.~(\ref{Diag}), 
i.e. there should be a linear relation between spread and market impact with a slope close to $2/(1-C_1)$. This prediction
has been tested on empirical data in \cite{Wyart06}, where different markets were considered. The prediction can be tested in two different ways 
-- for a given stock across time, and across all different stocks. In both cases, a rather convincing agreement with the theory is 
obtained. We show for example in Fig.~(\ref{cac}) the cross-sectional test of Eq.(\ref{mmmg}) over 68 different stocks of the {\sc pse} in 2002.
The relative values of the spread and the average impact varies by a factor 5 between the different stocks, 
which makes it possible to test the linear relations (\ref{cc3},\ref{mmmg}). A linear fit with zero intercept gives a slope of $2.86$,\footnote{
The intercept of a two-parameter regression is in fact found to be slightly negative.} while the average of $2/(1-C_1)$ over all stocks is found 
to be $\approx 2.64$. 

\begin{figure}[tbp]
\begin{center}
\rotatebox{270}{\resizebox{7.0cm}{!}{\includegraphics{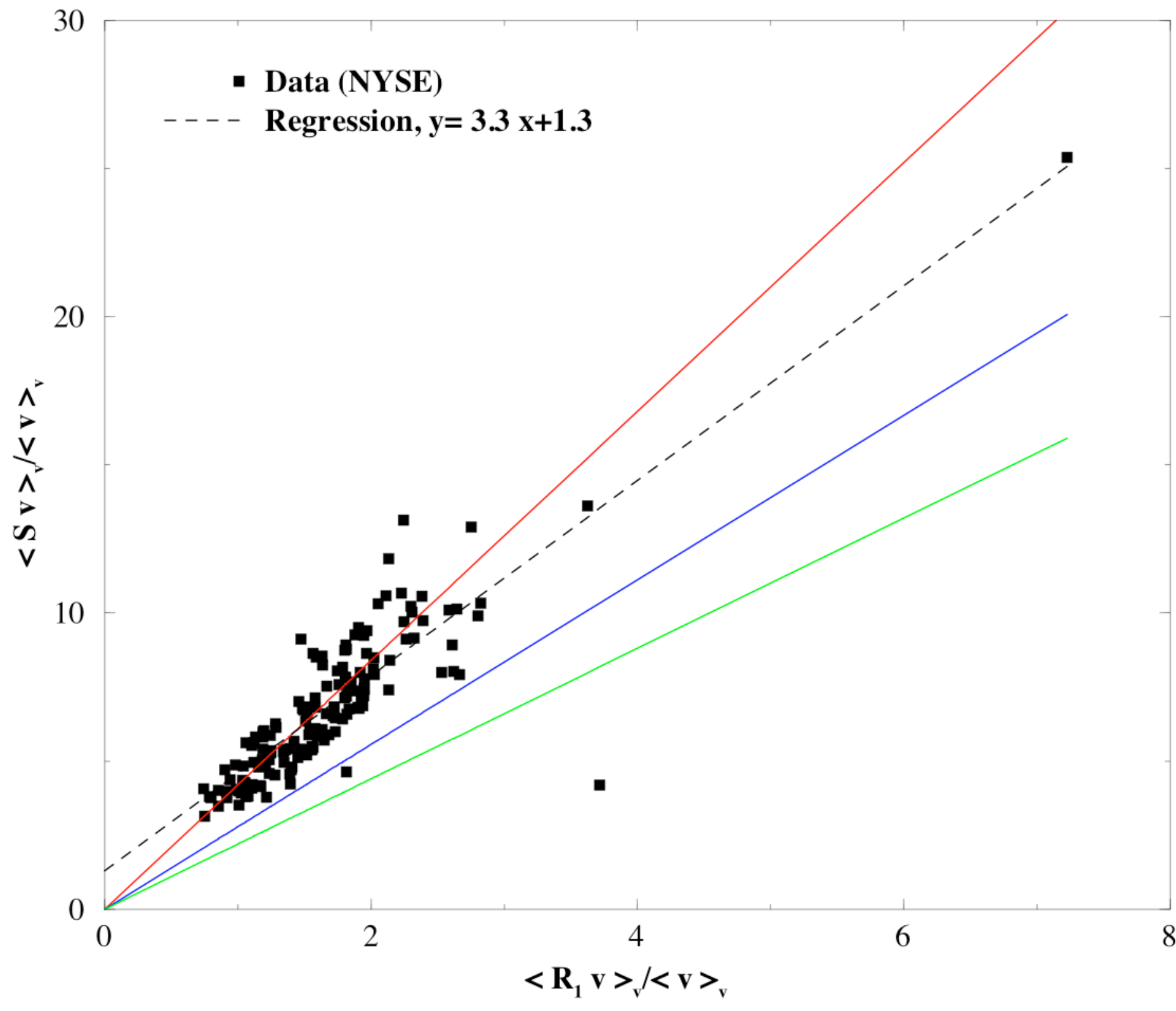}}}
\caption{\small{155 stocks of the {\sc nyse} 2005. Each point corresponds to a pair 
($y=\langle v S \rangle/\langle v\rangle$, $x=\langle v {\cal R}_1 \rangle/\langle v\rangle$), computed by averaging 
over the year. Both quantities are expressed in basis points. We also show our bounds, Eqs. (\ref{cc},\ref{cc3},\ref{mmmg}). 
The data shows clearly that market orders are less favorable than in the electronic Paris Bourse. The regression now has a 
positive intercept of $1.3$ bp with an $R^2=0.87$.}}
\label{nyser1}
\end{center}
\end{figure}

However, the situation appears to be different on the {\sc nyse}, where specialists are present.
Plotting the data corresponding to the 155 most actively traded stocks on the {\sc nyse} in 2005 
in the spread-impact plane, one now finds that the empirical results cluster around the upper red line 
limit where market orders become costly -- see Fig.~(\ref{nyser1}). The regression has a significantly larger slope of $3.3$, larger than $2/(1-C_1) \approx 2.78$,  
and a positive intercept $2 \phi \approx 1.3$ basis points.\footnote{This is five times smaller than the average spread, leading to 
$\phi/\theta \sim 0.25$, much smaller than the result $\phi/\theta \sim 1 - 2$ found within {\sc mrr} model in 1990, 
or a similar value reported in \cite{Stoll00}.} This suggests the existence of monopoly rents on {\sc nyse}: even if there is some competition to provide 
liquidity with other market participants.  Market makers post spreads that are systematically over-estimated compared to the situation 
in electronic markets, with a non-zero extrapolated spread $2 \phi$ for zero market impact. 
This result is in agreement with older studies on the {\sc nyse}: \cite{Harris96} used data from the early 90's to show that limit orders were more favorable than market orders; and \cite{Handa96} showed that pure limit order strategies were indeed profitable. We refer to \cite{Wyart06} for more discussion.

The empirical analysis therefore shows that on liquid markets, an approximate symmetry between limit and market orders holds, in the sense that neither market orders nor limit orders are systematically unfavorable. Markets operate in the `neutral wedge' of Fig.~(\ref{Diag}).  In fully electronic markets, competition for providing liquidity is efficient in keeping the spread close to its lowest value. On markets with specialists, such as the {\sc nyse}, spreads appears to be significantly larger market orders are now marginally costly on average. 

Note that the above analysis does not require any model specific assumptions such as the nature of order flow correlations or the fraction of informed trades, etc. In fact, the above results hold even if trades are all uninformed but still mechanically impact the price. 

\subsection{Spread dynamics after a temporary liquidity crisis}\label{spreaddyn}

The above analysis has shown the existence of relations between market impact and the unconditional value of the spread. The spread, however, is a variable with interesting temporal dynamics. Several studies have characterized the statistical properties of spread. Generally these studies have found that the spread distribution is fat tailed and the time correlation properties are consistent with a long memory process (\cite{Plerou05,Mike05,Gu07}). 

It is also interesting to ask how the spread responds after a temporary liquidity crisis. As we will describe in more detail in Section \ref{gapsec}, even at the scale of individual transactions, price returns are heavy tailed, i.e it is not unfrequent to observe individual transactions triggering large price changes. This often happens because a market order removes all the volume at the best, and the next to best occupied price level has a price very different from the price at the best (\cite{Farmer04b}). As a consequence even a small order can create a large price change creating a very large spread. A large spread is what we mean here by a ``temporary liquidity crisis".  

\begin{figure}[ptb]
\begin{center}
\includegraphics[scale=0.35,angle=-90]{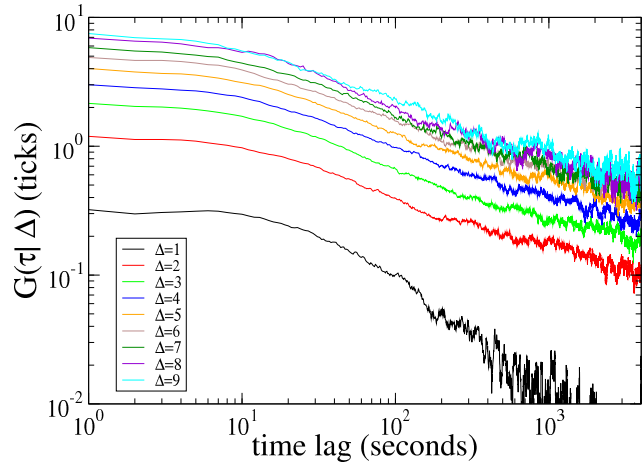}
\end{center}
\caption{Conditional spread decay $G(\tau|\Delta)$ defined in
Eq.~\ref{g} for the stock AZN. The figure shows $G(\tau|\Delta)$ for
different positive values of $\Delta$ (in ticks) corresponding to an
opening of the spread at time lag $\tau=0$. Adapted from Ref. \cite{Ponzi06}.}
\label{AZNcondspread}
\end{figure}

We will now describe the average dynamics followed by the spread as it converges to its "typical" value. First of all a large spread is a strong incentive for limit orders inside the spread and a strong disincentive for market orders. Direct measurements of the order flow conditional on the spread value confirm this intuition (\cite{Mike05,Ponzi06}). The limit order flow inside the spread has a limit price distribution which is roughly independent of spread size and monotonically decreasing when one moves from the same best toward the opposite best. This suggests that the typical spread dynamics is not a fast reversion to its typical value, but rather it is a slow process where each liquidity provider competes with the others to close the spread.  Each player tries to do this as slowly as possible in order to get a more favorable price from the incoming market orders, but at the same time competition prevents this from being too slow.  Empirically this slow decay has been measured in \cite{Zawadowski06,Ponzi06}. One way of quantifying the average dynamics is by computing the quantity, \cite{Ponzi06},
\begin{equation}
G(\tau|\Delta)=E(S_{t+\tau}|S_t-S_{t-1}=\Delta)-E(S_t)
\label{g}
\end{equation}
where $S_t$ is the spread at time $t$ (in seconds).  This quantity is the expected value of the spread at time $t+\tau$ conditional to the fact that at time zero there is a spread change of size $\Delta$. Figure~\ref{AZNcondspread} shows this quantity for the stock AZN traded at the LSE as a
function of $\tau$ for different positive and negative values of
$\Delta$. The decay of $G(\tau|\Delta)$ as a function of $\tau$ is
very slow and for large values of $\tau$ is compatible with a power
law decay with a fitted exponent in the range $0.4-0.5$. A similar slow decay of the volatility after a shock has been reported in 
\cite{Lillo03,Zawadowski06,Bouchaud08}.

\section{Liquidity and volatility}
\label{liquidityVolatility}

\subsection{Liquidity and large price changes}\label{gapsec}

One of the best known statistical regularities of financial time series is the fact that the empirical distribution of asset price changes is heavy tailed, i.e. there is a higher probability of extreme events than in a Gaussian distribution. This property has been verified by many authors on many different financial time series (for example, \cite{Mandelbrot63,Lux96,Gopikrishnan98}). Extensive empirical analyses have shown that the distribution of price change over time intervals ranging from few minutes to one or few trading days is asymptotically distributed in a way that is approximately independent of the time interval size. Many estimates indicate that the part of the distribution describing large price changes is a power-law. For larger time intervals the asymptotic behavior of the return distribution becomes slowly consistent with a Gaussian tail in accordance with Central Limit Theorem.   The heavy tailed property of large price change is important for financial risk, since it means that large price fluctuations are much more common than one might expect under a Gaussian hypothesis. 

There have been several conjectures about the origin of heavy tails in prices. Two theories that make testable hypotheses about the detailed underlying mechanism are the subordinated random process theory of \cite{Clark73} and the recent theory of \cite{Gabaix03}.  
The first model has its origins in a proposal of \cite{Mandelbrot67} that was developed by Clark. Mandelbrot and Taylor proposed that prices could be modeled as a subordinated random process $Y(t) = X(\tau(t))$, where $Y$ is the random process generating returns, $X$ is Brownian motion and $\tau(t)$ is a stochastic time clock whose increments are independent and identically distributed and uncorrelated with $X$. Clark hypothesized that the time clock $\tau(t)$ is the cumulative trading volume in time $t$. In simple terms, the subordination hypothesis states that price changes would be Gaussian if one measured them in equal intervals of volume (or number of trades) rather than in real time intervals.  

Gabaix et al.'s proposal, in contrast, is that high volume orders cause large price movements. They argue that the distribution of large trade size scales as $P(V>x)\sim x^{-\alpha}$, where $v$ is the volume of the trade and $\alpha\approx 1.5$. Based the assumption that agents maximize a first order utility function, with a risk penalty term that is proportional to standard deviation rather than variance, they claim that the average market impact function has the form $\Delta p\propto V^\psi$, where $\psi\approx 0.5$. From this follows that large price changes have a power law distribution with exponent $\alpha/\psi\approx 3$.  For a critique of the empirical results and a rebuttal see \cite{Farmer04,Plerou04}  

Both the Clark and the Gabaix theories emphasize the role of trading volume as the determinant of large price changes. Even if it is clear that volume has some role in determining price changes, recent studies show that trading volume could not be the key factor. In a recent paper \cite{Farmer04b} considered the distribution of returns generated by individual market orders. They showed that at even at this microscopic time scale price returns are heavily tailed, and more importantly, the size of price moves is essentially independent of the volume of the orders. Both these facts seriously challenge the explanation of fat tails based on volume fluctuations. In this paper Farmer et al. showed that price returns associated with individual transactions are driven by liquidity fluctuations. The authors proposed and tested a mechanism for explaining how liquidity fluctuations determine large price changes. Even for the most liquid stocks in the London Stock Exchange,  the limit order book often contains large gaps, corresponding to a block of adjacent price levels containing no quotes. When such  a gap exists next to the best price, a new market order can remove the best quote and generate a large price change. At this time scale the distribution of large price changes merely reflects the distribution of gap sizes in the order book. The LSE data indicate that approximately $85\%$ of the trades having a non zero price impact have a volume equal to the volume at the best. Moreover $97\%$ of the trades having a non zero price impact generate a price change equal to the first gap. In summary the fluctuations of the gap sizes in the book are a key determinant of large price changes. The gap size is a measure of the liquidity available in the market as limit orders. Thus  fluctuations of liquidity, i.e. in the market's ability to absorb new market orders, are the origin of large price changes, while the trading volume plays a minor role. 

The above proposed mechanism raises the question of the importance of temporary liquidity crises, evidenced by large gaps in the book, for price changes over long time intervals. Although a definite answer is not available, there are three indications that short time scale and long time scale price fluctuations may be related.
First, the gap size displays long memory properties in time, \cite{Lillo05}. This means that the gap size, i.e. the liquidity availability, is strongly correlated in time. Periods when the typical gap size is large are likely to be followed by periods of large gaps, i.e. liquidity availability is a persistent quantity. Second, it has been recently shown that the permanent component of the price impact is roughly proportional to the immediate impact caused by the trade (\cite{Ponzi06}). Thus the distribution of permanent price impacts, which is closely related to the distribution of price changes over relatively long time intervals, is approximately the same as the distribution of temporary price impacts, i.e. of gaps in the order book.  The third indication concerns the relative importance of volume and liquidity in explaining aggregate price changes, which is discussed in more detail in the next section.

\subsection{Volume vs. liquidity fluctuations as proximate causes of volatility}

\begin{figure}[ptb]
\includegraphics[scale=0.35,angle=0]{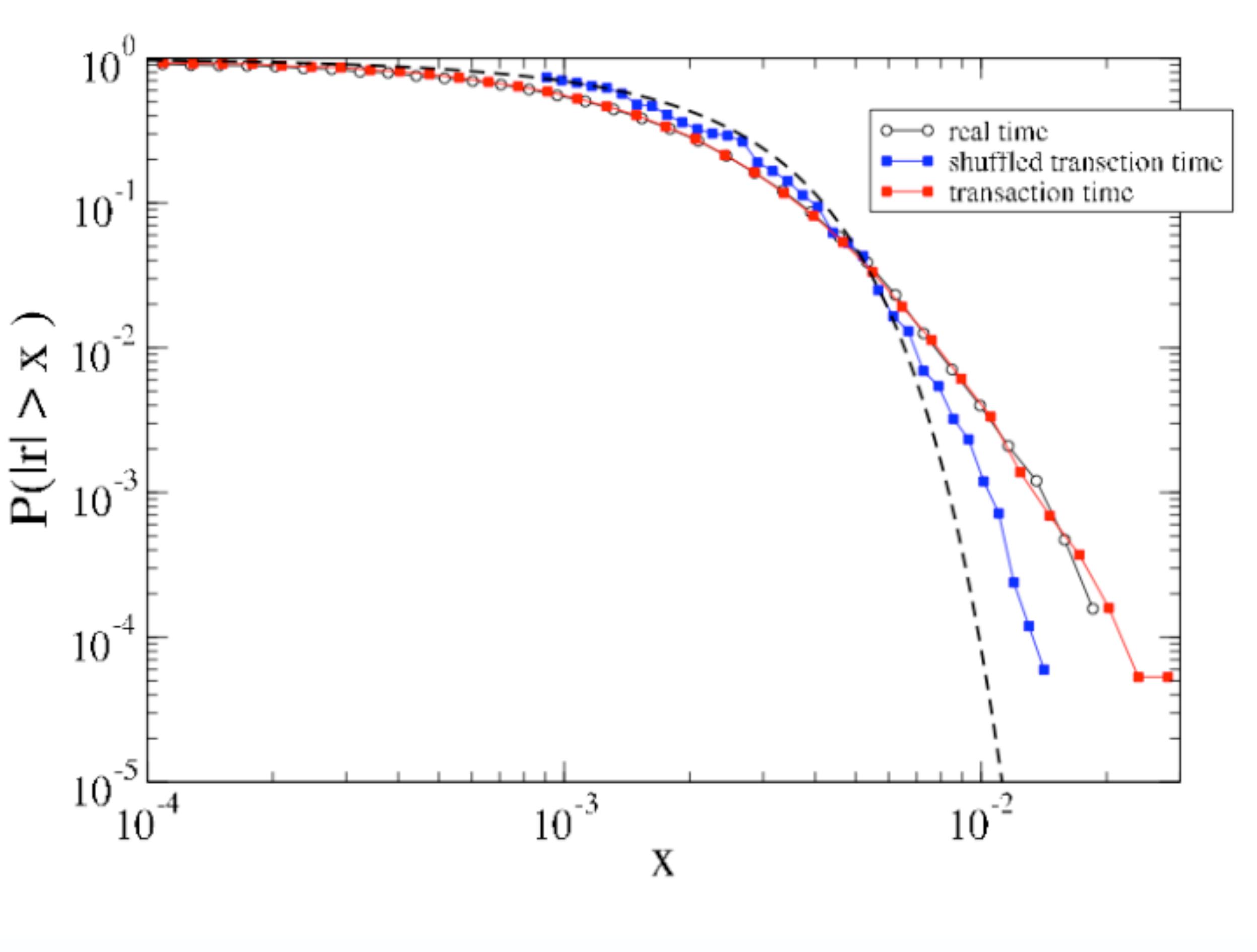}
\caption{Cumulative distribution of absolute (log)returns $P(|r| >x)$ for the NYSE stock Procter \& Gamble under different time clocks, plotted on double logarithmic scale.
The black circles refer to $15$ min returns, the red squares refer to returns aggregated with a fixed number of transactions, and the blue square shows the cumulative distribution obtained by randomly shuffling individual transaction returns and then aggregating them in a way that matches the number of transactions in each real time interval.  The dashed black line corresponds to a normal distribution.}
\label{markettimes}
\end{figure}

The existence of a relation between volume and volatility has been known for a long time. This relation has been often interpreted as a causal relation, suggesting that volume (or number of transactions) is the driving factor determining volatility (\cite{Ane00}). In the previous section we discussed the subordination hypothesis, which states that returns would be Gaussian if  measured in equal intervals of volume rather than in equal intervals of real time.  The recent theory by Gabaix et al. (2003,2006) reaches the same conclusion. Here we present some evidence challenging this view and indicating that liquidity fluctuations may be more important than volume in explaining volatility fluctuations.  The question can be posed in terms of Equation~\ref{liquidityAndVolume}, i.e. $\Delta p = \mathcal{T}(I)/\lambda$:  Which is more important in determining the size of price movements, $\mathcal{T}(I)$ or $\lambda$?

In a recent paper, \cite{Gillemot05} have presented evidence based on several different tests, involving comparisons of long-memory and regressions of the volatility in specific time intervals, showing that liquidity is a more important determinant of volume.  Even when one aggregates returns over a fixed number of transactions (or volume) the return probability density function remains heavy tailed with properties very similar to those in fixed intervals of time. A simple way to see this effect is given in Figure~\ref{markettimes}, which shows the empirical probability $P(|r| > x)$ as a function of $x$ for the NYSE stock Procter \& Gamble.  Here $r$ is the price return over a 15 minute time interval. Suppose returns are measured in transaction time, i.e. every $87$ transactions rather than every 15 minutes, where $87$ is chosen because it is the average number of transactions in 15 minutes (during the period from Jan 29, 2001 to December 31, 2003). The empirical distribution of transaction time returns matches that of real time returns very well. Since in this case the number of transaction is held constant, the shows that the heavy tail of the return distribution is not due to variations in the number of transactions. The same effect is seen by aggregating transactions with volume rather than the number of transactions fixed (see \cite{Gillemot05} for details).  

This result shows that the fluctuation in number of trades or volume associated with a fluctuating trading activity is not the main determinant of the heavy tails of the return distribution. To highlight this effect, Figure~\ref{markettimes} also shows the distribution of returns obtained from a surrogate distribution, constructed by randomly shuffling the returns of individual transactions and by aggregating them in a way that matches the number of transactions in each real time interval. In doing so the unconditional distribution of returns of individual transactions is preserved, as well as the fluctuation properties of trading frequency, but any temporal correlations of individual trade returns are destroyed. The figure shows that the tail of the surrogate distribution is less heavy than the real one, indicating that fluctuations and the time correlation properties of the reaction of prices to trades, i.e. liquidity, are more important than fluctuations in trading frequency. 

More supporting evidence for the importance of liquidity in determining volatility comes from a recent paper testing the microscopic random walk hypothesis against real data (\cite{LaSpada08}). The price dynamics can be described as a random walk in which the increments are  due to individual transactions. Under the assumption that the sign and the size of the price increments are mutually independent stochastic processes, it is possible to derive an exact expression for the volatility expected in a time interval with a given number of transactions. When one tests this expression on real data, it is found that for one hour intervals the model consistently over-predicts the volatility of real price by about $70\%$ and that this effect becomes stronger as the length of the time interval increases.  This fact suggests that the assumption of independence of size and sign of price changes is wrong. However data show that the contemporaneous correlation between size and sign of returns is non statistically significant. By performing a series of shuffling experiments,  \cite{LaSpada08} show that the discrepancy between the volatility of the model and of the data is caused by a subtle but long-memory non-contemporaneous correlation between the signs and sizes of individual returns. Therefore, even after controlling for the number of transactions and the order imbalance in a given time interval, the random walk model has a strong bias in predicting the volatility, which is caused by the long-memory of liquidity.  This once against indicats that volume is not the key factor in explaining volatility. The neglected subtle relation between return signs and sizes shows that fluctuating liquidity is an important factor in explaining volatility\footnote{ In Section \ref{impactTemporal} we discuss how such a correlation is a consequence of the long memory of order flow and of market efficiency. The asymmetric liquidity models described in Section \ref{impactTemporal} predict a reduction of volatility relative to what one would expect under an unconditional permanent impact model.}. 

Finally, the correlation between large volumes and large returns was directly studied in \cite{Bouchaud08}, both for trade by trade data and for one-minute 
bins, with the conclusion that such a correlation is totally absent from the data.

\subsection{Spread vs. volatility}
 
 It is worth investigating the relation between spread and volatility in the framework of the {\sc mrr} model discussed above. In fact this model predicts a simple relation between volatility and impact, as can be seen from  Eq. (\ref{spreadmrr}). Together with the relation between spread and impact discussed at length above, 
this suggests a direct link between volatility per trade and spread, which we motivate and test in this section. 
 
\begin{figure}[tbp]
 \begin{center}
\rotatebox{270}{\resizebox{7.0cm}{!}{\includegraphics{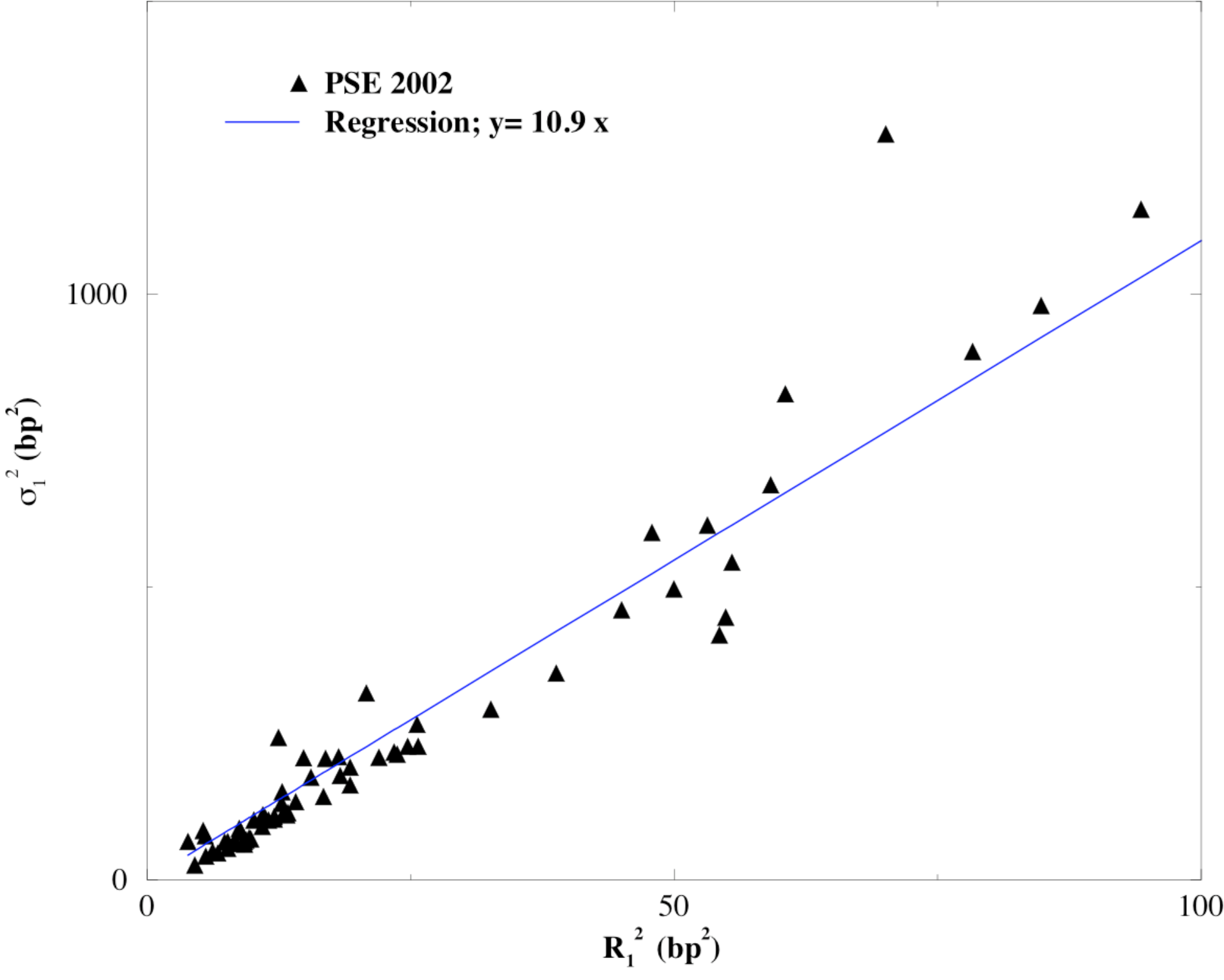}}}
 \caption{\small{Plot of $\sigma_1^2$ vs.  ${\overline{\cal R}}_1^2$, showing that the linear relation Eq. (\ref{dd2}) holds quite 
 precisely with $\Sigma^2 = 0$ and $a \approx 10.9$. (The intercept of the
 best affine regression is even found to be slightly negative). Data here corresponds to the 68 stocks of the {\sc pse} in 2002. 
 The correlation is very high: $R^2=0.96$.
 }}
 \label{sigvsR}
 \end{center}
 \end{figure}

By definition of the volatility per trade $\sigma_1^2 = E[(m_{\ell+1}-m_\ell)^2]$ and of the
instantaneous impact $r_{i,i+1} \equiv (m_{i+1}-m_i).\epsilon_i$, one has as an identity:\footnote{Neglecting the extremely 
small drift contribution.}
\be 
\label{dd}
\sigma _1^2 \equiv E[r_{i,i+1}^2].
\ee
The instantaneous impact $r_{i,i+1}$ is expected to fluctuate over time for several reasons. First, the volume of the trade, 
the volume in the book and the spread strongly fluctuate with time (\cite{Mike05,Wyart06}). Large impact fluctuations may also 
arise from quote revisions due to addition or cancellation of limit orders. Second, there might also be important news affecting 
the `fundamental price' of the stock. These may result in large, instantaneous jumps of the mid-point with virtually no trade at all. 
In order to account for both effects, one may  generalize the above {\sc mrr} relation (Eq. \ref{spreadmrr}) as in \cite{Bouchaud04,Rosenow02,Wyart06}:
\be\label{dd2}
\sigma _1^2 = A {\cal R}_1^2 + \Sigma^2,
\ee
where ${\cal R}_1 \equiv E[{\cal R}_1(v)]$ is the average impact after one trade, $A$ is a coefficient accounting for the variance 
of impact fluctuations and $\Sigma^2$ is the news component of the volatility (see Section \ref{temporal1}). This relation holds quite precisely across different
stocks of the {\sc pse}, with a correlation of $R^2=0.96$ (see Fig. (\ref{sigvsR})). Perhaps surprisingly, the exogenous `news volatility' contribution
$\Sigma^2$ is found to be small. (The intercept of the best affine regression is even found to be slightly negative). 
This could be related to the observation made in \cite{Farmer04b} that for most price jumps, some limit orders are cancelled too slowly and 
get `grabbed' by fast market orders. This means that most of these events also contribute to the impact component ${\cal R}_1$.\footnote{One could 
argue that our results simply show that the news volatility $\Sigma$ itself is proportional to ${\overline{\cal R}}_1$ and thus to the spread $S$. 
However, there is no reason why this should {\it a priori} be the case. For example, a model where rare jumps of typical amplitude $J$
and probability per trade $p \ll 1$ leads to $\Sigma = \sqrt{p} J$, whereas the cost of such jumps, contributing to $S$, is $p J \ll \Sigma$.} 
We can neglect $\Sigma^2$ in the above equation; in this sense the volatility of the stocks can be mostly attributed to market activity and 
trade impact. This is in agreement with the conclusions of Evans and Lyons on currency markets (\cite{Evans02}); 
see also the discussion in \cite{Bouchaud04,Hopman02}).

A final important assumption is that of {\it universality}. When the tick size is small enough and the typical number of shares 
traded is large enough, all stocks within the same market should behave identically up to a rescaling of the average spread and 
the average volume. In particular we assume that the statistics of (i) the volume of market orders (ii) the spread S and 
(iii) the impact ${\cal R}_1$, and the various correlations between these quantities are independent of the stock when these quantities
are normalized by their average value. Empirical evidence for (at least approximate) universality can be found in \cite{Lillo03d} and \cite{Bouchaud02}.
However, one expects that universality holds only for large cap, small tick stocks -- large tick stocks are not covered by the analysis below.
  
Universality then implies that:
\be\label{univ1}
E[v S] = B E[v] E[S], \qquad E[v {\cal R}_1(v)] = B' E[v]{\cal R}_1,
\ee
where $B,B'$ are stock independent numbers. Eq. (\ref{univ1}) accounts well for the Paris Stock Exchange data studied in \cite{Wyart06},
where it was found that: $B \approx 1.02$ and $B' \approx 1.80$: the incoming volume and the spread are nearly uncorrelated, 
whereas the volume traded and the impact are correlated ($B' > 1$), as expected.

Therefore, using Eq. (\ref{mmmg}) as an equality and Eqs. (\ref{dd2},\ref{univ1}) with $\Sigma^2=0$, we obtain the main result of this section:
\be
\label{rgr}
E[S]  = C \, \sigma _1,
\ee
where $C$ is a stock independent numerical constant, which can be expressed using the constants introduced above 
as $C=2\lambda B'/\sqrt{A}B$. This very simple relation between volatility {\it per trade} and average spread was noted in \cite{Bouchaud04,Zumbach04,Wyart06},
and we present further data  to support this conjecture. 
Therefore, the fact that the cost of limit and market orders should be nearly equal on average [Eqs.(\ref{cc},\ref{mmmg})] and the 
absence of a specific contribution of news to the volatility lead to a particularly simple relation between liquidity and volatility. 
As an important remark, we note that the above relation is not expected to hold for the volatility {\it per unit time} 
$\sigma$, since it involves an extra stock-dependent and time-dependent quantity, namely the the trading frequency $f$, through:
\be\label{sigmanu}
\sigma =  \sigma_1 \sqrt{f}.
\ee

\begin{figure}[tbp]
 \begin{center}
 \rotatebox{270}{\resizebox{7.0cm}{!}{\includegraphics{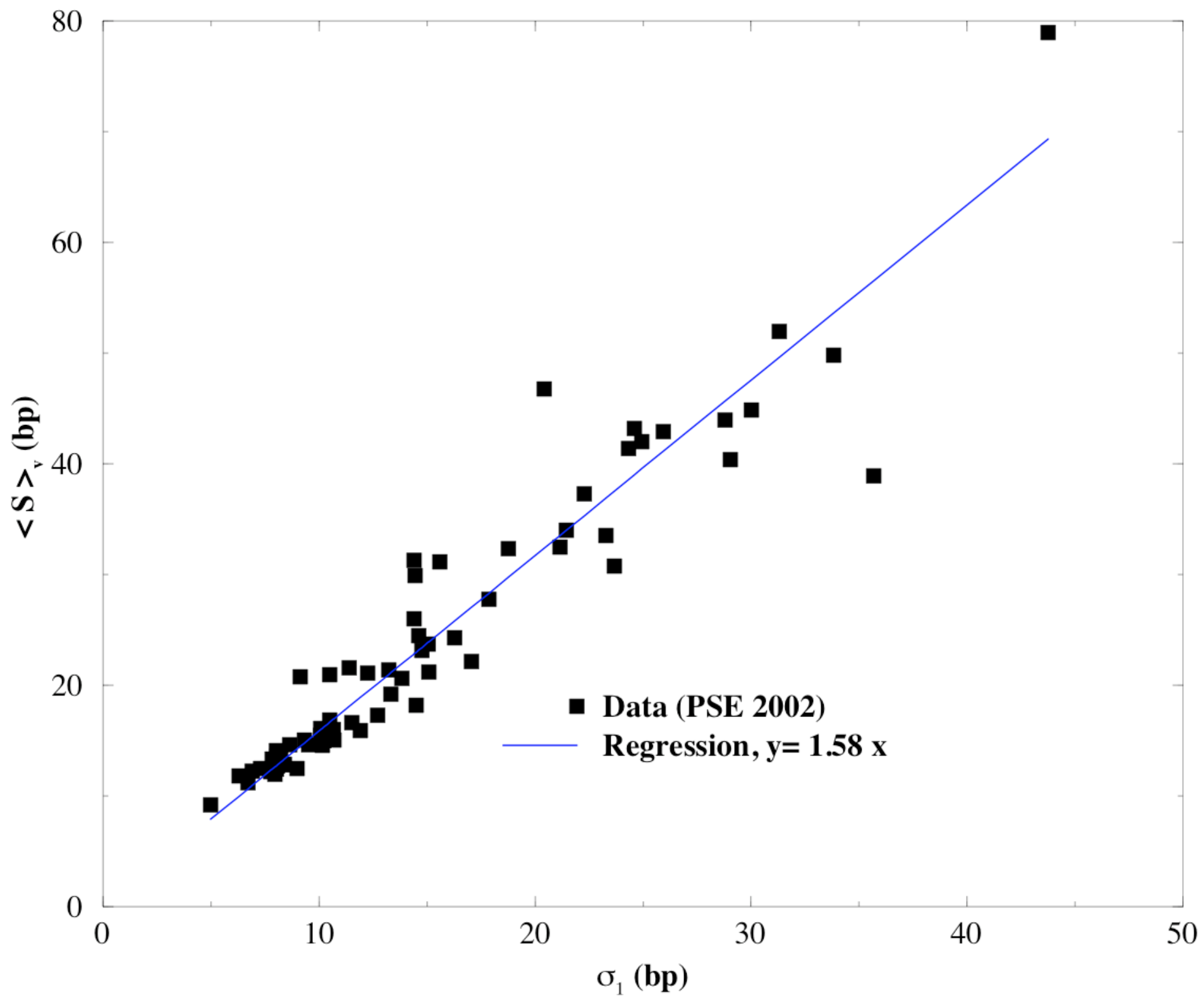}}}
 \caption{\small{Test of Eq. 
 (\ref{rgr}) for 68 stocks from the Paris Stock Exchange in 2002, averaged over the entire year. The value of 
 the linear regression slope is $c \approx 1.58$, with $R^2=0.96$}}
 \label{fte2}
 \end{center}
 \end{figure}

The above predicted linear relation between spread and volatility per trade was tested empirically in \cite{Wyart06} on small tick stocks. 
For example, the results for Paris Stock Exchange are shown in Fig. \ref{fte2}.  One finds that Eq. (\ref{rgr}) describes the data very well, 
with $R^2$ values over 0.9. One can also check that there is an average intra-day pattern which is followed in close correspondence both by $E[S]$ 
and $\sigma_1$: spreads are larger at the opening of the market and decline throughout the day. Note that the trading frequency $f$ 
increases as time elapses, which, using Eq. (\ref{sigmanu}), explains the familiar U-shaped pattern of the volatility per unit time.

Note that there are two complementary economic interpretations of the relation $\sigma_1 \sim S$ in small tick markets: 
 \begin{itemize}
 \item (i) Since the typical available liquidity in the order book is quite small, market orders tend to grab a significant fraction 
 of the volume at the best price; furthermore, the size of the `gap' above the ask or below the bid is observed to be 
 on the same order of magnitude as the bid-ask spread itself which therefore sets a natural scale for price variations. Hence both the impact and 
 the volatility per trade are expected to be of the order of $S$, as observed. 
 \item (ii) The relation can 
 also be read backward as $S \sim \sigma_1$: when the volatility per trade is large, the risk of placing limit orders is large and therefore
 the spread widens until limit orders become favorable. 
 \end{itemize}
Therefore, there is a clear two-way feedback that imposes the relation $\sigma_1 \sim S$, and that can in fact lead to liquidity instabilities: 
large spreads create large volatilities, which in turn may open the spread more. A detailed study of such effects would be highly valuable. 
On average, however, any deviation from the balance between spread and volatility tends to be corrected by the resulting relative flow of limit 
and market orders. The result $\sigma_1 \sim S$ therefore appears as a fundamental property of the market organization, which should be satisfied 
within any theoretical description of the micro-structure. This is an important constraint on models of order flow; however, none of the simple models studied in the 
past (zero intelligence models \cite{Daniels03}, bounded-range models \cite{Foucault05,Luckock03,Rosu05}, or diffusion-reaction models \cite{Slanina01}) are able to predict the above structural relation between $S$ and $\sigma_1$ (see however \cite{Mike05} for
recent developments using a ``low intelligence" model, as discussed in Section~\ref{Mike}). 

\subsection{Market cap effects}

It is interesting to study the systematic dependence of the volatility and spread as a function of market 
capitalisation $M$. Across stocks, the volatility per unit time shows a systematic slow decrease with $M$, 
$\sigma \propto M^{-\varphi}$, where $\varphi$ is small. The trading frequency $f$, on the other hand, increases 
with $M$ as $f \propto M^\zeta$. For stocks belonging to the {\sc ftse}-100, Zumbach 
 finds $\zeta \approx 0.44$ (\cite{Zumbach04}), while for US stocks the scaling for $f$ is less clear (\cite{Eisler06}),
 with apparently two regimes, one for $M > 10$ B\$, where $\zeta \approx 0.44$ and the other for $M < 10$ B\$, for
 which $\zeta \approx 0.86$. The average amount per trade $v_m$, on the other hand, also increases with $M$, 
 in such a way that $f \times v_m$ is directly proportional to $M$. This last scaling holds with
 rather good accuracy and merely states that the total volume of transactions is proportional to market capitalisation, which is somewhat expected a priori. What is interesting is that this is insured by having both the frequency of trades {\it and} the volume per trade increase with $M$, and not, for example, the transaction frequency at fixed amount per trade. The constant of proportionality is such that $\sim 10^{-3}$ of the total market cap is exchanged per day, on average, both in London and in New-York (\cite{Zumbach04,Eisler06}). 
 
 Combining the above two relations for the volatility per trade $\sigma_1=\sigma/\sqrt{f}$ results in the following scaling law for the spread $S$,
 \be\label{spreadMKT}
 S \sim \sigma_1 \propto M^{-\omega}, \qquad \omega = \varphi-\frac{\zeta}{2} \approx 0.22.
 \ee
 The average spread therefore decreases with market capitalisation. This result is in good agreement 
 with data from the {\sc lse}, \cite{Zumbach04}, and from the {\sc pse}, \cite{Wyart06}. It can also be
 directly be compared with the impact data of \cite{Lillo03d} in the {\sc nyse}, where it was established that:
 \be
 {\cal R}_1(v) \approx M^{-0.3} F\left(M^{0.3}\frac{v}{\overline v}\right),
 \ee
 where $\overline v$ is the average volume per trade for a given stock, and $F$ a master curve that behaves 
 approximately as a power-law with exponent $b$. 
  Since spread and impact are proportional, this last result is directly comparable to Eq. (\ref{spreadMKT}).
 The average over $v$ of the above result then leads to
 $E[{\cal R}_1] \sim M^{-\omega}$ with $\omega \approx 0.3(1-b)$, which is in the range $0.15 - 0.25$ (see section \ref{impact1} above for a discussion of the 
value of $b$).

\section{Order book dynamics}
 \label{orderBook}
 
The previous section stresses the key role that liquidity plays in price formation.  In double auction markets prices are formed in the limit order book.  Thus one obvious approach to understanding liquidity is to investigate the causes of liquidity fluctuations in the limit order book.  Although the dynamics of liquidity is still very much an open question, several studies have identified statistical regularities in the behavior of limit order books and give some insight into the relationship between order flow and liquidity.

\subsection{Heavy tails in order placement and the shape of the order book\label{orderbookHeavyTails}}

 
There are several statistical regularities of limit orders placement. First of all, as mentioned above, limit order signs are also well described by a long memory process with an Hurst exponent very close to the one for market order signs. \cite{Lillo03c} reported a value of $H =0.69$ for market orders and of  $H =0.71$ for limit orders. 

Limit orders are characterized also by the limit price. The absolute value of the difference between the limit price and the best available price is a measure of the patience of the trader. Patient (impatient) traders submit limit orders very far from (close to) the spread.   
One of the statistical regularities recently observed in the microstructure of financial markets is the power law distribution of limit order price in continuous double auction financial markets (\cite{Bouchaud02,Zovko02}). Let $b(t)-\Delta$ denote the price of a new buy limit order and $a(t)+\Delta$ the price of a new sell limit order. Here $a(t)$ is the best ask price and $b(t)$ is the best sell price. The $\Delta$ is measured at the time when the limit order is placed. It is found that $\rho(\Delta)$ is very similar for buy and sell orders. Moreover for large values of $\Delta$ the probability density function is well fitted by a single power-law
\begin{equation}
\rho(\Delta)\sim\frac{1}{\Delta^{1+\mu}}
\label{pow}
\end{equation}
There is no consensus on the value of the exponent $\mu$. \cite{Zovko02} estimated the value $\mu=1.5$ for stocks traded at the London Stock Exchange, whereas \cite{Bouchaud02} estimated the value $\mu=0.6$ for stocks traded at the Paris Stock Exchange.  More recently \cite{Mike05} fitted the limit order distribution for LSE stocks with a Student's distribution with $1.3$ degrees corresponding to a value $\mu=1.3$. This power-law extends from $1$
tick to over $100$ ticks (sometimes even $1000$ ticks), corresponding to a relative change of price of $5 \%$ to $50 \%$.
Such a broad distribution of limit order prices tells us that the opinion of market participants about the price of the stock in a
near future could be anything from its present value to $50 \%$ above or below this
value, with all intermediate possibilities. This means that market participants, quite oddly, anticipate the existence of large price jumps that
would lead to trading opportunities.

A heavy tail in the distribution of relative limit price $\Delta$ indicates that there is a large heterogeneity in the limit price, i.e. in the patience associated with each limit order. Patience is in turn related to the time scale the investor is willing to wait before her order is filled.  The typical time to fill\footnote{
The mean time to fill of a limit order is infinite if the price process can be approximated by a random walk. "Typical" above means some other measure such as the median time to fill.}
of a limit order grows with $\Delta$. In a recent study \cite{Lillo07b} suggested that the origin of the  heavy tails in the distribution of the relative limit price $\Delta$ can be attributed to a heterogeneity of time scales characterizing the trading behavior of individual utility maximizers investors, and tested this theory by using brokerage data from the LSE.   

\begin{figure}[ptb]
\begin{center}
\includegraphics[scale=0.25]{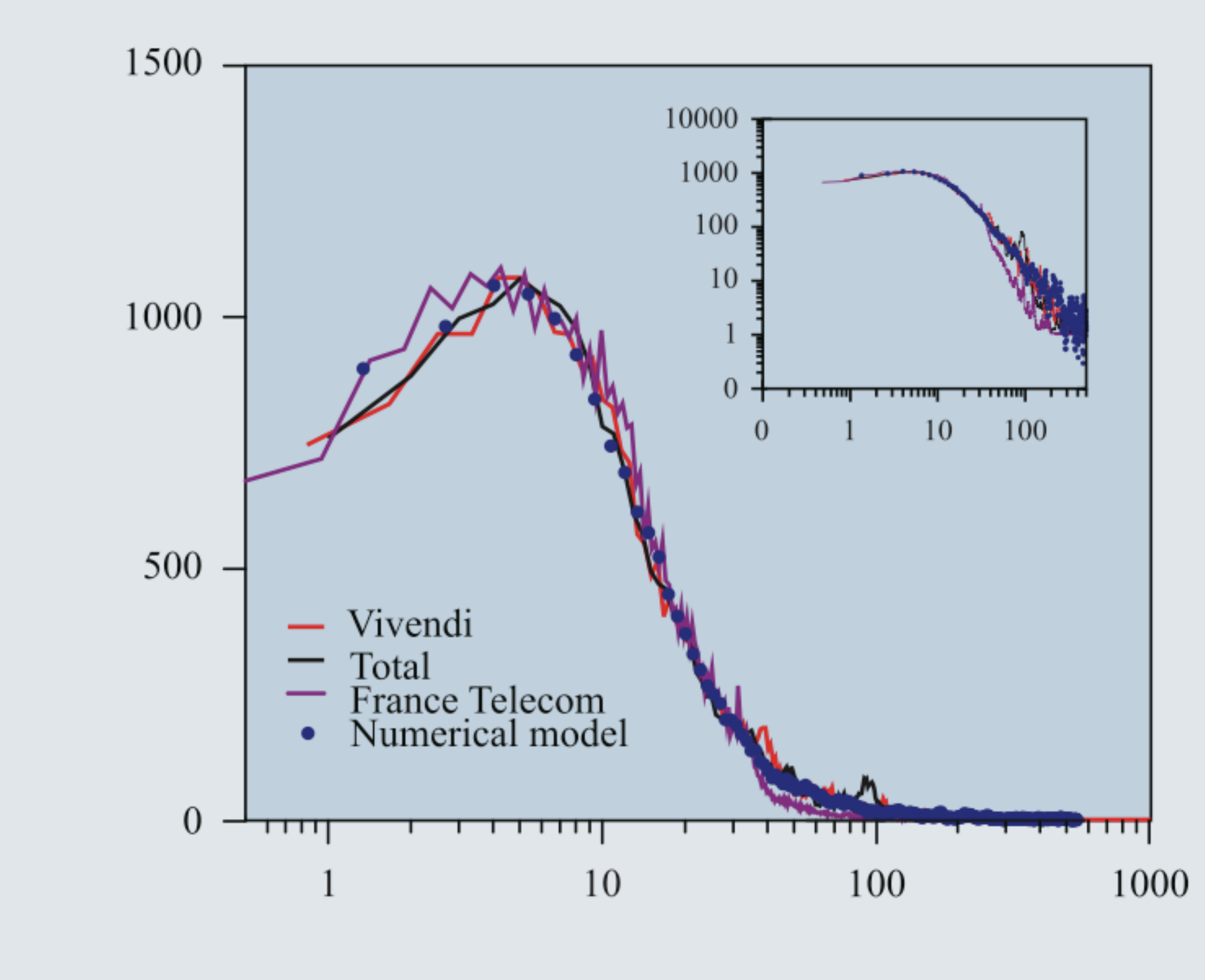}
\includegraphics[scale=0.25]{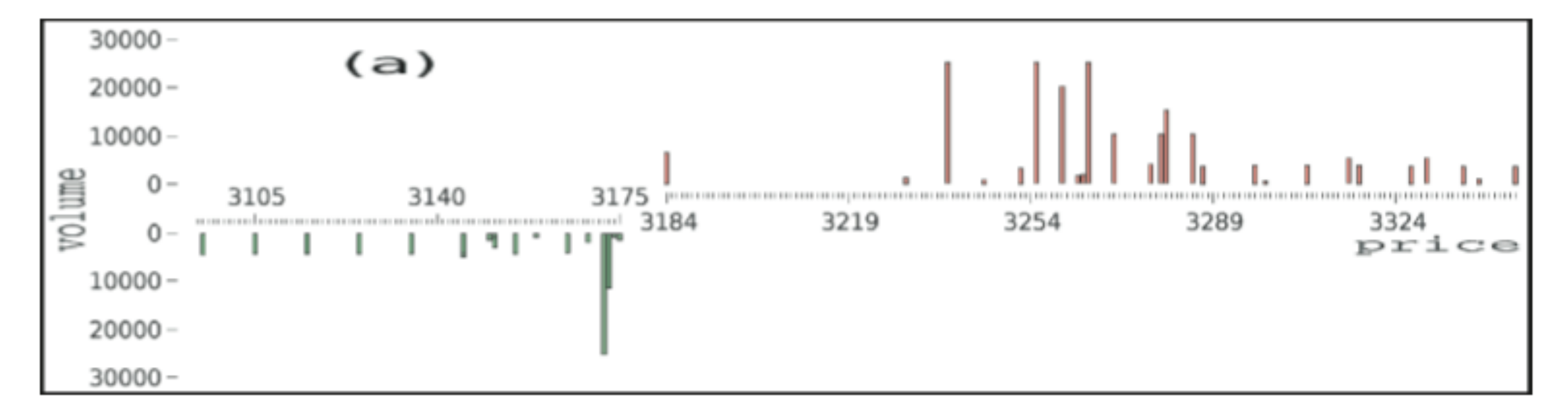}
\end{center}
\caption{Left. Average shape of the order book. Right. Instantaneous shape of the order book.}
\label{book}
\end{figure}

The order flow and the interaction of orders determine the instantaneous state of the book $\Omega_t$. By averaging over time empirical studies consistently show that the average shape of the order book is roughly symmetric between the bid and offer side of the book and is consistent across different stocks (\cite{Bouchaud02,Zovko02,Mike05}). They show that the maximum of the averaged book is not the best price, as shown in the left panel of Fig. \ref{book}, even though this is the most likely place for an order to be placed.  In Section \ref{statmodels} we will present statistical models explaining this fact.

It is important to stress that the average shape of the book is very different from the``typical" shape of the book. As \cite{Farmer04b} showed,  for most LSE stocks  the typical shape of the book is extremely sparse (see the right panel of Fig. \ref{book}).  This occurs when the ratio between tick size and price is small, so that there are often many unoccupied price levels.  As we discussed in Section \ref{gapsec}, this fact has important consequences for the price impact of individual transactions and on the origin of large price fluctuations.  

\subsection{Volume at best prices: the Glosten-Sandas model}

The distribution of available volumes at the best can be fitted by a gamma distribution with an exponent less than unity, meaning that the most probable value of the volume is much
smaller than the average value. Both the value of the spread $S$ and the quantity available at the bid and the ask, $\Phi_b, \Phi_a$, give
information on the willingness of liquidity providers to enter a trade. One would like to understand the relation between these quantities --
intuitively, large spreads are more favorable to liquidity providers and should attract larger volumes. More generally, it would be 
extremely interesting to have a theory for the shape of the whole order book, i.e. the relation between the available volume and the 
distance from the best price.

The approach of Glosten and Sandas attempts to answer the above questions, within a framework where market orders are informed trades (\cite{Glosten94,Sandas01}). The idea is now 
that information is time dependent and modelled as a random variable that gives the predicted future variation of the midpoint, which 
we call (in conformity with the above notation) $\epsilon_n r(n,n+\ell)$. Just before the $n$th trade, a liquidity provider considers the volume of 
the queue at the ask, $\Phi_{a,n}$ and decides to add an extra (infinitesimal) limit order if its expected gain, conditional on execution, 
is greater than some minimum value ${\cal G}_{\min} \geq 0$. This reads:
\be\label{Sandas}
E[m_{n+\ell}-m_n|\epsilon_n=1,v_n \geq \Phi_{a,n}] \leq \frac{S_n}{2} - {\cal G}_{\min}.
\ee
At this stage, Glosten and Sandas add several questionable assumptions. A crucial one is that the volume that the informed trader chooses to
trade is proportional to the information he has: $v_n = \alpha r(n,n+\ell)$, {\it independently} of the shape of the book at that moment, and
in particular of the available volume at the ask. He is prepared to walk up the book if necessary, which occurs with only a very
small probability in practice:  as discussed in Section \ref{whyConcave}, trading is, in fact, discretionary. Introducing the probability of information 
content $P_\ell(r)$, and dropping the index $n$ for convenience, the above conditional expectation inequality reads:
\be
\int_{\Phi_{a}/\alpha}^{+\infty} r P_\ell(r) {\rm d}r \leq \left[\frac{S}{2}-{\cal G}_{\min}\right] \int_{\Phi_{a}/\alpha}^{+\infty}  P_\ell(r) {\rm d}r,
\ee
where we have used the fact that information is assumed to be reliable, i.e. the expected mid-point change is indeed given by the informed trader
prediction. In order to achieve a quantitative prediction, Sandas further assumes that $P_\ell(r)$ has an exponential shape:\footnote{This exponential
assumption is in fact not so important. For example, a pure power-law distribution $P_\ell(r) \propto r^{-1-\mu}$ when $r > r_0$ would lead to the
following result instead: ${\Phi_a}/{\alpha} \leq (1-\mu^{-1})[{S}/{2}+{\cal G}_{\min}]$ ($\mu > 1$).}
\be
P_\ell(r) = \beta e^{-\beta r} \longrightarrow \frac{\Phi_a}{\alpha} + \frac{1}{\beta} \leq \frac{S}{2}+{\cal G}_{\min}.
\ee
In fact, this calculation can be reinterpreted to give the total volume of orders available $\Phi_<$ at a price less or equal to $p=m + S/2$, 
and therefore makes a prediction for the shape of the order book:
\be
\Phi_<(p)=\alpha (p-m) - \alpha {\cal G}_{\min} - \frac{\alpha}{\beta},
\ee
i.e. a linear order book with slope $\alpha$ and, in principle, a {\it negative} intercept. 
(The prediction for the buy side of the book is obvious by symmetry). Note that within this framework, the volume dependent impact of market orders
is by assumption linear: ${\cal R}_\ell(v)=v/\alpha$, which we already know is quite a bad representation of real data, where impact is always
strongly sublinear (see Section \ref{impact1}). Altogether, this model fares quite badly when compared with empirical data:
\begin{itemize}
\item The order book intercept, which should be negative according to the model, is found to be positive when the model is fitted to empirical data.
suggesting negative costs for placing limit orders;
\item The slope $\alpha$, when obtained from the slope of the order book, is found to be ten times larger than when obtained from direct impact estimates.
\item As mentioned above, the empirical shape of order books is non-monotonic, exhibiting a maximum away from the best price. This is not accounted for
by the model.
\end{itemize}
The reason for such a failure is essentially that, as discussed in Section \ref{whyConcave}, as shown by \cite{Farmer04b}, the volume of the incoming market order is in fact strongly correlated with the available volume at
the best price. This is in fact why impact is sublinear in volume, and is at the heart of the liquidity 
game we have been detailing in the previous pages. One cannot consider that the market order flow is an exogenous process to which the limit order
flow must adapt -- rather, the two coevolve in a strongly intertwined manner.

\begin{figure}
\begin{center}
\includegraphics{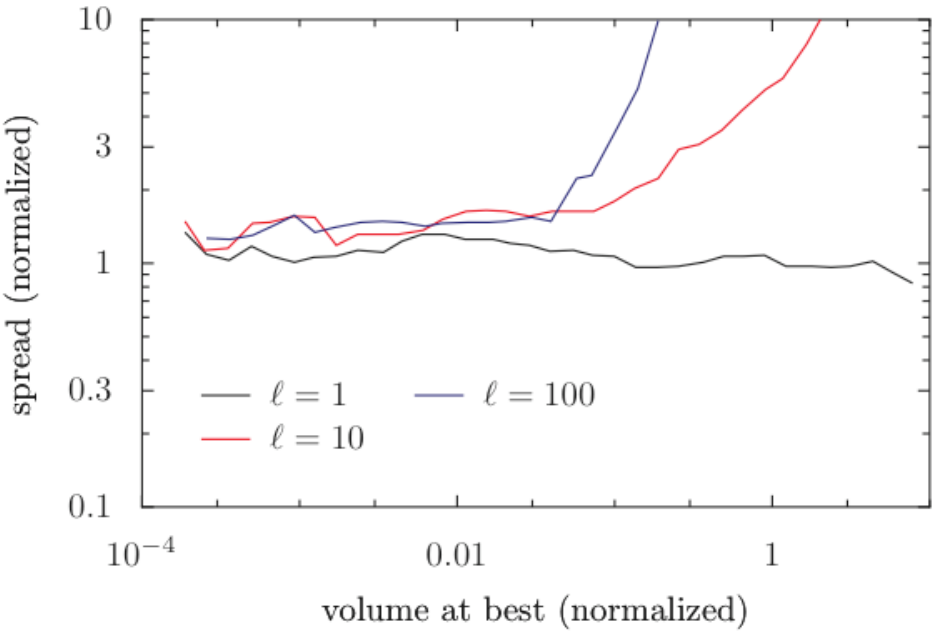}
\end{center}
\caption{Neutral line for the profitability of adding a new limit order at the best price, for three different values of the horizon $\ell$. The profit is positive above and negative below the curves. The indicated time horizons are given in number of transactions. The curves were gained by averaging for the $10$ most liquid stocks traded in the Paris Stock Exchange during the year 2002. Both volume and spread were normalized by their means for each stock before averaging. From \cite{Eisler08}}
\label{Zoltan}
\end{figure}

One can however directly test Eq.~(\ref{Sandas}) on empirical data, without any further theoretical assumptions, much as we did in the previous section.
We choose $\ell=1,10,100$ and identify a ``neutral line'' in the $S,\Phi$ plane separating the region (above that line) where executed limit orders are profitable from a region where they are costly (see Fig.~(\ref{Zoltan}), and \cite{Eisler08}). One sees that after the $\ell=1$ trade the separation line is flat and is located
around the value of the average spread. This means that the value of the spread is such that limit orders and markets order break even on average at 
high frequencies, as discussed in section \ref{phasediagram}. However, judged on longer time scales, the profitability of a limit order behind a 
large preexisting order only becomes positive for spreads significantly larger than the average. In other words, correlations between spread and volume, of the type predicted by the Glosten-Sandas model (Eq.\ref{Sandas}) indeed appear on longer time scales. 

\subsection{Statistical models of order flow and order books} \label{statmodels}

\subsubsection{Zero intelligence models}
\label{zeroIntelligence}
An alternative point of view is to model the order flow directly as a stochastic process, decomposed into three components: market orders, addition of limit orders, and cancellation of limit orders.  There is a long literature developing models of this type\footnote{
Examples of stochastic process models of limit order books include (\cite{Mendelson82,Cohen85,Domowitz94,Bak97,Bollerslev97,Eliezer98,Maslov00,Slanina01,Challet01,Daniels03,Chiarella02,Bouchaud02,Smith03,Farmer05,Mike05}). See \cite{Smith03} for a more detailed survey.}.
We will describe the approach of \cite{Daniels03} (see also \cite{Smith03}), which has the advantage that it make predictions that can be tested against real data.   They assume that each elementary event is independent and concerns a fixed `quantum' of volume $v$. Buy and sell market orders are described by two Poisson processes of rate $\mu$. Limit orders have a constant probability per unit time  
$\rho$ to land anywhere they will not generate an immediate transaction, and existing limit orders have a probability $\nu$ to be cancelled. This
model is of course highly schematic, since it neglects all correlations between market and limit orders, in particular, the ``stimulated refill'' effect
that we argued to be so important. Another important effect that is neglected is the dependence of the cancelling rate on the size of the queue: one
can actually observe that the probability of cancellation decreases as the number of orders at that price increases. 

A simple self-consistent argument makes it possible to estimate the size of the spread $S$ in this model. The total flux of limit orders between the mid-point 
and $S/2$ is by definition $\int_0^{S/2} d\Delta \rho(\Delta)$, where $\Delta$ is the distance from the midpoint and we are allowing here for the possibility that $\rho$ might depend on $\Delta$.  If we assume that $S$ is sufficiently small so that $\rho$ is approximately 
constant, one finds that this incoming flux is $\approx \rho(0)S/2$. Whenever $\mu > \rho(0)S/2$, the rate of market order eats up the limit
orders that appear within the spread completely, and the average volume present is close to zero. The cancellation term can be safely neglected if removal by
market orders is more efficient, i.e. when $\mu \gg \nu(0)$. But the argument breaks down when $S \sim 2 \mu/\rho(0)$, which sets the typical 
position of the best price, provided the tick size is small compared to $S$. The spread is therefore larger for larger market order rates, 
and smaller when the flow of limit orders is larger, as expected intuitively. The above ``scaling'' result for the spread has been derived more 
quantitatively when $\rho$ and $\nu$ are independent of $\Delta$. One finds for the average spread:
\be
E[S] = \frac{\mu}{\rho} F(\frac{\nu}{\mu}),
\label{spreadEOS}
\ee
where $F(u)$ is a monotonically increasing function that can be approximated as $F(u) \approx 0.28+1.86 u^{3/4}$. Therefore, in the limit where
cancellation can be neglected, one recovers the above result $S \approx 0.28 \mu/\rho(0)$. This prediction can be compared with empirical data 
by independently measuring the spread and the rates of the various processes (\cite{Farmer05}). In view of the simplicity of the model, the agreement with data is quite good, but systematic deviations remain. In view of the importance of feedback mechanism that are neglected, this is hardly surprising.

The results above are interesting because they demonstrate that some properties of the limit order book are dictated more or less automatically by the structure of the continuous double auction itself.  In particular, Equation~\ref{spreadEOS} is an ``equation of state'' relating statistical properties of price formation to those of order flow.  This equation of state is clearly inaccurate due to the extreme assumptions that must be made to derive it. However, it has some reasonable level of empirical validity suggesting that such a relationship indeed exists for real markets.  See the discussion concerning attempts to find a more realistic equation of state in Section~\ref{Mike}.

\subsubsection{Statistical model of order book}

The above model can also explain the hump shape of the average order book. From a theoretical point of view, however, the problem is difficult to handle: if one chooses a fixed reference frame, the rates of incoming orders and cancellations change with the mid-point, while if one chooses the
reference frame where the mid-point is fixed, limit orders that are already present get shifted around. The main difficulty comes from the fact that the 
motion of the mid-point is dictated by the order flow. In order to make progress, one can artificially decouple the motion of the mid-point and 
impose that it follows a random walk. An approximate quantitative theory of the volume in the book $\Phi(\Delta)$ can then be written as follows. 
Sell orders at distance $\Delta$  from the current midpoint at time $t$ are those which were placed there at a time $t' < t$, and have survived 
until time $t$. These orders (i) have not been cancelled; (ii) have not been crossed by the ask at any intermediate time $t''$ 
between $t'$ and $t$. 

An order at distance $\Delta$ at time $t$ in the reference frame of the midpoint $m(t)$ appeared in the order book at time $t'$ at 
a distance $\Delta+m(t)-m(t')$. The average order book can thus be written, in the long time limit $t \to \infty$, as
\be
\Phi_{\rm{st}}(\Delta) = \lim_{t\to \infty} \Phi(\Delta,t) = \int_{-\infty}^t {\rm d} t' \int du \rho\left(\Delta+u\right) {\cal P}
\left(u | {\cal C}(t,t')\right) {\rm e}^{-\nu(t-t')},
\ee
where ${\cal P}\left(u | {\cal C}(t,t')\right)$ is the conditional probability that the time evolution of the price produces
a given value of the mid-point difference $u=m(t)-m(t')$, given the condition  that the path always satisfies $\Delta+m(t)-m(t'') \geq 0$ 
at all intermediate times $t'' \in [t',t]$.\footnote{We neglect here the fluctuations of the spread. The condition should in fact 
read $\Delta+a(t)-a(t'') = \Delta + m(t)-m(t") + (S(t)-S(t"))/2 \geq 0$.}
The evaluation of ${\cal P}$ requires the knowledge of the statistics of the price process, which we assume to be purely diffusive. 
In this case, ${\cal P}$ can be calculated using the method of images. One finds:
\be
{\cal P}\left(u | {\cal C}(t,t')\right)=\frac{1}{\sqrt{2\pi D \tau}}
\left[\exp\left(-\frac{u^2}{2D\tau}\right)
-\exp\left(-\frac{(2 \Delta +u)^2}{2D\tau}\right) \right],
\ee
where $\tau = t-t'$ and $D$ is the diffusion constant of the price process.

After a simple computation, one finally finds, up to a multiplicative constant which only
affects the overall normalisation,
\be
\Phi_{\rm{st}}(\Delta)=\Phi(\Delta,t \to \infty)={\rm e}^{-\alpha \Delta} \int_0^\Delta {\rm d} u \rho(u) \sinh(\alpha u)
+ \sinh(\alpha \Delta) \int_\Delta^\infty {\rm d} u \rho(u) {\rm e}^{-\alpha u}, 
\ee
where $\alpha^{-1}=\sqrt{D/2\nu}$ measures the typical variation of price during the lifetime of an order $\nu^{-1}$. 

The above formula depends on the statistics of the incoming limit order flow, modeled by $\rho(u)$. When
$\rho(u)=e^{-\beta u}$, all integrals can be perfomed explicitly and one finds:
\be
\Phi_{\rm{st}}(\Delta) = \Phi_0 \frac{\alpha \beta}{\alpha - \beta} \left[e^{-\beta \Delta} -e^{-\alpha \Delta}\right],
\ee
which can easily be seen to be zero for $\Delta=0$, reach a maximum and decay back to zero exponentially at
large $\Delta$. Here $\Phi_0$ is the total volume in the sell side of the book.

\begin{figure}
\begin{center}
\includegraphics[width=8cm,angle=270]{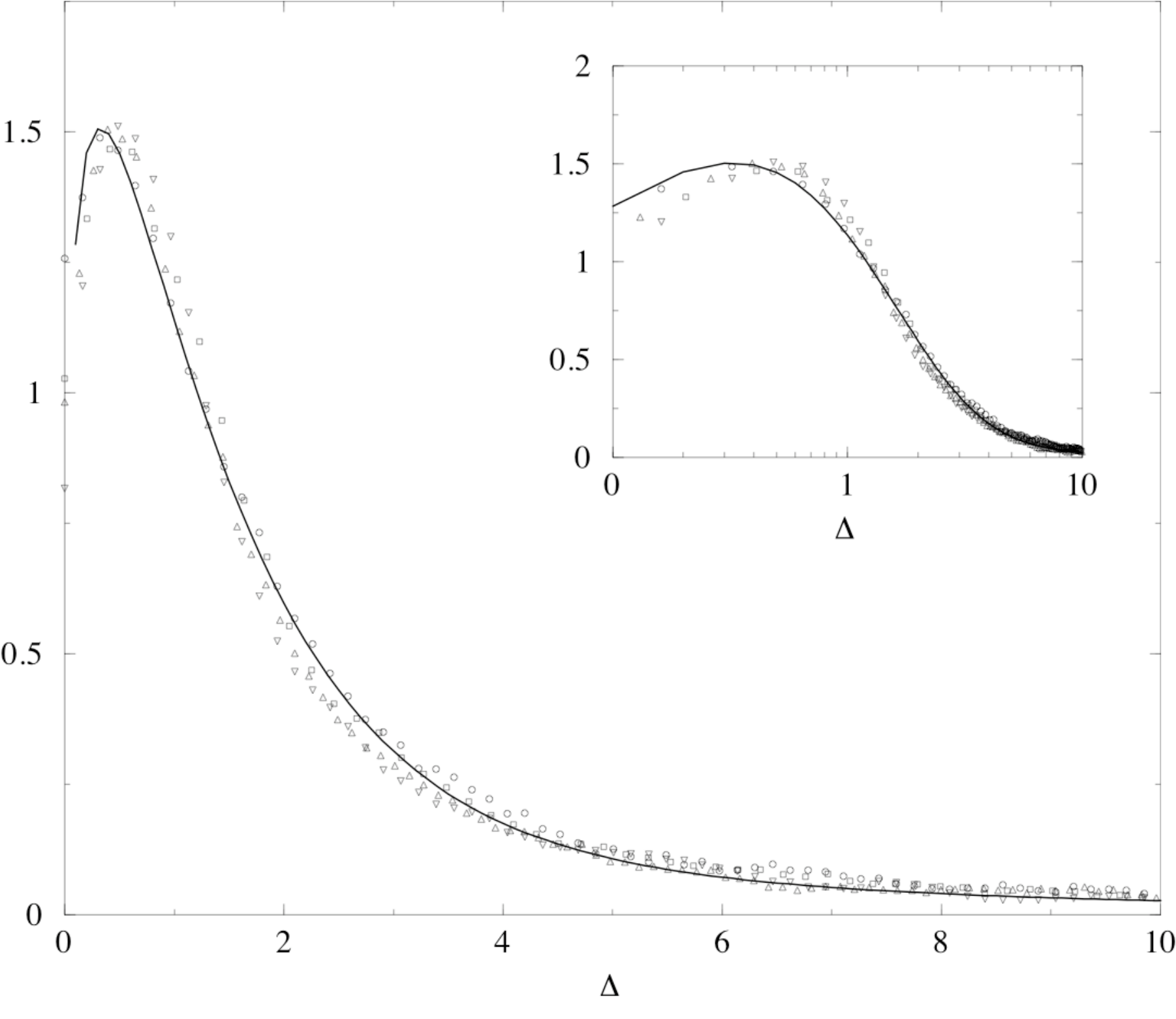}
\end{center}
\caption{The average order book for a Poisson rate model
with various choices of parameters (see \cite{Bouchaud02}) and $\mu=.6$. After rescaling the axes, the various results roughly fall on the same curve,
which is well reproduced by the simple analytic approximation leading to Eq.~(\ref{final}), shown as the full line.
\label{avbook} }
\end{figure}

We have seen above that the limit order price distribution is characterized by a power law with exponent $\mu$ (see Eq. \ref{pow}). When $\mu < 1$, the parameter $\alpha$ in the above formula can be rescaled away in the limit where 
$\alpha^{-1}$ is much larger than the tick size (this is relevant for small tick stocks, where $\alpha^{-1} \sim 10$ ticks). In this case,
the shape of the average order book only depends on $\mu$. In rescaled units $\delta=\alpha \Delta$, it is given by the following
convergent integral:
\be\label{final}
\Phi_{\rm{st}}(\Delta)={\rm e}^{-\delta} \int_0^{\delta} {\rm d} u \,u^{-1-\mu} \sinh(u)
+ \sinh(\delta) \int_{\delta}^\infty {\rm d} u \, u^{-1-\mu} {\rm e}^{-u}.
\ee
For $\Delta \to 0$, the average available volume vanishes in a singular way, as
$\Phi_{\rm{st}}(\Delta) \propto \Delta^{1-\mu}$, whereas for
$\Delta \to \infty$, the average volume simply reflects the incoming flow of orders:
$\Phi_{\rm{st}}(\Delta) \propto \Delta^{-1-\mu}$. We have shown in Fig.~\ref{avbook}
the average order book obtained numerically from the above Poisson model with a power-law order flow, and compared it with Eq.~(\ref{final}), 
for various choices of parameters and $\mu=0.6$, as found for various stocks of the Paris Stock Exchange. After rescaling the two axes, the
numerical models lead to very similar average order books, and the analytic approximation, although crude, appears rather effective.
The average shape of the order book therefore reflects the competition between a power-law flow of limit orders with
a finite lifetime, and the price dynamics that removes the orders close to the current price.

\subsubsection{A simple empirical agent based model for liquidity fluctuations \label{Mike}}

We now return to discuss the problem of the relationship between order flow and liquidity.  The pure zero intelligence model of \cite{Daniels03} was limited by its extreme assumptions of Poisson processes and the use of a highly stylized simplified model for order placement.  A model based on more realistic assumptions was made by \cite{Mike05}.  They made simple econometric models for order placement and cancellation, and showed that by simulating this model it was possible to reproduce many of the empirical features of prices, including a quantitative match for the distribution of returns and the distribution of the spread.

In Section~\ref{orderbookHeavyTails} we discussed the remarkable heavy tails in order placement.  This result applied only to orders placed inside the same best\footnote{
E.g. for buy orders the same best is the best bid; the power law applies to orders placed at prices less than the best bid.  The ``opposite best" for buy orders is the best ask.}.
Mike and Farmer also studied the distribution of order prices for orders placed inside the spread or crossing the opposite best (i.e. those generating immediate transactions).    Remarkably they found, in a certain sense described below, that the same power law behavior applied there as well.  The frequency of order placement peaks at the same best and dies out on either side, and can be reasonably well fit by a Student distribution (which has a power law tail).  Under the rule that orders that cross the opposite best price are executed, this simple rule does a reasonably good job of explaining execution frequency.  One of the predictions that emerges automatically is that when the spread is small it is more likely for an order to cross the opposite best, i.e. market orders become more likely.   This at least partially explains the `stimulated refill' process mentioned earlier, since when the spread is large, orders chosen at random are more likely to fall inside the spread (and therefore accumulate in the limit order book), whereas when the spread is small executions are more likely.  In fact, the model based on this relied on this effect to preserve stability in the number of orders accumulating in the order book.

In this model the rate of cancellation was empirically found to depend on factors such as the number of orders in the order book, the imbalance in the order book, and the position of a given order relative to the opposite best price.  Finally, it takes as an exogenous input the long-memory of order signs discussed in Section~\ref{longMemory}.  When these three elements are put together (order placement, order sign and cancellation) it is possible to simulate this model, generating a time series of order books with the corresponding prices.  Note that it is critical that there is feedback between price formation and the order placement process.  The resulting series of prices are not efficient, which is not surprising given that no effort was made to make them so and there are no agents who can take advantage of inefficiencies.  

Nonetheless, for a subset of stocks whose properties are similar to those that were used to build the model, which were called ``type I" stocks, it does a good job of reproducing many of the properties of real prices\footnote{
Type I stocks are those with reasonably low volatility and small tick size.  Type II stocks are those with high volatility, and Type III were stocks with large tick sizes.  At this stage the model only performs well for type I stocks.}.
In particular it provides a good quantitative match with the distribution of returns and the distribution of the spread.  This match includes not just the shape of the distributions, but their scale, including the absolute level of volatility.  That is, for type I stocks a simulation of prices based on the measured parameters of the order flow produces forecasts of volatility that make a good match in absolute terms, i.e. the predictions and measured values lie along the identity line.  This provides further evidence for the existence of an equation of state relating order flow and prices.  (It remains to extend this model so that it also works well for types II and III).

To summarize, the interesting point about this model is that it suggests that volatility is directly related to fairly simple properties of the order flow.
 
 \section{Impact and optimized execution strategies}\label{execution}
 
 The fact that trades impact prices is obviously detrimental to trading strategies: since, again, liquidity is
 so small, trades must typically be divided in small chunks and spread throughout the day. But because of impact, the price paid for the last lot is on average higher than the price for the first lot. This poses a well defined
 problem: what is the optimal trading profile as a function of time of day, such that the average execution price
 is as low as possible compared to the decision price (a quantity often called `implementation shortfall'). 
 
 Assume that a trader has a total volume $V$ to execute; he decides to cut his order in $N$ trades, each of size $v$, with $n v =V$.
 His trading profile $\phi(t)$ is such that the number of trades between $t$ and $t+dt$ is $\phi(t) dt$. His own impact on the price 
 of the stock at time $t' \geq t$ is modeled as:
 \be
 p(t') - p_0(t') = P(0) \int_0^{t'} dt \phi(t) G_0(t'-t) \ln v,
 \ee
 where $G_0$ is the continuous time version of the single trade impact discussed in section \ref{transient}. Using all the results obtained above, one has:
 \be\label{G0cont}
 G_0(t-t')= \frac{g_0 S}{f^\beta|t'-t|^\beta},
 \ee
 where $g_0$ is a number of order unity (since impact and spread are proportional) and $f$ the number of trades per unit time. We neglect here the possible dependence of the spread $S$ and of $f$ on time of day.
 
 The total extra cost due to impact for a given profile $\phi(t)$ is therefore given by:
 \be
 \int_0^T dt \int_0^t dt' \phi(t) G_0(t-t') \phi(t') \equiv \frac12 \int_0^T dt \int_0^T dt' \phi(t) G_0(|t-t'|) \phi(t'),
 \ee
 where $T$ is the trading period (say one day). The above quantity should be minimized with the constraint that the total trading 
 volume is fixed, i.e.:
 \be\label{total}
 \int_0^T dt \phi(t) v = V.
 \ee
 This problem can easily solved using the method of functional derivatives with a Lagrange multiplier $z$ to enforce the constraint. This 
 leads to the following linear equation for the profile:
 \be\label{profile}
 \int_0^T dt'  G_0(|t-t'|) \phi(t') = z,
 \ee
 where $z$ is such that Eq. (\ref{total}) is satisfied. 
 
 As a pedagogical example, let us assume that the impact decays exponentially as:
 \be
 G_0(\tau) = G_0 \exp(-\alpha \tau) + G_\infty 
 \ee
 Thanks to the constraint Eq. (\ref{total}), the value of $G_\infty$ can be reabsorbed in $z$ and drops out of the computation: the permanent part of the impact is irrelevant to the optimisation of execution costs (although the resulting implementation shortfall, of course, depends on $G_\infty$). The solution of the optimisation problem then reads:
 \be
 G_0 \phi^*(t)=z \delta(t) + z \delta(T-t) + \frac{z \alpha}{2},
 \ee
 and the constraint is:\footnote{Note that the two delta functions only contribute to half of their ``area" to the total volume, since they are at
 the edge of the integration range.}
 \be
 \frac{1}{G_0} \left[ 2 \frac{z}{2} + \frac{z \alpha T}{2}\right] = V \longrightarrow z = \frac{G_0 V}{1 + \alpha T/2},
 \ee
 so finally :
 \be
 \phi^*(t)= \frac{V}{1 + \alpha T/2} \left[\delta(t) + \delta(T-t) + \frac{\alpha}{2}\right]:
\ee
 the optimal profile is composed of two peaks at the open and at the close of the day, and a flat profile in between. The ratio of the volume traded
 within the day to the volume traded at the open and at the close is $\alpha T/2$: for a fast decaying impact ($\alpha T$ large), most of the volume
 should be spread out evenly intraday, whereas for a slowly decaying impact, trading should mostly concentrate at the open and at the close.
 
 More generally, it can be shown that the solution to Eq. (\ref{profile}) is symmetric around $T/2$ and U-shaped (this is also mentioned in \cite{Hasbrouck07}, ch. 15). 
 In particular, when $G_0(\tau)$ is given by Eq. (\ref{G0cont}), one finds that the optimal profile diverges both at $t=0$ and $t=T$, respectively as $t^{\beta-1}$ and $(T-t)^{\beta-1}$.
 An approximate solution to Eq. (\ref{profile}) in that case reads:
 \be
 \phi^*(t) \approx V \frac{\Gamma[2\beta]}{T^{2\beta-1} \Gamma^2[\beta]} t^{\beta-1} (T-t)^{\beta-1}.
 \ee
 It is interesting to note that none of the parameters $g_0, S, f$ entering in the numerical evaluation of $G_0$ appear in the shape of the profile, 
 since again these can be reabsorbed in the definition of $z$ at an early stage of the computation. 
 
 A generic U-shape solution for the optimized execution profile suggests an interesting interpretation of the observed U-shaped total traded volume as
 a function of the time of day.

\section{Toward an empirical characterization of a market ecology}\label{ecologyemp}

The description of financial markets we have depicted above is based on the assumption of the existence of different degrees of heterogeneity among market participants. The first level of heterogeneity is due to the existence of a broad distribution of scales among market participants. Here scale refers to any quantity that measures the typical size of the trades of an investor.
Moreover the size of the hidden order determines the time horizon over which the order is worked and the number of transactions needed to complete the order.

As described in Section \ref{ecology}, the second degree of heterogeneity is due to the existence of (at least) two classes of agents acting systematically on opposite sides of the market.  One group corresponds to liquidity providers and the other to liquidity takers.  It would be extremely valuable to have a comprehensive empirical study that connects the heterogeneity of market participants with their strategy and with the properties of price dynamics. Unfortunately it is not easy to obtain databases containing this level of information.  Some data providers are starting to release datasets containing information about the financial institutions involved in the transaction and/or  submission or cancellation of orders from the book.  It is important to stress that such financial institutions are not individual traders or agents, but rather are usually credit entities and investment firms which are members of the stock exchange and are entitled to trade at the exchange. Very often these institutions are both acting as brokers for other clients and trading for their own account.   Although an institution may act on behalf of many individuals and institutions having different strategies, recent findings show that in most cases it is possible to characterize an institution with an overall strategy, corresponding to that of the bulk of their trades.  In the following two sections we present the results of two recent papers investigating the behavior of institutions in the Spanish Stock Exchange.

\subsection{Identifying hidden orders} \label{hidordemp}
In a recent paper \cite{Vaglica07} used brokerage data on transactions in the Spanish Stock Exchange to identify hidden orders and to characterize their statistical properties. The identification of hidden orders is done by using an algorithm designed to identify segments of the inventory time series of an institution characterized by an approximately constant and statistically significant drift term. The working hypothesis is that these segments are associated with hidden orders. A hidden order is characterized by the traded volume $V$, the number of transactions $N$, and the (real) time period $T$ needed to complete the order\footnote{
In \cite{Vaglica07} the investigated variables  are the volume and the number of trades associated with those transactions characterizing the hidden order as a buy or a a sell hidden order.}.
It is found that the distribution of these quantities scale asymptotically for large values as 
\be
P(V>x)\sim\frac{1}{x^2}~~~~~~~P(N>x)\sim\frac{1}{x^{1.8}}~~~~~~~P(T>x)\sim\frac{1}{x^{1.3}}.
\ee
These relations show that the size of the hidden orders is asymptotically Pareto distributed in accordance with the hidden order model described in Section \ref{lmf}. It should be noted that the value of the exponent for $V$ and for $N$ is slightly larger than the value $1.5$ expected by the theory described in Section \ref{lmf} and a more careful testing of the theory is needed. The low value of the exponents indicates that the size of hidden orders is a very heterogeneous quantity, probably reflecting the heterogeneity of market participants. To test this hypothesis, \cite{Vaglica07}, have considered the distributional properties of hidden orders of individual brokerage codes. It is found that for the distribution of hidden order size of individual brokers is consistent with a lognormal distribution, whereas the pool of the hidden orders of all brokers is not consistent with a lognormal. This indicates that investor size heterogeneity is at the origin of the power law distribution of hidden order size.

The size variables of an hidden orders are clearly related to each other.  If the volume $V$ is large we expect that the number of transactions $N$ and the time needed to complete the orders will also be large.  The relation between the size variables reflects the strategic behavior chosen by the trader to work the order. By performing a principal component analysis to the hidden orders Vaglia et al. find that 
\begin{equation}
N\sim V^{1.1},~~~~~~T\sim V^{1.9},~~~~~~~N\sim T^{0.66}.
\label{scaling2}
\end{equation}
The fraction of variance explained by the first eigenvalue is of the order of $88\%$, so these characterizations are reasonably sharp.  
The first relation indicates that the number of transactions of a hidden order is roughly proportional to it size. This means that even if a trader needs to trade a large hidden order, she will not split the order in larger chunks. This observation is consistent with the empirical finding that it is rare that the size of market orders is larger than the volume available at the opposite best (see Section \ref{gapsec} and \cite{Farmer04b}). The other two relations indicate that the larger the volume of the hidden order, the slower the trading rate. This result has also been verified by using other statistical hidden order detection algorithms and still needs to be properly understood. Finally it is worth noting that  the relations of Eq \ref{scaling2} also hold approximately true when one considers hidden orders belonging to a single brokerage code. In other words the scaling relations of Eq. \ref{scaling2} are not the effect of  heterogeneity among traders.

\subsection{Specialization of strategies}

\begin{figure}[ptb]
\includegraphics[scale=0.35]{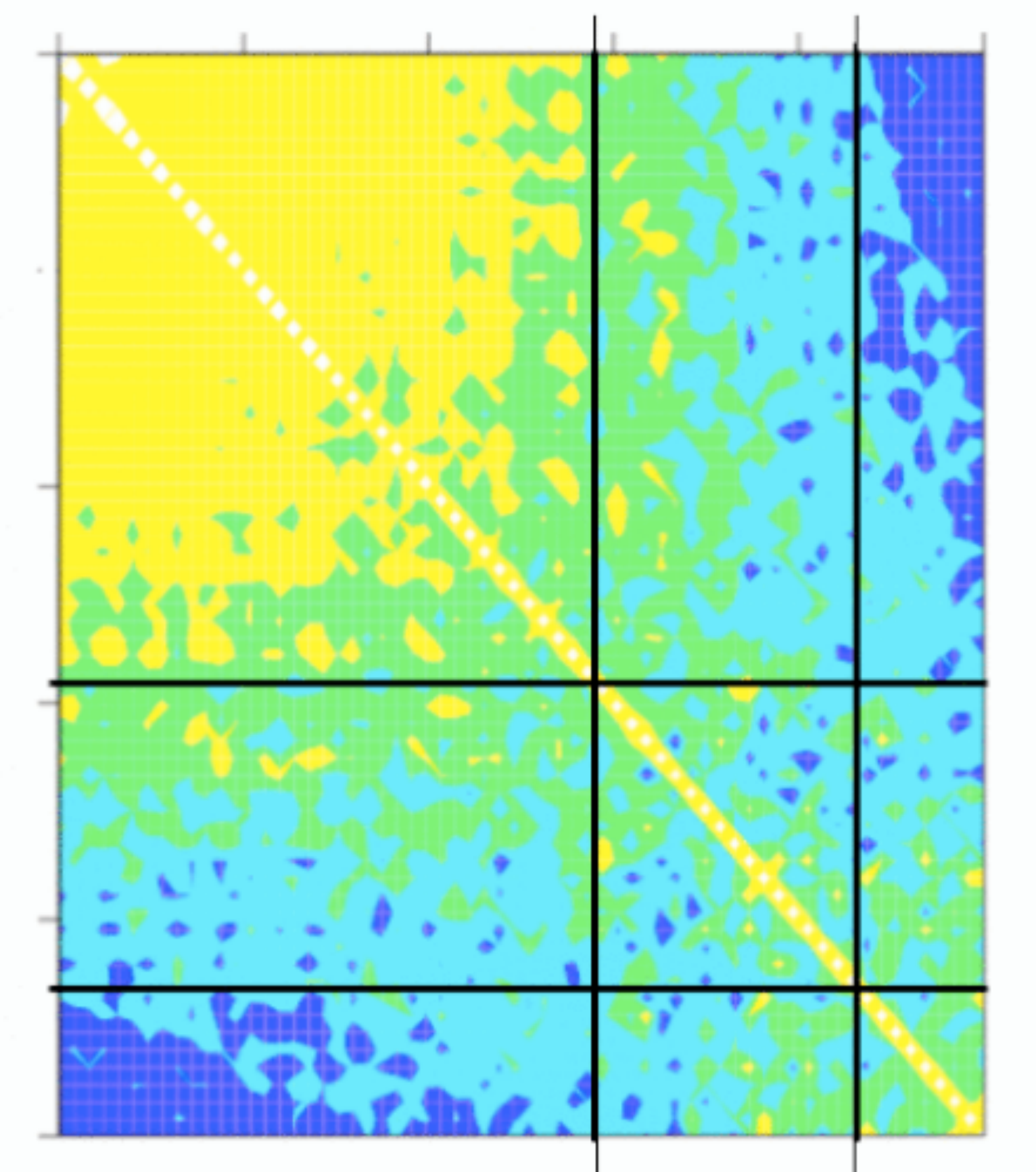}
\caption{ 
Contour plot of the correlation matrix of daily inventory variation of institutions trading the stock BBVA in 2003.  This is plotted by sorting the firms in the rows and columns according to 
the strength of the correlation of their inventory variation with the return of the price of BBVA during the same period.  Colors are chosen to highlight positive or negative firm daily inventory 
cross correlation values above a given significance level. Specifically, yellow (blue) indicates positive (negative) cross correlation with a significance of $2\sigma$, whereas green (cyan) indicates positive (negative) cross correlation below $2\sigma$. The thick black lines in the matrix are obtained from the bottom panel by partitioning the firms in three groups according to the value of the correlation between returns and inventory variation  (smaller than $-2\sigma$, between $-2\sigma$ and $2\sigma$ and larger than $2\sigma$). Adapted from \cite{Lillo07}.  
}
\label{matrix}
\end{figure}

The presence of distinct classes of institutions and their mutual interaction has been investigated in a  recent work by \cite{Lillo07}. This study clearly identifies classes of institutions that are characterized by having a similar trading behavior. Specifically the study has focused on the cross correlation between the inventory variation of different institutions. In general it is found that the cross correlation of the inventory variation of different institutions is often statistically significant, for both positive and negative values.  Principal component analysis reveals that the first eigenvalue of the correlation matrix is associated with a factor that is strongly correlated with price return.
To give an idea of the level of correlation of the activity between different institutions, in Fig. \ref{matrix} we show the contour plot of the correlation matrix of daily inventory variation of the institutions trading the stock BBVA in the Spanish Stock Exchange in 2003. Different colors refer to different levels of correlation (see caption). The institutions are sorted according to the value of the correlation of their inventory variation with the price return of BBVA.   
Two groups of firms are seen, one on the top left corner and the other on the bottom right corner.

The figure shows a clear block structure that makes it possible to identify communities of institutions characterized by a similar trading behavior. Specifically, the trading institutions can be partitioned in three subsets. The first (see the bottom right corner in Fig. \ref{matrix}) is composed by institutions with an inventory variation positively correlated with price return, i.e. these institutions buy when the price goes up and sell when the price goes down. Moreover they are typically large institutions and have strongly autocorrelated order flow, probably because of order splitting. The second subset (see the top left corner in Fig. \ref{matrix}) is composed of institutions whose inventory variation is negatively correlated with price return; these institutions buy when the price goes down and sell when the price goes up. The size of these institutions is very heterogeneous, as is the autocorrelation of their order flow. Finally the third group is made up of uncategorized firms. As Figure \ref{matrix} shows, the cross correlation between the inventory variation of an institution belonging to the first group and an institution belonging to the second group is typically negative (blue areas in the top right and bottom left corners). This and other more direct evidence suggests that institutions belonging to these two groups are often trading counterparties.

\section{Conclusions} 
\label{conclusions}

In this review paper we discussed market impact on two different, but overarching, levels. The first level deals with ultra-high frequency scales: that
of elementary transactions (a level that in physics is called ``microscopic''). It is concerned
with the phenomenological description and mathematical modelling of empirical observations on order flow, impact, order book, bid-ask spread, etc., which are of direct interest for high frequency trading, execution and slippage control. Results on that front are surprisingly rich and to some extent unexpected. Among the most salient points, one 
finds that impact of individual trades is nonlinear (concave) in volume and has a nontrivial lag dependence, that can alternatively be thought of as a history-dependent impact. This is at variance with many simple models, including the famous Kyle model, where impact is assumed to be linear and permanent. The subtle temporal 
structure of impact can be traced back to the long-memory in the fluctuations of supply and demand. The compatibility of 
the long-memory in order flow and the absence of predictability of asset returns has profound consequences on price formation,

The second level deals with phenomena on a longer ``coarse-grained'' time scale, and is more in line with the questions economists like to ask about
markets and prices, such as: ``Are prices in equilibrium?", ``What is the information content of these prices?", or ``Why is the volatility so high?".   
Much as in physics, where the detailed understanding of the microscopic world provides invaluable insight on macroscopic phenomena, we believe that a consistent picture of the microstructure mechanisms will help put in perspective some of these traditional questions about markets and prices. From the set of results presented above, two concepts seem to emerge with possible far-reaching theoretical consequences: 
\begin{itemize}
\item (a) Because the 
outstanding liquidity of markets is always very small,  trading is inherently an incremental process, and prices cannot be instantaneously 
in equilibrium, and cannot instantaneouly reflect all available information. There is nearly always a substantial offset between latent offer and latent demand, that only slowly gets incorporated in prices. Only on an aggregated level does one recover an approximately linear impact with a permanent component. 
\item (b) On anonymous, electronic markets, there
cannot be any distinction between ``informed'' trades and ``uninformed'' trades. The average impact of all trades must be the same, which means 
that impact must have a mechanical origin: if everything is otherwise held constant, the appearance of an extra buyer (seller) must on average 
move the price up (down).  
\end{itemize}
The theory of rational expectations and efficient markets has increasingly emphasized information and belittled the role of supply and demand, in contradiction with the intuition of traders and of the layman.\footnote{
On this point, see the lucid discussion in \cite{Lyons01}, from which we
reproduce the following excerpt: {\it Consider an example, one that clarifies how economist and practitioner worldviews
differ. The example is the timeworn reasoning used by practitioners to account for
price movements. In the case of a price increase, practitioners will assert, ``there
were more buyers than sellers''. Like other economists, I smile when I hear this. I
smile because in my mind the expression is tantamount to ``price had to rise to
balance demand and supply''.}} The work we reviewed above underlines the role of fluctuations in supply and demand, which may or may not be exogenous,
and may or may not be informed in a traditional sense -- it does not really matter. Attempts to estimate ex-post the fraction of truly informed trades
leads to very small numbers, at least judged on a short time basis, meaning that the concept of informed trades is not very useful to understand 
what is going on in markets at high frequencies. But still, prices manage to be almost perfectly unpredictable, even on very short time scales. The
conclusion is that any useful notion of information must be internal to the market: trades, order flow, cancellations {\it are} information, whatever
the final cause of these events may be. 

We are aware that some of these ideas go strongly against the prevailing view in market microstructure theory,
and entail a rather abrupt change of paradigm. We hope that this work will help clarify the issues  and motivate further work to 
reconstruct a fully rigourous modelling framework, deeply rooted in the empirical data.  Such data is now widely available and so abundant that it should be possible to raise the achievements of microstructure theory to the level of precision achieved in the physical sciences. 

\section*{Acknowlegments}

We want to thank Klaus Schenk-Hoppe and Thorsten Hens for giving us the opportunity to write this review article, and for their patience. We also want to thank our collaborators on these matters, who helped us to shape our understanding of the subtle world of market microstructure and impact: R. Almgren, S. Ciliberti, C. Deremble, Z. Eisler, A. Gerig, L. Gillemot, E. Henry, A. Joulin, J. Kockelkoren, E. Moro, G. La Spada,Y. Lemperiere, R.N. Mantegna, S. Mike, A. Ponzi, M. Potters, G. Vaglica, H, Waelbroeck, J. Wieslinger, and M. Wyart.  JDF gratefully acknowledges support from Bill Miller, Barclays Bank, and NSF grant HSD-0624351.

\section*{Appendix 1: Mechanical vs. non-mechanical impact}
\label{mechanicalImpact}

As we summarized in Section~\ref{kindsOfImpact}, there are two very different views of what causes impact.  The standard view is that it is essentially driven by information:  the arrival of a trade signals new information, which causes market participants to update their valuations.  But suppose a trade arrives that is really not based on any information?  Does such a trade have a purely mechanical effect on prices?  If so, what is the nature of that effect?

In Section~\ref{kindsOfImpact} we already introduced one such notion of mechanical impact, imagining a standard market clearing framework in which agents randomly alter their excess demand functions asynchronously.  As each agent alters her demand function she makes trades that generate market impact.  Whether or not these are permanent depends on whether the alternations are permanent.  Insofar as such alternations are permanent, the effect on prices will also be permanent.

In this Appendix we examine another notion of mechanical impact for continuous double auctions. We define a mechanical impact as what happens if someone places an order in the order book if this order has no effect on any future orders.  We are essentially asking the question of what happens to the price if an order is injected into the order book at random, but no one pays any attention to it.  We describe a method for analyzing order book data to answer this question (\cite{Farmer07}).  The essential result is that while there is a significant instantaneous mechanical impact, due to the simple fact that such an order can consume the best quotes and move the midprice, but this impact decays to zero.  This decay seems to follow a power law, decaying very fast initially and very slowly later on. The reason for this decay is that orders are continually being removed from the order book, and as this happens the mechanical impact decays away.  The the mechanical impact as defined in this sense largely reflects the rate at which orders are flushed out of the order book.


\subsection{Definition of mechanical impact for order books}

The following definition of mechanical impact makes the convenient simplifying assumption that the market framework is a continuous double auction.  Consider the order flow $\Omega = (\omega_1, \omega_2, \ldots, \omega_t, \ldots)$ consisting of individual orders $\omega_t$, which can be either new trading orders or cancellations of existing trading orders.  Each individual order could be originated because of information relating to the value of the asset, or it could be originated ``at random", e.g. due a demand for liquidity driven by events having no bearing on the asset being traded.

Under the rules of the continuous double auction any initial limit order book and subsequent order flow generates a unique sequence of limit order books, which correspond to a unique sequence of midprices.  The auction $A$ can be regarded as a deterministic function
\[
b_{t+1} = A(b_t, \omega_{t+1} )
\]
that maps an order $\omega_t$ and a limit order book $b_t$ onto a new limit order book $b_{t+1}$.    For a given order flow $\Omega_{t}^{t + \tau} = \{\omega_{t}, \omega_{t+1}, \ldots, \omega_{t + \tau}\}$ the auction $A$ is applied to each successive order to generate the limit order book $b_{t + \tau}$ at time $t + \tau$,
\[
b_{t + \tau} = A^{\tau}(b_t, \Omega_{t}^{t + \tau}).
\]
The continuous double auction can thus be thought of as a deterministic dynamical system with initial condition $b_0$ and exogenous input $\Omega$.

Each limit order book $b_t$ defines a unique logarithmic midprice $p_t = p(b_t)$.  The midpoint price at time $t+1$ can be written in term of the composition of the price operator $p$ and the auction operator $A$ as $p_{t+1} = p \circ A(b_t, \omega_{t+1})$. Thus, any initial limit order book $b_t$ and subsequent order flow $\Omega_{t}^{t + \tau}$ will generate a series of future prices $p_{t+1}, p_{t+2}, \ldots, p_{t + \tau}$, where for example the last price $p_{t + \tau}$ is
\begin{equation}
p_{t + \tau} = p \circ A^\tau(b_t, \Omega_{t}^{t + \tau}).
\label{auctionDef}
\end{equation}

To give a precise meaning to mechanical impact, suppose we modify a particular order $\omega_t$ and replace it by a new order $\omega'_t$, while leaving the rest of the order flow unaltered.  Since by assumption this modification  does not affect the rest of the order flow, we can freely assume that it occurred for purely mechanical reasons.   We can then compare the future stream of prices generated by the order flow $\Omega_{t}^{t + \tau} = \{\omega_{t}, \omega_{t+1}, \ldots, \omega_{t + \tau}\}$ to that generated by the altered order flow ${\Omega'}_{t}^{t + \tau} = \{\omega_{t}, \omega_{t+1}, \ldots, \omega_{t + \tau}\}$.  E.g. for time $t + \tau$, $p'_{t + \tau} = p \circ A^\tau (b_t, {\Omega'}_{t}^{t + \tau})$.

This can be used to measure the mechanical impact of any existing order $\omega_t$ by comparing the prices that are generated when $\omega_t$ is present to those that  would have been generated if were were absent.  We thus replace $\Omega_t^{t + \tau}$ by
${\Omega'}_{t}^{t + \tau} = \{0, \omega_{t+1}, \ldots, \omega_{t + \tau}\}$, where $0$ in this case represents a null order, i.e. one that does not change the order book.
We can then define the {\it mechanical impact} $\Delta p^M_\tau (t)$ of the order $\omega_t$ as
\begin{equation}
\Delta p^M_\tau (t) = p \circ A^\tau (b_t, \Omega_{t}^{t + \tau}) - p \circ A^\tau (b_t, {\Omega'}_{t+1}^{t + \tau}).
\label{mechanicalImpactDef}
\end{equation}
The real price $p$ contains both the informational and mechanical impact of order $\omega$, while in the hypothetical price $p'$ the mechanical impact is missing, i.e. it contains only the informational impact.  Under subtraction only the mechanical impact remains.  This isolates the part of the price impact that is ``purely mechanical", in the sense that it is generated solely by the effect of placing an order in the book and observing its effect under the deterministic operation of the continuous double auction.  The {\it information impact} can be defined as the portion of total impact that cannot be explained by mechanical impact, i.e.
\[
\Delta p^I_\tau = \Delta p^T_\tau - \Delta p^M_\tau,
\]
where $\Delta p^T_\tau$ is the total impact.
Whatever components of the total impact not explained by mechanical impact must be due to correlations between the order $\omega_t$ and other events.  With the data we have it is impossible to say whether the placement of the order $\omega_t$ causes changes in future events $\Omega_{t+1}$, or whether the properties of $\Omega_{t+1}$ are correlated with those of $\omega_t$ due to a common cause.  In either case, changes in price that are not caused mechanically must be due to information -- either the information contained in $\omega_t$ affecting $\Omega_{t+1}$, or external information affecting both $\omega_t$ and $\Omega_{t+1}$.
These ideas can be extended to apply to arbitrary modifications of the order stream, e.g. infinitesimal modification of order $\omega_t$, and to define mechanical generalization of elasticity (\cite{Zamani08}).

\subsection{Empirical results}

This definition has been applied to several stocks in the London Stock Exchange and studied as a function of the order sequence number $t$ (which as above simply labels the temporal sequence in which the orders are received).   It is clear that the mechanical impact is highly variable. In some cases there is an initial burst of mechanical impact, which dies to zero and then remains there. In some cases there are long gaps in which the impact remains at zero and then takes on nonzero values after more than a thousand transactions. In other cases there is no mechanical impact at all.

\begin{figure}[ptb]
\includegraphics[scale=0.55]{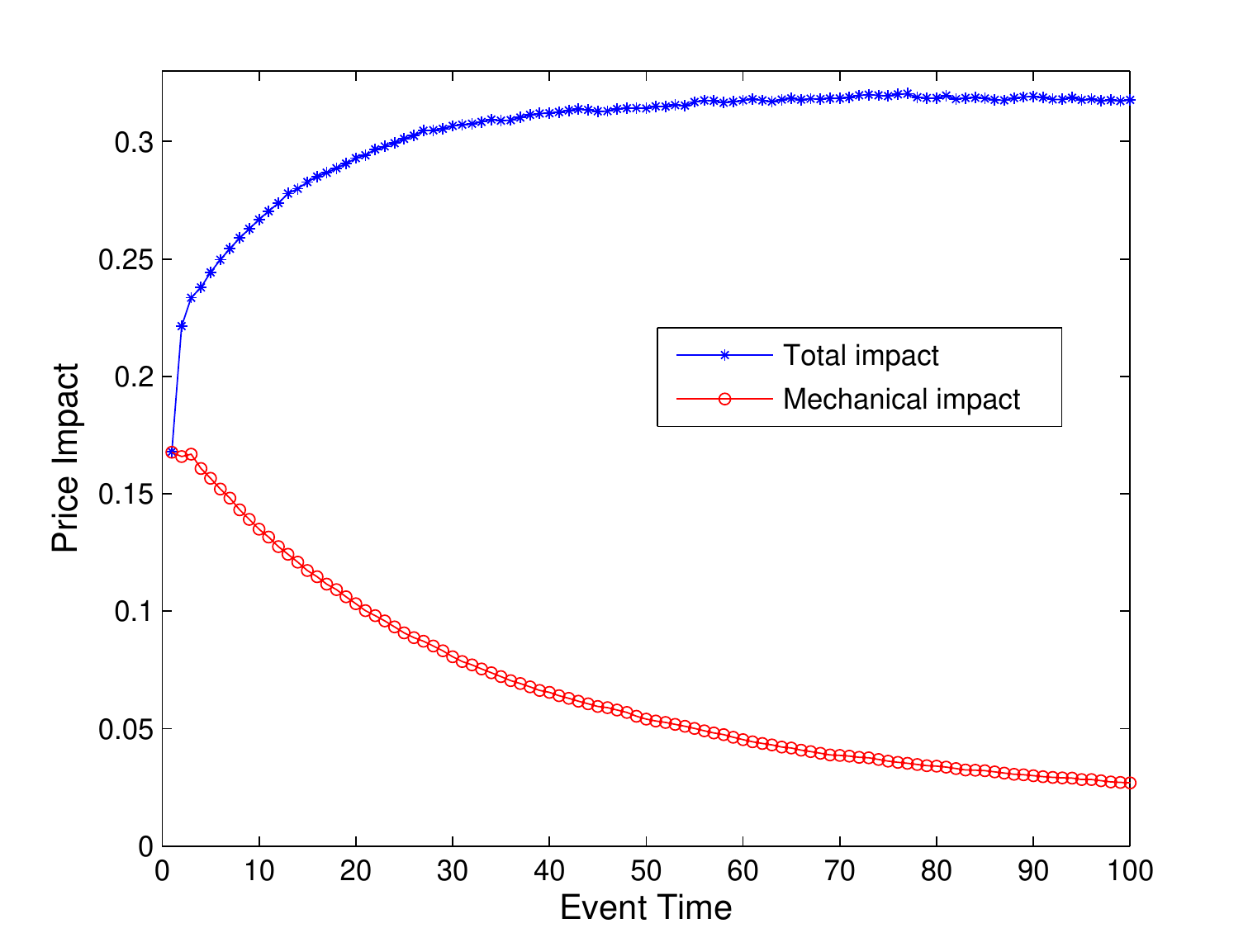}
\caption{Average mechanical impact $ E_t[\epsilon_t \Delta p^M_\tau (t)]$ (red squares) and total impact $E_t[\epsilon_t \Delta p^T_\tau (t)]$ (blue stars), in units of the average spread, plotted as a function of number of the order sequence  \label{avImpact} for the LSE stock AZN.}
\end{figure}

Despite the extreme variability, when an average is taken over time a consistent pattern emerges.  By definition for $\tau = 1$ the impact is entirely mechanical, since the only order that can affect the price is the reference event $\omega_t$.   After $\tau = 1$, however, the mechanical impact decays, so that on average by the time of the next transaction it is roughly half of its initial value, i.e. it is half of the total impact.   In the limit as $\tau \to \infty$, for the stocks investigated in the LSE, to a good approximation for large $\tau$ the mechanical impact decays to zero as a power law $\Delta p^M_\tau (t) \sim \tau^\alpha_M$, with exponent $\alpha_M \approx 1.6$.   See the example given in Figure~\ref{avImpact}.
For event time the total impact and mechanical impact are by definition the same at $\tau = 1$.  This is because in moving from $\tau = 0$ to $\tau = 1$ the only event that affects the price is the reference event $\omega_t$ --  alterations in $\Omega_{t+1}$ cannot effect $\Delta p^T_1$.  For larger values of $\tau$ the mechanical impact decreases and the informational impact increases.  As the example shows, over the timescale shown here (100 orders), when measured in units of the average spread, the mechanical impact is initially about $0.17$, and then decays monotonically toward zero. In contrast the total impact increases toward what appears to be an asymptotically constant value slightly greater than $0.3$.  This the source of our statement that the initial value of mechanical impact is about half the asymptotic value of the total impact.

In thinking about this we should stress a few points.  By requiring that any associated alterations of orders be considered information, we have taken a very strict definition of mechanical impact.  Within our definition of informational there are two fundamentally different ways in which placing or removing an order can be correlated with the placement or removal of other orders.  One is that placing or removal of an order causes a change in the placement or removal of another order.  The alternative, however, is that the placement or removal of the two orders are caused by the same external event, and are therefore correlated.  From this point of view it can appear as if one order causes the other, simply because it happens to occur a bit earlier.

While it might be surprising that the mechanical effect of order placement is completely temporary, in fact this has a trivial explanation:  Once all the orders that were originally in the book when the reference order was placed, by definition all trace of the original order's presence is gone, and so the mechanical impact is zero.  Thus the power law decay of market impact that is a observed for mechanical impact is just a reflection of the rate at which orders turn over in the order book, and is not related to the decay of the total impact discussed in section \ref{impactTemporal}.

\section*{Appendix 2: Volume fluctuations}
 
 How should one take into account volume fluctuations in the formalism developed in Section \ref{transient}? Since the volume of trades $v$ is rather broadly distributed, 
 the impact of trades could itself be highly fluctuating as well. This is not so, because large trade volumes mostly occur when a
 comparable volume is available at the opposite best price, in such a way that the impact of large trades is in fact quite similar to that
 of small trades. Mathematically, we have seen that the average impact is a slow power law function $v^\psi$, 
 or even a logarithm $\log v$. As a simplifying limit, we postulate an logarithmic impact and a broad, log-normal distribution of $v$. The resulting 
 impact of the n$^{th}$ trade $q_n=\epsilon_n \ln v_n$ is then a (zero mean) Gaussian random variable, which inherits long range correlations from the sign 
 process. Suppose, as in the {\sc mrr} model, that only the surprise in $q_n$ moves the price -- this insures {\it by construction} 
 that the price returns are uncorrelated. An elegant way to write this down mathematically is to express the (correlated) Gaussian variables $q_n$
 in terms of a set of auxiliary uncorrelated Gaussian variables $\eta_m$, through:
 \be\label{app1}
 q_n =  \sum_{m \leq n} K(n-m) \eta_m, \qquad E[\eta_m \eta_{m+\ell}]=\delta_{\ell,0}
 \ee   
 where $K(.)$ a certain kernel such that the $q_n$ have the required correlations:\footnote{The following
 equation can be uniquely solved to extract $K(\ell)$ from ${\cal C}_\ell$, using the so-called Levinson-Durbin
 algorithm for solving Toepliz systems (see e.g. \cite{Percival92}).}
 \be
 {\cal C}_\ell = E[q_n q_{n+\ell}] \equiv  \sum_{m \geq 0} K(m+n)K(m). 
 \ee
 In the case where $\cal C$ decays as $c_0\ell^{-\gamma}$ with $0 < \gamma <1$,
 it is easy to show that the asymptotic decay of $K(n)$ should also be 
 a power-law $k_0 n^{-\delta}$ with $2 \delta -1 = \gamma$ and $k_0^2=c_0 \Gamma(\delta)/\Gamma(\gamma)\Gamma(1-\delta)$. Note that 
 $1/2 < \delta < 1$. Inverting Eq. (\ref{app1}) leads to:
 \be\label{app2}
 \eta_n =  \sum_{m \leq n} Q(n-m) q_m,
 \ee 
 where $Q$ is the matrix inverse of $K$, such that $\sum_{m=0}^\ell K(\ell-m) Q(m)=\delta_{\ell,0}$. 
 For a power-law kernel $K(n) \sim k_0 n^{-\delta}$, one obtains $Q(n) \sim (\delta-1) \sin \pi \delta/(\pi k_0) n^{\delta-2} < 0$ 
 for large $n$. Note that whenever $\delta < 1$, one can show that $\sum_{m=0}^{\infty} Q(m) \equiv 0$. 
 
 When all $q_m, m \leq n-1$ are known, the corresponding $\xi_m, m \leq n-1$ can be computed; the predicted
 value of the yet unobserved $q_n$ is then given by:
 \be
 E_{n-1}[q_n] = \sum_{m < n} K(n-m) \eta_m,
 \ee
 and the surprise in $q_n$ is simply:
 \be
 q_n - E_{n-1}[q_n] = K(0) \eta_n.
 \ee
 The generalization of the price equation of motion [Eq. (\ref{surprise})] is therefore:
 \be\label{surprise2}
 m_{n+1}-m_n = \xi_n + \theta K(0) \eta_n, 
 \ee
 which, again by construction, removes any predictability in the price returns. From this equation of motion
 one can derive $G_0(\ell)$ and ${\cal R}_\ell$.\footnote{We now define $G_0$ as the impact of the $q$'s
 on the price.}
 From the expression of the $\eta$s in terms of the $q$s, one finds:
 \be\label{exactG0}
 G_0(\ell) \equiv \theta K(0) \sum_{m=0}^{\ell-1} Q(m) = - \theta K(0) \sum_{m=\ell}^{\infty} Q(m).
 \ee
 Using the above asymptotic estimate of $Q(.)$,we finally obtain
 \be
 G_0(\ell) \sim_{\ell \gg 1} \theta \frac{\sin (\pi \delta) K(0)}{\pi k_0} \ell^{\delta - 1} \equiv \Gamma_0 \ell^{-\beta}.
 \ee
 Identifying the exponents  leads to $\beta = 1 -\delta = {1-\gamma}/{2}$, recovering the above equality. 
 The quantity $\theta$, relating surprise in order flow to price changes, measures the so-called ``information content'' of the trades.
 It can be measured from empirical data using the above relation between prefactors.
 
 Finally, from Eq.(\ref{surprise2}), one finds the full impact function:
 \be
 {\cal R}_\ell= E[(m_{n+\ell}-m_n) q_n] = \theta K(0)^2, \quad \forall \ell
 \ee
 i.e. a completely {\it flat} impact function, independent of $\ell$, as in the simplified {\sc mrr} model decribed above. However, if we 
 assume with {\sc mrr} that the fundamental price, rather than the midpoint, is impacted by the surprise in $q_n$, we find that the full 
 impact function is again given by Eq. (\ref{RellMRR}): ${\cal R}_\ell=\theta[1 - C_\ell]$, which increases with $\ell$.

\section*{Appendix 3: The Bid-Ask spread in the MRR model}

A complementary point of view to that given in the main text is to analyze the cost of limit orders within the {\sc mrr} model. The following argument is interesting because it can be, in essence, generalized to more complex cases as well. Suppose one wants to trade at a random instant in time. Compared to the initial mid-point value, the average execution cost of an infinitesimal buy limit order is given by:
\be\label{limitmrr}
{\cal C}_L = \frac{1}{2} \left(-\frac{S}{2}\right) + \frac{1}{2} \left({\cal R}_1+{\cal C}_L^+ \right):
\ee
with probability $1/2$, the order is executed right away, $S/2$ below the mid-point; otherwise, the mid-point moves on average by a quantity ${\cal R}_1$, to which must be added the cost of a limit order conditioned to the last trade being a buy,  ${\cal C}_L^+$, for which a similar equation can be obtained:
\be
{\cal C}_L^+ = \frac{1-\rho}{2} \left(-\frac{S}{2}\right) + \frac{1+\rho}{2} \left({\cal R}_1^+ +{\cal C}_L^{++}\right),
\ee
with obvious notations. Since the {\sc mrr} model is Markovian, one has ${\cal R}_1^+={\cal R}_1$ and ${\cal C}_L^{++}={\cal C}_L^{+}$, so that:
\be
{\cal C}_L^+ = -\frac{S}{2} +  \frac{1+\rho}{1-\rho} {\cal R}_1.
\ee
Plugging this last relation in Eq. (\ref{limitmrr}), we finally find:
\be
{\cal C}_L = -\frac{S}{2} +  \frac{1}{1-\rho} {\cal R}_1.
\ee
Imposing that ${\cal C}_L\equiv 0$, one recovers the {\sc mrr} relation between the spread and the asymptotic impact (see Eq. \ref{spreadmrr}).

\section*{Bibliography}
\bibliographystyle{abbrvnat}
\bibliography{jdf}

\begin{thebibliography}{123}
\expandafter\ifx\csname natexlab\endcsname\relax\def\natexlab#1{#1}\fi
\expandafter\ifx\csname url\endcsname\relax
  \def\url#1{{\tt #1}}\fi

\bibitem[Almgren et~al.(2005)Almgren, Thum, Hauptmann, and Li]{Almgren05}
R.~Almgren, C.~Thum, H.~L. Hauptmann, and H. Li.
\newblock Equity market impact.
\newblock {\em Risk}, July, 2005.

\bibitem[Ane and Geman(2000)]{Ane00}
T.~Ane and H.~Geman.
\newblock Order flow, transaction clock, and normality of asset returns.
\newblock {\em The Journal of Finance}, 55\penalty0 (5):\penalty0 2259--2284,
  2000.

\bibitem[Bak et~al.(1997)Bak, Paczuski, and Shubik]{Bak97}
P.~Bak, M.~Paczuski, and M.~Shubik.
\newblock Price variations in a stock market with many agents.
\newblock {\em Physica A-Statistical Mechanics and Its Applications},
  246\penalty0 (3-4):\penalty0 430--453, 1997.

\bibitem[Barber et~al.(2004)Barber, Lee, Liu, and Odean]{Barber04}
B.~M. Barber, Y.-T. Lee, Y.-J. Liu, and T.~Odean.
\newblock Do individual day traders make money? evidence from taiwan.
\newblock Technical report, U.C. Davis, 2004.

\bibitem[Barclay and Warner(1993)]{Barclay93}
M.~J. Barclay and J.~B. Warner.
\newblock Stealth trading and volatility.
\newblock {\em Journal of Financial Economics}, 34:\penalty0 281--305, 1993.

\bibitem[Beran(1994)]{Beran94}
J.~Beran.
\newblock {\em Statistics for Long-Memory Processes}.
\newblock Chapman \& Hall, New York, 1994.

\bibitem[Bessembinder(2003)]{Bessembinder03}
H.~Bessembinder.
\newblock Issues in assessing trade execution costs.
\newblock {\em Journal of Financial Markets}, 6\penalty0 (3):\penalty0
  233--257, 2003.

\bibitem[Biais et~al.(1997)Biais, Foucault, and Hillion]{Biais97}
B.~Biais, T.~Foucault, and P.~Hillion.
\newblock {\em Microstructure des marches financiers}.
\newblock Presses Universitaires de France. 1997.

\bibitem[Black(1971)]{Black71}
F.~Black.
\newblock Towards a fully automated exchange.
\newblock {\em Review of Financial Analysts}, 27:\penalty0 29, 1971.

\bibitem[Black(1986)]{Black86}
F.~Black.
\newblock Noise.
\newblock {\em Journal of Finance}, 41\penalty0 (3):\penalty0 529--543, 1986.

\bibitem[Bollerslev et~al.(1997)Bollerslev, Domowitz, and Wang]{Bollerslev97}
T.~Bollerslev, I.~Domowitz, and J.~Wang.
\newblock Order flow and the bid-ask spread: An empirical probability model of
  screen-based trading.
\newblock {\em Journal of Economic Dynamics and Control}, 21\penalty0
  (8-9):\penalty0 1471--1491, 1997.

\bibitem[Bouchaud and Cont(1998)]{Bouchaud98}
J.-P. Bouchaud and R.~Cont.
\newblock A langevin approach to stock market fluctuations and crashes.
\newblock {\em European Physics Journal B}, 6:\penalty0 543--550, 1998.

\bibitem[Bouchaud et~al.(2004)Bouchaud, Gefen, Potters, and Wyart]{Bouchaud04}
J.-P. Bouchaud, Y.~Gefen, M.~Potters, and M.~Wyart.
\newblock Fluctuations and response in financial markets: The subtle nature of
  ``random" price changes.
\newblock {\em Quantitative Finance}, 4\penalty0 (2):\penalty0 176--190, 2004.

\bibitem[Bouchaud et~al.(2006)Bouchaud, Kockelkoren, and Potters]{Bouchaud04b}
J.-P. Bouchaud, J.~Kockelkoren, and M.~Potters.
\newblock Random walks, liquidity molasses and critical response in financial
  markets.
\newblock {\em Quantitative Finance}, 6\penalty0 (2):\penalty0 115--123, 2006.

\bibitem[Bouchaud et~al.(2002)Bouchaud, Mezard, and Potters]{Bouchaud02}
J.-P. Bouchaud, M.~Mezard, and M.~Potters.
\newblock Statistical properties of the stock order books: empirical results
  and models.
\newblock {\em Quantitative Finance}, 2\penalty0 (4):\penalty0 251--256, 2002.

\bibitem[Campbell and Shiller(1989)]{Campbell89}
J.~Y. Campbell and R.~J. Shiller.
\newblock The dividend-price ratio and expectations of future dividends and
  discount factors.
\newblock {\em Review of Financial Studies}, 1\penalty0 (3):\penalty0 195--228,
  1989.

\bibitem[Casdagli et~al.(1991)Casdagli, Eubank, Farmer, and Gibson]{Casdagli91}
M.~Casdagli, S.~Eubank, J.~D. Farmer, and J.~Gibson.
\newblock State space reconstruction in the presence of noise.
\newblock {\em Physica D}, 51:\penalty0 52--98, 1991.

\bibitem[Challet(2007)]{Challet07}
D.~Challet.
\newblock So you are making money in financial markets. should you tell your
  friends how?
\newblock Technical report, 2007.

\bibitem[Challet and Stinchcombe(2001)]{Challet01}
D.~Challet and R.~Stinchcombe.
\newblock Analyzing and modeling 1+1d markets.
\newblock {\em Physica A}, 300\penalty0 (1-2):\penalty0 285--299, 2001.

\bibitem[Chan and Lakonishok(1993)]{Chan93}
L.~K. Chan and J.~Lakonishok.
\newblock Institutional trades and intraday stock price behavior.
\newblock {\em Journal of Financial Economics}, 33:\penalty0 173--199, 1993.

\bibitem[Chan and Lakonishok(1995)]{Chan95}
L.~K. Chan and J.~Lakonishok.
\newblock The behavior of stock prices around institutional trades.
\newblock {\em The Journal of Finance}, 50\penalty0 (4):\penalty0 1147--1174,
  1995.

\bibitem[Chiarella and Iori(2002)]{Chiarella02}
C.~Chiarella and G.~Iori.
\newblock A simulation analysis of the microstructure of double auction
  markets.
\newblock {\em Quantitative Finance}, 2:\penalty0 346--353, 2002.

\bibitem[Chordia and Subrahmanyam(2004)]{Chordia04}
T.~Chordia and A.~Subrahmanyam.
\newblock Order imbalance and individual stock returns: Theory and evidence.
\newblock {\em Journal of Financial Markets}, 72:\penalty0 485--518, 2004.

\bibitem[Clark(1973)]{Clark73}
P.~K. Clark.
\newblock Subordinated stochastic process model with finite variance for
  speculative prices.
\newblock {\em Econometrica}, 41\penalty0 (1):\penalty0 135--155, 1973.

\bibitem[Cohen et~al.(1985)Cohen, Conroy, and Maier]{Cohen85}
K.~J. Cohen, R.~M. Conroy, and S.~F. Maier.
\newblock Order flow and the quality of the market.
\newblock In Y.~Amihud, T.~Ho, and R.~A. Schwartz, editors, {\em Market Making
  and the Changing Structure of the Securities Industry}, pages 93--110. Rowman
  \& Littlefield, Lanham, 1985.


\bibitem[P. Curty and M. Marsili (2006)]{Curty06}
P. Curty and M. Marsili.
\newblock Phase coexistence in a forecasting game
\newblock {\em Journal of Statistical Mechanics}, P03013, 2006.

\bibitem[Cutler et~al.(1989)Cutler, Poterba, and Summers]{Cutler89}
D.~M. Cutler, J.~M. Poterba, and L.~H. Summers.
\newblock What moves stock prices?
\newblock {\em The Journal of Portfolio Management}, 15\penalty0 (3):\penalty0
  4--12, 1989.

\bibitem[Daniels et~al.(2003)Daniels, Farmer, Gillemot, Iori, and
  Smith]{Daniels03}
M.~G. Daniels, J.~D. Farmer, L.~Gillemot, G.~Iori, and D.~E. Smith.
\newblock Quantitative model of price diffusion and market friction based on
  trading as a mechanistic random process.
\newblock {\em Physical Review Letters}, 90\penalty0 (10):\penalty0
  108102--108104, 2003.

\bibitem[Delong et~al.(1990)Delong, Shleifer, Summers, and Waldmann]{Delong90}
J.~B. Delong, A.~Shleifer, L.~H. Summers, and R.~J. Waldmann.
\newblock Positive feedback and destabilizing rational speculation.
\newblock {\em Journal of Finance}, 45:\penalty0 379--395, 1990.

\bibitem[Ding et~al.(1993)Ding, Granger, and Engle]{Ding93}
Z.~Ding, C.~W.~J. Granger, and R.~F. Engle.
\newblock A long memory property of stock returns and a new model.
\newblock {\em Journal of Empirical Finance}, 1\penalty0 (1):\penalty0 83--106,
  1993.

\bibitem[Domowitz and Wang(1994)]{Domowitz94}
I.~Domowitz and J.~Wang.
\newblock Auctions as algorithms: computerized trade execution and price
  discovery.
\newblock {\em Journal of Economic Dynamics and Control}, 18\penalty0
  (1):\penalty0 29--60, 1994.

\bibitem[Eisler et~al.(2008)Eisler, Bouchaud, and Kockelkoren]{Eisler08}
Z.~Eisler, J.-P. Bouchaud, and J.~Kockelkoren.
\newblock Technical report, 2008.
\newblock in preparation.

\bibitem[Eisler and Kertecz(2006)]{Eisler06}
Z.~Eisler and J.~Kertecz.
\newblock Size matters, some stylized facts of the market revisited.
\newblock {\em European Journal of Physics}, B51:\penalty0 145--154, 2006.

\bibitem[Eliezer and Kogan(1998)]{Eliezer98}
D.~Eliezer and I.~I. Kogan.
\newblock Scaling laws for the market microstructure of the interdealer broker
  markets.
\newblock Technical report, http://www.arxiv.org/abs/cond-mat/9808240, 1998.

\bibitem[Engle and Rangel(2005)]{Engle05}
R.~Engle and J.~Rangel.
\newblock The spline {GARCH} model for unconditioanl volatlity and its global
  macroeconomic causes.
\newblock Technical report, NYU and UCSD, 2005.

\bibitem[Evans and Lyons(2002)]{Evans02}
M.~D.~D. Evans and R.~K. Lyons.
\newblock Order flow and exchange rate dynamics.
\newblock {\em Journal of Political Economy}, 110\penalty0 (1):\penalty0
  170--180, 2002.

\bibitem[Farmer et~al.(2006)Farmer, Gerig, Lillo, and Mike]{Farmer06}
J.~Farmer, A.~Gerig, F.~Lillo, and S.~Mike.
\newblock Market efficiency and the long-memory of supply and demand: Is price
  impact variable and permanent or fixed and temporary?
\newblock {\em Quantitative Finance}, 6\penalty0 (2):\penalty0 107--112, 2006.

\bibitem[Farmer(2002)]{Farmer02}
J.~D. Farmer.
\newblock Market force, ecology and evolution.
\newblock {\em Industrial and Corporate Change}, 11\penalty0 (5):\penalty0
  895--953, 2002.

\bibitem[Farmer et~al.(2004)Farmer, Gillemot, Lillo, Mike, and Sen]{Farmer04b}
J.~D. Farmer, L.~Gillemot, F.~Lillo, S.~Mike, and A.~Sen.
\newblock What really causes large price changes?
\newblock {\em Quantitative Finance}, 4\penalty0 (4):\penalty0 383--397, 2004.

\bibitem[Farmer and Lillo(2004)]{Farmer04}
J.~D. Farmer and F.~Lillo.
\newblock On the origin of power laws in financial markets.
\newblock {\em Quantitative Finance}, 314:\penalty0 7--10, 2004.

\bibitem[Farmer et~al.(2005)Farmer, Patelli, and Zovko]{Farmer05}
J.~D. Farmer, P.~Patelli, and I.~Zovko.
\newblock The predictive power of zero intelligence in financial markets.
\newblock {\em Proceedings of the National Academy of Sciences of the United
  States of America}, 102\penalty0 (6):\penalty0 2254--2259, 2005.

\bibitem[Farmer and Zamani(2007)]{Farmer07}
J.~D. Farmer and N.~Zamani.
\newblock Mechanical vs. informational components of price impact.
\newblock {\em European Physical Journal B.}, 55:\penalty0 1899--2000., 2007.

\bibitem[Fisher(1983)]{Fisher83}
F.~M. Fisher.
\newblock {\em Disequilibrium Foundations of Equilibrium Economics}.
\newblock Cambridge University Press, Cambridge, 1983.

\bibitem[Foucault et~al.(2005)Foucault, Kadan, and Kandel]{Foucault05}
T.~Foucault, O.~Kadan, and E.~Kandel.
\newblock Limit order book as a market for liquidity.
\newblock {\em The Review of Financial Studies}, 18\penalty0 (4):\penalty0
  1171--1217, 2005.

\bibitem[Gabaix et~al.(2006)Gabaix, Gopikrishnan, Plerou, and
  Stanley]{Gabaix06}
X.~Gabaix, P.~Gopikrishnan, V.~Plerou, and H.~Stanley.
\newblock Institutional investors and stock market volatility.
\newblock {\em Quarterly Journal of Economics}, 121:\penalty0 461--504, 2006.

\bibitem[Gabaix et~al.(2003)Gabaix, Gopikrishnan, Plerou, and
  Stanley]{Gabaix03}
X.~Gabaix, P.~Gopikrishnan, V.~Plerou, and H.~E. Stanley.
\newblock A theory of power-law distributions in financial market fluctuations.
\newblock {\em Nature}, 423:\penalty0 267--270, 2003.

\bibitem[Gallagher and Looi(2006)]{Gallagher06}
D.~Gallagher and A.~Looi.
\newblock Trading behavior and the performance of daily institutional trades.
\newblock {\em Accounting and Finance}, 46:\penalty0 125--147, 2006.

\bibitem[Gerig(2007)]{Gerig07}
A.~Gerig.
\newblock {\em A Theory for Market Impact: How Order Flow Affects Stock Price}.
\newblock PhD thesis, University of Illinois, 2007.

\bibitem[Gillemot et~al.(2006)Gillemot, Farmer, and Lillo]{Gillemot05}
L.~Gillemot, J.~D. Farmer, and F.~Lillo.
\newblock There's more to volatility than volume.
\newblock {\em Quantitative Finance}, 6\penalty0 (5):\penalty0 371--384, 2006.

\bibitem[Glosten(1994)]{Glosten94}
L.~Glosten.
\newblock Is the electronic limit order book inevitable?
\newblock {\em Journal of Finance}, 49\penalty0 (4):\penalty0 1127--1161, 1994.

\bibitem[Glosten and Milgrom(1985)]{Glosten85}
L.~R. Glosten and P.~R. Milgrom.
\newblock Bid, ask, and transaction prices in a specialist market with
  heterogeneously informed traders.
\newblock {\em Journal of Financial Economics}, 14\penalty0 (1):\penalty0
  71--100, 1985.

\bibitem[Gopikrishnan et~al.(1998)Gopikrishnan, Meyer, Amaral, and
  Stanley]{Gopikrishnan98}
P.~Gopikrishnan, M.~Meyer, L.~Amaral, and H.~Stanley.
\newblock Inverse cubic law for the probability distribution of stock price
  variations.
\newblock {\em European Physical Journal B.}, 3:\penalty0 139--140, 1998.

\bibitem[Gopikrishnan et~al.(2000)Gopikrishnan, Plerou, Gabaix, and
  Stanley]{Gopikrishnan00}
P.~Gopikrishnan, V.~Plerou, X.~Gabaix, and H.~E. Stanley.
\newblock Statistical properties of share volume traded in financial markets.
\newblock {\em Physical Review E}, 62\penalty0 (4):\penalty0 R4493--R4496,
  2000.
\newblock Part A.

\bibitem[Granger and Joyeux(1980)]{Granger80}
C.~W.~J. Granger and R.~Joyeux.
\newblock An introduction to long-range time series models and fractional
  differencing.
\newblock {\em Journal of Time Series Analysis}, 1:\penalty0 15--30, 1980.

\bibitem[Grossman(1989)]{Grossman89}
S.~J. Grossman.
\newblock {\em The Informational Role of Prices}.
\newblock MIT Press, Cambridge, 1989.

\bibitem[Grossman and Stiglitz(1980)]{Grossman80}
S.~J. Grossman and J.~E. Stiglitz.
\newblock On the impossibility of informationally efficient markets.
\newblock {\em The American Economic Review}, 70\penalty0 (3):\penalty0
  393--408, 1980.

\bibitem[Gu, Chen, and Zhou(2007)]{Gu07}
G.-F. Gu, W. Chen, and W.-X. Zhou. 
\newblock Quantifying bid-ask spreads in the Chinese stock market using limit-order book data - Intraday pattern, probability distribution, long memory, and multifractal nature
\newblock {\em European Physical Journal B} 57:\penalty0 81-87, 2007.

\bibitem[Guedj and Bouchaud(2005)]{Guedj05}
O.~Guedj and J.-P. Bouchaud.
\newblock Experts earnings forecasts: Bias, herding and gossamer information.
\newblock {\em International Journal of Theoretical and Applied Finance},
  8:\penalty0 933, 2005.

\bibitem[Handa and Schwartz(1996)]{Handa96}
P.~Handa and R.~A. Schwartz.
\newblock Limit order trading.
\newblock {\em Journal of Finance}, 51\penalty0 (5):\penalty0 1835--61, 1996.

\bibitem[Harris and Hasbrouck(1996)]{Harris96}
L.~Harris and J.~Hasbrouck.
\newblock Market vs. limit orders: The superdot evidence and order submission
  strategy.
\newblock {\em The Journal of Financial and Quantitative Financial Analysis},
  31\penalty0 (2):\penalty0 213--231, 1996.

\bibitem[Hasbrouck(1988)]{Hasbrouck88}
J.~Hasbrouck.
\newblock Trades, quotes, inventories, and information.
\newblock {\em Journal of Financial Economics}, 22\penalty0 (2):\penalty0
  229--52, 1988.

\bibitem[Hasbrouck(2007)]{Hasbrouck07}
J.~Hasbrouck.
\newblock {\em Empirical market microstructure: The institutions, economics and
  econometrics of securities trading}.
\newblock Oxford University Press, Oxford, 2007.

\bibitem[Hopman(2006)]{Hopman02}
C.~Hopman.
\newblock Do supply and demand drive stock prices?
\newblock {\em Quantitative Finance}, 2006.
\newblock To appear.

\bibitem[Hosking(1981)]{Hosking81}
J.~R.~M. Hosking.
\newblock Fractional differencing.
\newblock {\em Biometrika}, 68:\penalty0 165--176, 1981.

\bibitem[Joulin et~al.(2008)Joulin, Lefevre, Grunberg, and
  Bouchaud]{Bouchaud08}
A.~Joulin, A.~Lefevre, D.~Grunberg, and J.-P. Bouchaud.
\newblock Stock price jumps: news and volume play a minor role.
\newblock Technical report, 2008.

\bibitem[Keim and Madhavan(1996)]{Keim96}
D.~B. Keim and A.~Madhavan.
\newblock The upstairs market for large-block transactions:analysis and
  measurement of price effects.
\newblock {\em The Review of Financial Studies}, 9\penalty0 (1):\penalty0
  1--36, 1996.

\bibitem[Kempf and Korn(1999)]{Kempf99}
A.~Kempf and O.~Korn.
\newblock Market depth and order size.
\newblock {\em Journal of Financial Markets}, 2\penalty0 (1):\penalty0 29--48,
  1999.

\bibitem[Kyle(1985)]{Kyle85}
A.~S. Kyle.
\newblock Continuous auctions and insider trading.
\newblock {\em Econometrica}, 53:\penalty0 1315--1335, 1985.

\bibitem[Laspada et~al.(2008)Laspada, Farmer, and Lillo]{LaSpada08}
G.~La Spada, J.~D. Farmer, and F.~Lillo.
\newblock The non-trivial random walk of stock prices.
\newblock {\em European Journal of Physics, B.}, 2008.
\newblock To appear.

\bibitem[LeBaron and Yamamoto(2007)]{Lebaron07}
B.~LeBaron and R.~Yamamoto.
\newblock Long-memory in an order-driven market.
\newblock {\em Physica A}, 383:\penalty0 85--89, 2007.

\bibitem[Lillo(2007)]{Lillo07b}
F.~Lillo.
\newblock Limit order placement as an utility maximization problem and the
  origin of power law distribution of limit order prices.
\newblock {\em European Journal of Physics}, 55:\penalty0 453--459, 2007.

\bibitem[Lillo and Farmer(2004)]{Lillo03c}
F.~Lillo and J.~D. Farmer.
\newblock The long memory of the efficient market.
\newblock {\em Studies in Nonlinear Dynamics \& Econometrics}, 8\penalty0 (3),
  2004.

\bibitem[Lillo and Farmer(2005)]{Lillo05}
F.~Lillo and J.~D. Farmer.
\newblock The key role of liquidity fluctuations in determining large price
  fluctuations.
\newblock {\em Fluctuations and Noise Letters}, 5:\penalty0 L209--L216, 2005.

\bibitem[Lillo et~al.(2008{\natexlab{a}})Lillo, Farmer, and A.]{Lillo08}
F.~Lillo, J.~D. Farmer, and G.~A.
\newblock A theory for aggregate market impact.
\newblock Technical report, Santa Fe Institute, 2008{\natexlab{a}}.
\newblock Unpublished research.


\bibitem[Lillo et~al.(2003{\natexlab{b}})Lillo, Farmer, and Mantegna]{Lillo03d}
F.~Lillo, J.~D. Farmer, and R.~N. Mantegna.
\newblock Master curve for price impact function.
\newblock {\em Nature}, 421:\penalty0 129--130, 2003{\natexlab{b}}.

\bibitem[Lillo and Mantegna(2003)]{Lillo03}
F.~Lillo and R.~N. Mantegna.
\newblock Power-law relaxation in a complex system: Omori law after a financial
  market crash.
\newblock {\em Physical Review E}, 68\penalty0 (1), 2003.
\newblock Part 2.

\bibitem[Lillo et~al.(2005)Lillo, Mike, and Farmer]{Lillo05b}
F.~Lillo, S.~Mike, and J.~D. Farmer.
\newblock Theory for long memory in supply and demand.
\newblock {\em Physical Review E}, 7106\penalty0 (6):\penalty0 066122, 2005.

\bibitem[Lillo et~al.(2008{\natexlab{b}})Lillo, Moro, Vaglica, and
  Mantegna]{Lillo07}
F.~Lillo, E.~Moro, G.~Vaglica, and R.~Mantegna.
\newblock Specialization and herding behavior of trading firms in a financial
  market.
\newblock {\em New Journal of Physics}, 10:\penalty0 043019,
  2008{\natexlab{b}}.

\bibitem[Lo(1991)]{Lo91}
A.~W. Lo.
\newblock Long-term memory in stock market prices.
\newblock {\em Econometrica}, 59\penalty0 (5):\penalty0 1279--1313, 1991.

\bibitem[Lobato and Velasco(2000)]{Lobato00}
I.~N. Lobato and C.~Velasco.
\newblock Long-memory in stock-market trading volume.
\newblock {\em Journal of Business \& Economic Statistics}, 18:\penalty0
  410--427, 2000.

\bibitem[Luckock(2003)]{Luckock03}
H.~Luckock.
\newblock A steady-state model of the continuous double auction.
\newblock {\em Quantitative Finance}, 39:\penalty0 385--404, 2003.

\bibitem[Lux(1996)]{Lux96}
T.~Lux.
\newblock The stable paretian hypothesis and the frequency of large returns: an
  examination of major german stocks.
\newblock {\em Applied Financial Economics}, 6\penalty0 (6):\penalty0 463--475,
  1996.

\bibitem[Lyons(2001)]{Lyons01}
R.~Lyons.
\newblock {\em The microstructure approach to foreign exchange rates}.
\newblock MIT Press, Cambridge, MA, 2001.

\bibitem[Madhavan(2000)]{Madhavan00}
Madhavan.
\newblock Market microstructure: A survey.
\newblock {\em Journal of Financial Markets}, 3:\penalty0 205, 2000.

\bibitem[Madhavan et~al.(1997)Madhavan, Richardson, and Roomans]{Madhavan97b}
A.~Madhavan, M.~Richardson, and M.~Roomans.
\newblock Why do security prices change? a transaction-level analysis of nyse
  stocks.
\newblock {\em The Review of Financial Studies}, 10\penalty0 (4):\penalty0
  1035--1064, 1997.

\bibitem[Mandelbrot(1963)]{Mandelbrot63}
B.~Mandelbrot.
\newblock The variation of certain speculative prices.
\newblock {\em The Journal of Business}, 36\penalty0 (4):\penalty0 394--419,
  1963.

\bibitem[Mandelbrot and van Ness(1968)]{Mandelbrot68}
B.~Mandelbrot and J.~W. van Ness.
\newblock Fractional brownian motions, fractional noises and applications.
\newblock {\em SIAM Review}, 10\penalty0 (4):\penalty0 422--437, 1968.

\bibitem[Mandelbrot and Taylor(1967)]{Mandelbrot67}
B.~B. Mandelbrot and H.~M. Taylor.
\newblock On distribution of stock price differences.
\newblock {\em Operations Research}, 15\penalty0 (6):\penalty0 1057--1062,
  1967.

\bibitem[Maslov(2000)]{Maslov00}
S.~Maslov.
\newblock Simple model of a limit order-driven market.
\newblock {\em Physica A-Statistical Mechanics and Its Applications},
  278\penalty0 (3-4):\penalty0 571--578, 2000.

\bibitem[Mendelson(1982)]{Mendelson82}
H.~Mendelson.
\newblock Market behavior in a clearing house.
\newblock {\em Econometrica}, 50\penalty0 (6):\penalty0 1505--1524, 1982.

\bibitem[Mike and Farmer(2008)]{Mike05}
S.~Mike and J.~Farmer.
\newblock An empirical behavioral model of liquidity and volatility.
\newblock {\em Journal of Economic Dynamics and Control}, 32:\penalty0
  200--234, 2008.

\bibitem[Milgrom and Stokey(1982)]{Milgrom82}
P.~Milgrom and N.~Stokey.
\newblock Information trade and common knowledge.
\newblock {\em Journal of Economic Theory}, 26\penalty0 (1):\penalty0 17--27,
  1982.

\bibitem[Odean(1999)]{Odean99}
T.~Odean.
\newblock Do investors trade too much?
\newblock {\em The American Economic Review}, 89\penalty0 (5):\penalty0
  1279--1298, 1999.

\bibitem[O'Hara(1995)]{Ohara95}
M.~O'Hara.
\newblock {\em Market Microstructure Theory}.
\newblock Blackwell, Cambridge, 1995.

\bibitem[Packard et~al.(1980)Packard, Crutchfield, Farmer, and Shaw]{Packard80}
N.~H. Packard, J.~P. Crutchfield, J.~D. Farmer, and R.~S. Shaw.
\newblock Geometry from a time seriees.
\newblock {\em Physical Review Letters}, 45\penalty0 (9):\penalty0 712--716,
  1980.

\bibitem[Percival(1992)]{Percival92}
D.~Percival.
\newblock Simulating gaussian random processes with specified spectra.
\newblock {\em Computing Science and Statistics}, 24:\penalty0 534, 1992.

\bibitem[Plerou et~al.(2002)Plerou, Gopikrishnan, Gabaix, and
  Stanley]{Plerou02}
V.~Plerou, P.~Gopikrishnan, X.~Gabaix, and H.~E. Stanley.
\newblock Quantifying stock price response to demand fluctuations.
\newblock {\em Physical Review E}, 66\penalty0 (2):\penalty0 article no.
  027104, 2002.

\bibitem[Plerou et~al.(2004)Plerou, Gopikrishnan, Gabaix, and
  Stanley]{Plerou04}
V.~Plerou, P.~Gopikrishnan, X.~Gabaix, and H.~E. Stanley.
\newblock On the origin of power laws in financial markets.
\newblock {\em Quantitative Finance}, 4:\penalty0 11--15, 2004.

\bibitem[Plerou et~al.(2005)Plerou, Gopikrishnan, and Stanley]{Plerou05}
V.~Plerou, P.~Gopikrishnan, and H.~Stanley.
\newblock Quantifying fluctuations in market liquidity: Analysis of the bid-ask
  spread.
\newblock {\em Physical Review E.}, 71:\penalty0 046131--9, 2005.

\bibitem[Ponzi et~al.(2008)Ponzi, Lillo, and Mantegna]{Ponzi06}
A.~Ponzi, F.~Lillo, and R.~Mantegna.
\newblock Market reaction to temporary liquidity crises and the permanent
  market impact.
\newblock preprint at arXiv:physics/0608032v1.

\bibitem[Roll(1984)]{Roll84b}
R.~Roll.
\newblock Orange juice and weather.
\newblock {\em American Economic Review}, pages 861--880, 1984.

\bibitem[Rosenow(2002)]{Rosenow02}
B.~Rosenow.
\newblock Fluctuations and market friction in financial trading.
\newblock {\em International Journal of Modern Physics C}, 13\penalty0
  (3):\penalty0 419--425, 2002.

\bibitem[Ross(2004)]{Ross04}
S.~Ross.
\newblock {\em Neoclassical Finance}.
\newblock Princeton University Press, 2004.

\bibitem[Rosu(2005)]{Rosu05}
I.~Rosu.
\newblock A dynamic model of the limit order book.
\newblock Technical report, 2005.

\bibitem[Sandas(2001)]{Sandas01}
P.~Sandas.
\newblock Adverse selection and comparative market making:empirical evidence
  from a limit order market.
\newblock {\em The Review of Financial Studies}, 14\penalty0 (3):\penalty0
  705--734, 2001.

\bibitem[Sebenius and Geanakoplos(1983)]{Sebenius83}
J.~K. Sebenius and J.~Geanakoplos.
\newblock Dont bet on it - contingent agreements with asymmetric information.
\newblock {\em Journal of the American Statistical Association}, 78\penalty0
  (382):\penalty0 424--426, 1983.

\bibitem[Shiller(1981)]{Shiller81}
R.~J. Shiller.
\newblock Do stock prices move too much to be justified by subsequent changes
  in dividends?
\newblock {\em American Economic Review}, 71\penalty0 (3):\penalty0 421--436,
  1981.

\bibitem[Shleifer(2000)]{Shleifer00}
A.~Shleifer.
\newblock {\em Clarendon Lectures: Inefficient Markets}.
\newblock Oxford University Press, Oxford, 2000.

\bibitem[Slanina(2001)]{Slanina01}
F.~Slanina.
\newblock Mean-field approximation for a limit order driven market model.
\newblock {\em Physical Review E}, 64\penalty0 (5):\penalty0 article no.056136,
  2001.
\newblock Part 2.

\bibitem[Smith et~al.(2003)Smith, Farmer, Gillemot, and Krishnamurthy]{Smith03}
E.~Smith, J.~D. Farmer, L.~Gillemot, and S.~Krishnamurthy.
\newblock Statistical theory of the continuous double auction.
\newblock {\em Quantitative Finance}, 3\penalty0 (6):\penalty0 481--514, 2003.

\bibitem[Stoll(2000)]{Stoll00}
H.~Stoll.
\newblock Friction.
\newblock {\em Journal of Finance}, 55:\penalty0 1479, 2000.

\bibitem[Stoll(1978)]{Stoll78}
H.~R. Stoll.
\newblock The supply of dealer services in securities markets.
\newblock {\em Journal of Finance}, 33\penalty0 (4):\penalty0 1133--51, 1978.

\bibitem[Takens(1981)]{Takens81}
F.~Takens.
\newblock Detecting strange attractors in turbulence.
\newblock In D.~Rand and L.-S. Young, editors, {\em Dynamical Systems and
  Turbulence}, volume 898, pages 366--381. Springer-Verlag, Berlin, 1981.

\bibitem[Torre(1997)]{Torre97}
N.~Torre.
\newblock {\em BARRA Market Impact Model Handbook}.
\newblock BARRA Inc., Berkeley, 1997.

\bibitem[Vaglica et~al.(2008)Vaglica, Lillo, Moro, and Mantegna]{Vaglica07}
G.~Vaglica, F.~Lillo, E.~Moro, and R.~Mantegna.
\newblock Scaling laws of strategic behavior and size heterogeneity in agent
  dynamics.
\newblock {\em Physical Review E.}, 77:\penalty0 0036110, 2008.

\bibitem[Weber and Rosenow(2005)]{Weber05}
P.~Weber and B.~Rosenow.
\newblock Order book approach to price impact.
\newblock {\em Quantitative Finance}, 5:\penalty0 357--364, 2005.

\bibitem[Weber and Rosenow(2006)]{Weber04}
P.~Weber and B.~Rosenow.
\newblock Large stock price changes: volume or liquidity?
\newblock {\em Quantitative Finance}, 6\penalty0 (1):\penalty0 7--14, 2006.

\bibitem[Wiesinger et~al.(2008)Wiesinger, Eisler, Joulin, and
  Bouchaud]{Wiesinger08}
Z.~Wiesinger, Z.~Eisler, A.~Joulin, and J.-P. Bouchaud.
\newblock Technical report, 2008.
\newblock In preparation.

\bibitem[Wyart et~al.(2006)Wyart, Bouchaud, Kockelkoren, Potters, and
  Vettorazzo]{Wyart06}
M.~Wyart, J.-P. Bouchaud, J.~Kockelkoren, M.~Potters, and M.~Vettorazzo.
\newblock Relation between bid-ask spread, impact and volatility in double
  auction markets.
\newblock Technical report, 2006.

\bibitem[Zamani and Farmer(2008)]{Zamani08}
N.~Zamani and J.~D. Farmer.
\newblock Decomposition of mechanical and informational components of
  orderflow.
\newblock Technical report, Unfinished manuscript, 2008.

\bibitem[Zawadowski et~al.(2006)Zawadowski, Andor, and Kertecz]{Zawadowski06}
A.~Zawadowski, G.~Andor, and J.~Kertecz.
\newblock Short-term market reaction after extreme price changes of liquid
  stocks.
\newblock {\em Quantitative Finance}, 4:\penalty0 283--295, 2006.

\bibitem[Zovko and Farmer(2002)]{Zovko02}
I.~Zovko and J.~D. Farmer.
\newblock The power of patience; a behavioral regularity in limit order
  placement.
\newblock {\em Quantitative Finance}, 2\penalty0 (5):\penalty0 387--392, 2002.

\bibitem[Zovko and Farmer(2007)]{Zovko07}
I.~Zovko and J.~D. Farmer.
\newblock Correlations and clustering in the trading of members of the {L}ondon
  {S}tock {E}xchange.
\newblock In S.~Abe, T.~Mie, H.~Herrmann, P.~Quarati, A.~Rapisarda, and
  C.~Tsallis, editors, {\em Complexity, Metastability and Nonextensivity: An
  International Conference}, AIP Conference Proceedings. Springer, 2007.

\bibitem[Zumbach(2004)]{Zumbach04}
G.~Zumbach.
\newblock How the trading activity scales with the company sizes in the ftse
  100.
\newblock {\em Quantitative Finance}, 4:\penalty0 441, 2004.

\end{thebibliography}

\end{document}